\documentclass[aps,pre,notitlepage,10pt,twocolumn,superscriptaddress,nofootinbib]{revtex4-2}
\usepackage{hyperref}
\usepackage{xcolor}
\hypersetup{
    colorlinks,
    linkcolor={red!50!black},
    citecolor={blue!50!black},
    urlcolor={blue!80!black}
}
\usepackage{graphicx}
\usepackage{amsmath}
\usepackage{amssymb}
\usepackage{multirow}
\usepackage{wrapfig,lipsum}
\usepackage{float}

\usepackage{enumitem}

\DeclareSymbolFont{rsfs}{U}{rsfs}{m}{n}
\DeclareSymbolFontAlphabet{\mathscrsfs}{rsfs}

\usepackage{graphicx}
\usepackage[]{epsfig}
\usepackage{bm}
\usepackage{color}
\usepackage{dsfont}

\newcommand{\beq}{\begin{equation}}
\newcommand{\eeq}{\end{equation}}
\newcommand{\beqd}{\begin{displaymath}}
\newcommand{\eeqd}{\end{displaymath}}
\newcommand{\beqa}{\begin{eqnarray}}
\newcommand{\eeqa}{\end{eqnarray}}

\renewcommand{\b}{\beta}

\newcommand{\sign}{{\rm sign}}

\newcommand{\comment}[1]{}

\newcommand{\hR}{\hat{R}}
\newcommand{\hT}{\hat{T}}
\newcommand{\hX}{\hat{\chi}}

\newcommand{\Tau}{\mathcal{T}}

\def\LL{{\cal L}}

\begin{document}
\author{Patrick Charbonneau}
\affiliation{Department of Physics, Duke University, Durham, North Carolina 27708, USA}
\affiliation{Department of Chemistry, Duke University, Durham, North Carolina 27708, USA}
\author{Giampaolo Folena}
 \affiliation{Institute for Cross-disciplinary Physics and Complex Systems IFISC (CSIC-UIB), Campus Universitat Illes Balears, 07122 Palma de Mallorca, Spain.}
\author{Enrico M. Malatesta}
 \affiliation{Department of Computing Sciences, Bocconi University, 20136 Milano, Italy}
 \affiliation{Institute for Data Science and Analytics, Bocconi University, 20136 Milano, Italy}
\author{Tommaso Rizzo}
\affiliation{Institute of Complex Systems (ISC) - CNR, Rome unit, Piazzale Aldo Moro 5, 00185 Rome, Italy}
\affiliation{Dipartimento di Fisica, Sapienza Università di Roma, 00185, Italy}
\author{Francesco Zamponi}
\affiliation{Dipartimento di Fisica, Sapienza Università di Roma, 00185, Italy}

\title{Rare Trajectories in a Prototypical Mean-field Disordered Model:\\ Insights into Landscape and Instantons}

\begin{abstract}
For disordered systems within the random first-order transition (RFOT) universality class, such as structural glasses and certain spin glasses, the role played by activated relaxation processes is rich to the point of perplexity. Over the last decades, various efforts have attempted to formalize and systematize such processes in terms of instantons similar to the nucleation droplets of first-order phase transitions. In particular, Kirkpatrick, Thirumalai, and Wolynes proposed in the late '80s an influential nucleation theory of relaxation in structural glasses. Already within this picture, however, the resulting structures are far from the compact objects expected from the classical droplet description. In addition, an altogether different type of single-particle hopping-like instantons has recently been isolated in molecular simulations. Landscape studies of mean-field spin glass models have further revealed that simple saddle crossing does not capture relaxation in these systems. We present here a landscape-agnostic study of rare dynamical events, which delineates the richness of instantons in these systems. Our work not only captures the structure of metastable states, but also identifies the point of irreversibility, beyond which activated relaxation processes become a fait accompli. An interpretation of the associated landscape features is articulated, thus charting a path toward a complete understanding of RFOT instantons.
\end{abstract}

\maketitle

\section{Introduction}

In the late '60s, Jim Langer proposed a first-principle description of the rate of decay of metastable states in $\phi^4$ field theories \cite{langer1967,langer1969} through processes later identified by Sidney Coleman as instantons~\cite{Coleman1979}. The resulting rare but nearly instantaneous jumps between two free energy minima thereby formalized and systematized the classical theory of nucleation, itself rooted in J.~Willard Gibbs' 19th-century droplet picture~(see, e.g., Ref.~\cite{GEBAUER2011564}). This description of saddle points along dynamical trajectories (or transition states) has since  had a marked impact in fields ranging from quantum field theory and gravity to chemistry.

While for simple systems time-dependent instantons are straightforwardly associated with real-space nucleation processes, for disordered systems their structure is significantly richer. For instance, in the random field Ising model instantons not only control nucleation, but also hysteresis~\cite{mueller2006,nandi2016}; in lightly disordered solids, instantons underlie the Lifshitz tail of localized states in the energy spectrum~\cite{yaida2016}; in Edwards--Anderson-like spin glass models, an instanton analysis predicts that the replica symmetry breaking transition presents a highly non-trivial lower critical dimension, $d_\ell=5/2$~\cite{Franz1994,Boettcher2005,Maiorano2018,Astuti2019}.

For disordered systems within the random first-order transition (RFOT) universality class, arguably the most complex of such classes, the role played by instantons is rich to the point of perplexity. In these models of spin and structural glasses, the number of metastable states below the temperature $T_\mathrm{d}$ of dynamical arrest is so large as to give rise to a finite contribution to the system entropy, i.e., a non-zero complexity (or configurational entropy). Identifying relaxation pathways entails finding one's way out of a particularly confusing (high-dimensional) labyrinth. 

\subsection{Instantons in glasses: state of the art}

\paragraph*{Real space droplet nucleation -} Realizing that this complexity could be a driving force for nucleation led Kirkpatrick, Thirumalai, and Wolynes~\cite{kirkpatrick1987,kirkpatrick1989} to propose an influential description of the sluggish, glassy relaxation of systems in this regime. 
This description of instantons was still based on a real-space nucleation picture, similar to ordered systems, but with entropy as the driving force for the formation of the critical droplet.
Various efforts have since attempted to formalize and systematize this picture. The Franz--Parisi potential~\cite{franz1995}, in particular, presents RFOT systems as being analogously driven toward nucleation as first-order phase transitions, and has hence served as starting point of
calculations reminiscent of Langer's field-theoretical approach~\cite{dzero2005activated,franz2005first,Dotsenko2006,dzero2009}. Interestingly, further analysis of droplet surfaces has found these structures to be markedly distinct from the compact objects described by classical nucleation theory~\cite{stevenson2010universal,biroli2017}. In parallel, an altogether different type of single-particle (or few-particle) hopping-like events, which look nothing like droplets, has been identified to play a very important role as a seed for more complicated processes~\cite{schweizer2003entropic,keys2011excitations,Charbonneau2014,Biroli2021,scalliet2022thirty}. 

\paragraph*{Free energy landscape: saddles and paths  -} Given the difficulty of nailing down a straightforward real-space qualitative description of RFOT instantons, alternate approaches have been considered. The general relationship between instantons and saddle points of the potential energy (or `transition states'), in particular, has motivated attempts at understanding relaxation processes in terms of the critical points of rough landscapes, either the energetic one for numerical simulations~\cite{angelani2000saddles,broderix2000energy}, or the free energetic one 
for analytical study of fully-connected (mean-field) spin glass models~\cite{cavagna2001role}. 
For truly minimal RFOT systems, like the random energy model (REM), the landscape analysis appears to be consistent with the physical expectation that instantons are equivalent to crossing typical saddle points~\cite{Baity-Jesi2018,baity2018activated,carbone2022competition,gayrard2019}. However, the simple barrier crossing hypothesis clearly breaks down for systems  only slightly more complex than the REM, such as spherical $p$-spin glass models~\cite{ros2021,Pacco2024}. For those models, crossing typical saddle points between two minima in the landscape does not result in the system transitioning from one metastable state to another. Only through multiple such crossings or through visiting atypical saddle points can relaxation seemingly proceed. This more involved analysis has yet to be completed, but a recent work finds that correlations between landscape features are likely of paramount importance~\cite{Pacco2024triplets}.
In summary, the saddle points identified by previous studies are not sufficient to achieve decorrelation.

\paragraph*{Dynamical methods -}
Ideally, activated dynamics would have been first studied by \emph{dynamical} methods, and only later recapitulated by means of \emph{static},  landscape-based descriptions. Analyzing the former, however, is significantly more involved than the latter. Only recently could one of us show that a dynamical investigation of activated relaxation processes in RFOT systems is even feasible, through complex path-integral techniques~\cite{rizzo2021}. The challenge is nevertheless amply worth the effort. By considering the probability that a system jumps from one equilibrium state to another, fully decorrelated state within a given time, Ref.~\cite{rizzo2021} indeed reaped two key insights. First, Ref.~\cite[Sec.~A.3]{rizzo2021} raises important questions about determining the most efficient pathway toward ergodicity. To see why that is, recall that in mean-field models the probability of transitioning within a finite time is exponentially small in the system size $N$, and hence becomes of order unity upon reaching a time that is exponentially large in $N$. In simple systems -- 
with few metastable states --
this time is commonly argued to be the inverse of the exponentially small transition rate. This reasoning, however, breaks down in presence of an exponentially large number of intermediate states. Instead of considering direct transitions between equilibrium states, one might then examine transitions that go through metastable states with a higher free energy. After accounting for the exponential number of such states, Ref.~\cite{rizzo2021} found that the total transition rate becomes exponentially larger than that for direct transitions to equilibrium states. 
Second, Ref.~\cite{rizzo2021} found that the very nature of the jumps from one state to another is anomalous. In simple models these jumps occur within a finite time, independent of the duration of the time window allotted for the transition, hence their name, {\it instantons}. For RFOT models, by contrast, the jumps extend over the entire available time window. Interestingly, this  suggests that the system then explores a sequence of marginally stable states, around which the dynamics becomes extremely slow. This association, however, is somewhat confounding for energies then reached are 
\emph{markedly higher} than those of typical marginal states. 
In summary, the dynamical paths identified by previous studies suggest an extremely complex barrier-crossing process, which contrarily to instantons happens in an extremely long time; yet, the paths identified by previous studies
do not seem to be the right ones, as they pass through states of unrealistically high energy.

\paragraph*{The problem -} 
In short, the program seeking to relate static and dynamic descriptions of relaxation processes in complex systems has offered new insights, but has not properly resolved either description. This acute theoretical tension motivates the current work. Even if such tension were but an intellectual challenge, tackling it would be worth ample effort, at least leisurely. The fact that RFOT instantons play a key role in the dynamics of structural glasses, optimization problems, and related models heightens the urgency of breaking through this roadblock toward a first-principle description. 

\subsection{Main results}
\label{sec:schematiclandscape}

\begin{figure}[t]
\centering
\includegraphics[width=\columnwidth]{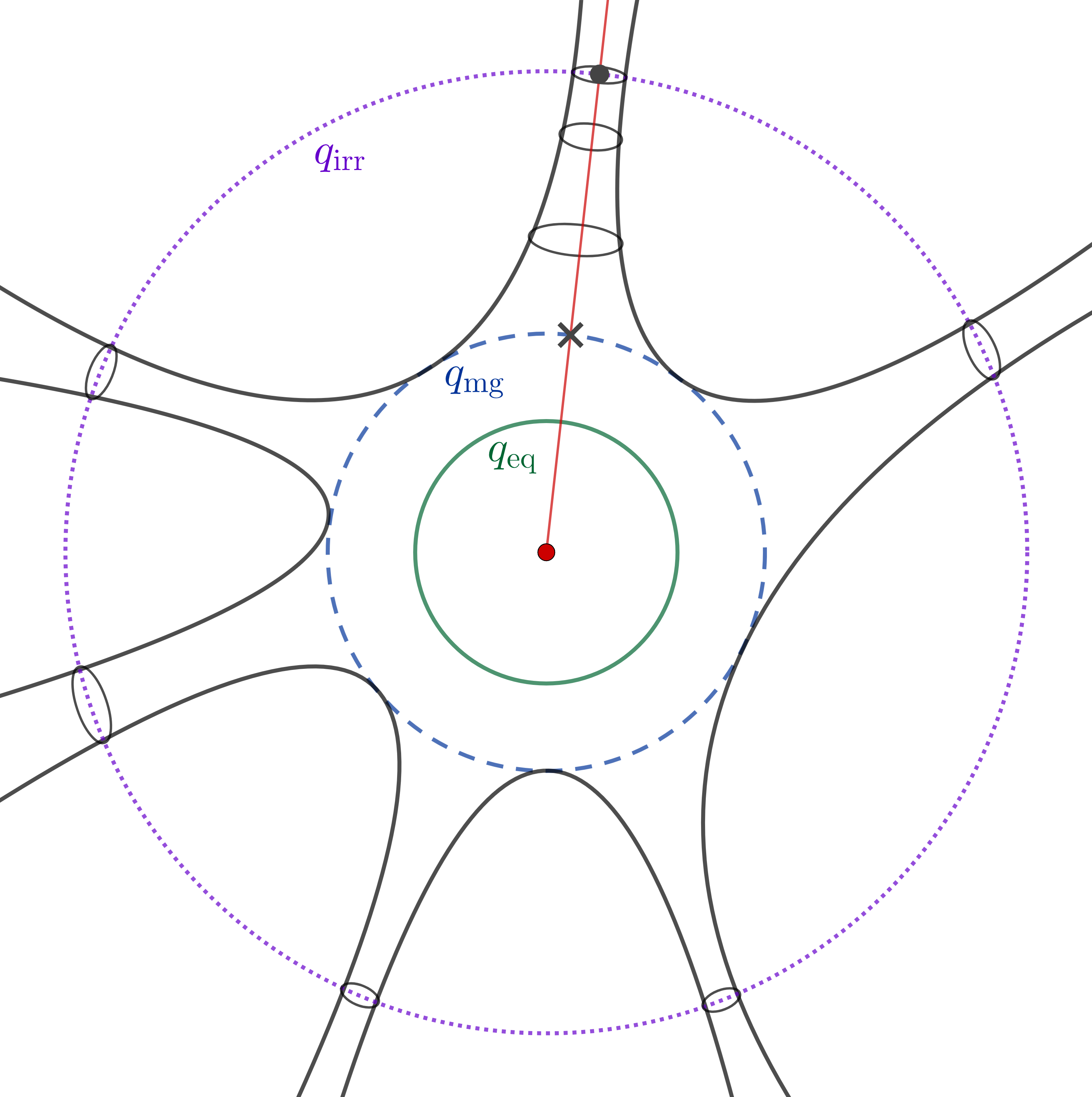}
\caption{Schematics of the basin of attraction in the free energy landscape around an equilibrium configuration, as obtained from our work. At the center lies the reference equilibrium configuration, ${\cal C}$ (red dot). A typical other configuration within the cluster, ${\cal C}'$, has an overlap $q_\mathrm{eq}$ with ${\cal C}$ (green circle). The free energy landscape remains \textit{convex} up to the marginal overlap $q_\mathrm{mg}$ (dashed blue circle). Beyond this point, the landscape becomes \textit{fibered} into numerous channels, but only a few of these channels contribute significantly to the measure. The free energy keeps increasing monotonically until $q_\mathrm{irr}$, at which point the dominant channels encounter their respective saddle points (dotted purple circle). Any trajectory starting within this basin that preserves $q> q_\mathrm{irr}$ typically returns to $q_\mathrm{eq}$ near ${\cal C}$. However, if a trajectory successfully crosses the \textit{irreversible} $q_\mathrm{irr}$, with high probability it never returns.}
\label{fig:Scheme}
\end{figure}

We here make progress on the above problem by analyzing the rare dynamical transitions below the dynamical arrest temperature $T_\mathrm{d}$ in mean-field models of the RFOT universality class, a regime characterized by an exponential number of metastable states. We specifically consider a dynamical process initiated from an equilibrium configuration ${\cal C}$, corresponding to one such state.
Although relaxation dynamics then remains trapped within that state for a time that grows exponentially with system size $N$, rare escape processes remain possible. 
A central quantity in their analysis is the {\it overlap} between the initial configuration ${\cal C}$ and another configuration ${\cal C'}$,
$q({\cal C},{\cal C}')$, which measures the similarity of configurations and is defined (precisely below) to be $q=1$ if the configurations
are identical, and $q=0$ if they are fully uncorrelated.
The main results of this work, schematically illustrated in Fig.~\ref{fig:Scheme}, are as follows.

\paragraph*{1. A new dynamical method for escape paths -}
Our work introduces a dynamical method to investigate the full richness of instantons in complex systems.
Instead of considering the transition rate from ${\cal C}$ to another independent equilibrium state as in Ref.~\cite{rizzo2021}, we consider the transition to the whole set of configurations ${\cal C'}$ that are partially uncorrelated with the initial one, i.e., with fixed overlap 
$q = q({\cal C},{\cal C}')$.
In principle, the system could then select an atypical set of such configurations that are more readily accessible, thus providing a more effective pathway to ergodicity.
We introduce a new dynamical potential $V_{t_{\mathrm{f}}}(q)$ that yields the exponentially small probability of finding the system in a configuration with a given overlap $q$ relative to the initial equilibrium configuration after a fixed time $t_{\mathrm{f}}$. This potential is the dynamical counterpart of the Franz--Parisi (FP) potential \cite{franz1995}.

\paragraph*{2. Phase space is fibered around metastable states -}
Our work identifies the importance of fibers in phase space\footnote{Real-space insights into the nature of the landscape transition from convex to non-convex can be gleaned by comparing the caging geometry of the purely convex hyperplane random Lorentz gas (hRLG) model with that of the random Lorentz gas (RLG)~\cite{bonnet2024,Folena2024}.}.
In Ref.~\cite{Barrat1998}, Barrat and Franz  studied the relaxational dynamics at $N\to \infty$ of a system initialized in a configuration
${\cal C}_0$ at overlap $q$ with the reference ${\cal C}$, as illustrated in Fig.~\ref{fig:Scheme}, where ${\cal C}$ lies at the center.
If $q$ is close enough to unity, i.e., the dynamics is initialized close enough to the center, 
it asymptotically reaches the sphere of equilibrium overlap $q_\mathrm{eq}$, 
which corresponds to the metastable minimum of the FP potential. 
Such relaxation characterizes a simple, convex landscape and takes place in a  time of order one.
 Upon further increasing the radial distance in overlap, Barrat and Franz identified a {\it marginal} overlap $q_\mathrm{mg}$, beyond which the 
free energy landscape is clustered into a multitude of states.
 While these states are stable at a fixed overlap with $q$ with ${\cal C}$, they become radially inward unstable once this constraint is removed,
 provided $q$ remains large enough, i.e., $q>q_\mathrm{irr}$.
    A Langevin dynamics then relaxes back\footnote{Interestingly, this picture is consistent with what one extracts from the energy landscape (instead of the free energy landscape) of spherical $p$-spin models \cite{ros2019b,ros2019}. A gradient descent initiated in these one-direction unstable minima then eventually reaches the reference configuration (also known as the \emph{spike}) \cite{sarao2019}.} to the state with $q=q_\mathrm{eq}$.
However, when the relaxational dynamics is initialized too far away from ${\cal C}$, i.e., with $q<q_\mathrm{irr}$, one lies outside the basin 
 of attraction of ${\cal C}$ and the dynamics wanders around the landscape~\cite{Barrat1998}.
 
 Our study considers the inverse process, i.e., starting from ${\cal C}$ and looking at the (exponentially small in $N$) probability of reaching overlap $q$ in a finite time. We show in Sec.~\ref{sec:RevReg}
 that provided $q>q_\mathrm{irr}$, the escape process
 is exactly the {\it time-reversed} process of the relaxational dynamics studied by Barrat and Franz. 
 This behavior is consistent with the usual expectation that instantons are the time-reverse of relaxational processes~\cite{lopatin1999instantons,Freidlin2012,grafke2019numerical}, but had not been previously considered for systems in the RFOT universality class.
 We thereby obtain two key insights (Fig.~\ref{fig:Scheme}):
 \begin{itemize}
 \item[2a.]  Instantonic processes are \emph{not}, in fact, instantaneous. Already for moderately large $q\in [q_\mathrm{irr}, q_\mathrm{mg}]$, and thus even more so for $q\to 0$, they do not happen on short time scales, but require instead very long ones. They are nevertheless fast relative to the relaxation dynamics of the system in the thermodynamic limit $N\rightarrow\infty$. The \emph{instanton} label should therefore be used with a certain caution when referring to these processes. 

\item[2b.]
The slowness of the process is due to the {\it fibered} nature of the landscape around each metastable state, i.e., the existence of a
clustered yet dynamically connected part of the basin for $q_\mathrm{irr} < q<q_\mathrm{mg}$. 
Note that each fiber has a different energy as a function of $q$, that each fiber starts and ends at different overlaps, and that, for each $q$, the fibers that minimize the free energy may differ. As a result, a great deal of dynamical heterogeneity is expected in this regime. 
A fiber ends when the free energy along it reaches a local maximum, which corresponds to a low-index saddle that acts as a dynamical pivot. Passing it means \emph{with very high probability} never coming back. The typical overlap at which the dominant fibers end defines the irreversible overlap $q_\mathrm{irr}$. 
 
 \end{itemize}

 \paragraph*{3. Escape from a state -}
 Because for $q<q_\mathrm{irr}$ dynamics is no longer centered around the reference configuration ${\cal C}$, our work shows that the static FP potential loses any meaningful predicting power for the forward dynamics. Dynamics is then dominated by \emph{instantonic} trajectories, and different states appear at the end of each fiber. Which of these fibers then dominate the escape dynamics? Unfortunately, a clear static answer to this question remains to be formulated. Here, we instead explore this question by means of our dynamical analogue to the static FP potential, which reaches the $q<q_\mathrm{irr}$ regime.
 
 A key feature of the RFOT landscape is the  {\it threshold energy}, $E_\mathrm{th}$, 
 at which most critical points are marginal, i.e., minima with one (or a few) zero mode. (For energies above $E_\mathrm{th}$, most critical points of the landscape are highly unstable saddles; below $E_\mathrm{th}$, most critical points are stable minima.)
As mentioned above, a previous study by one of us~\cite{rizzo2021} investigated escape paths connecting two randomly chosen equilibrium configurations, and found that such paths reach energies well above $E_\mathrm{th}$ before relaxing back to low-energy equilibrium states. This result contradicts the intuitive idea that once $E_\mathrm{th}$ is reached decorrelation is easy, and that $E_\mathrm{th}$ is an upper bound for the energy needed for relaxation.  
Here, we consider one equilibrium configuration and the set of states that reach overlap $q$ with it, which is a much less stringent constraint. However, we obtain similar results to Ref.~\cite{rizzo2021}. While the origin of this finding is still unclear, we speculate that the observed behavior reflects fiber contributions that are missed when assuming a convex equilibrium-state geometry (see Sec.~\ref{sec:analyticalresults}).

We can nevertheless provide some insight into the process by relying on the following arguments.
Both our path-integral calculation and the static Barrat--Franz calculation~\cite{Barrat1998} 
show that the energy remains below the threshold at $q=q_\mathrm{irr}$. 
It should therefore be possible to escape any given state while remaining below the threshold energy.
To validate this picture, we obtain the following results:
\begin{itemize}
\item[3a.]  We show that the static solution selects the deepest free-energy fibers at each $q>q_\mathrm{irr}$, and that the first fibers to end (i.e. to turn into saddles) are the deepest ones (see Appendix \ref{sec::CompFib}). We should therefore expect the dynamics to proceed along the deepest fibers and reach a saddle point
at $q=q_\mathrm{irr}$ where escape happens with energy below threshold.
This interpretation is supported by considering a free energy potential that describes a third replica ${\cal C}''$ constrained to be at fixed distance
to ${\cal C}, {\cal C}'$~\cite{cavagna1997}.
For $q<q_\mathrm{irr}$, such potential presents a distinct local minimum (called the $M_2$ minimum of the three replica potential in Ref.~\cite{cavagna1997}), thus indicating that the third replica remains distinct from the second, while for $q>q_\mathrm{irr}$ the second and third replica are in the same state.
This supports the idea that an equilibrium Langevin dynamics initiated at $q<q_\mathrm{irr}$ no longer relaxes to the reference state~\cite{Barrat1998}. 

\item[3b.] We present numerical results (in Fig.~\ref{fig:Fig3Sim}) suggesting that if $q$ is small enough, escape trajectories most probably fully decorrelate after crossing $q<q_\mathrm{irr}$. 

\end{itemize}

\paragraph*{4. A minimal picture of relaxation -}

Our results show that even in the mean-field RFOT description of glasses, which presents no spatial structure, relaxation remains structured, heterogeneous, and size-dependent. Although many escape paths (“fibers”) exist, the dynamics is largely controlled by a few long-lived ones. These paths proceed through nearly flat metastable states  before leading to other equilibrium states.

More specifically, our analysis shows the existence of three distinct regimes, depending on the overlap $q$ from the reference equilibrium configuration~${\cal C}$:
\begin{itemize}
    \item \textbf{convex} ($q_\mathrm{mg}<q<1$): the free energy landscape is \textit{convex} and the dynamics is essentially indistinguishable from that of a simple ferromagnetic model.
    \item \textbf{fibered} ($q_\mathrm{irr}<q<q_\mathrm{mg}$): the free energy landscape is \textit{fibered} and dominated by the deepest fibers for each $q$. Although the dynamics is then still centered around ${\cal C}$, it is expected to become strongly heterogeneous, at marked variance from ferromagnetic models. 
    \item \textbf{instantonic}  $(0<q<q_\mathrm{irr})$, the distance from ${\cal C}$ is large enough that the static FP potential no longer meaningfully describes the equilibrium dynamics in the long-time limit. The deepest fibers each end in different states, from which escape can then happen.
\end{itemize}
In the discussion section we explain these various results through what we believe to be a minimal yet generic model, in which the lowest fibers connect typical metastable states to “hub” states, from which dynamics can easily decorrelate.

\subsection{Structure of the paper}

The rest of this article is structured as follows. Section~\ref{sec:RFOTsummary} briefly reviews the equilibrium phenomenology of models within the RFOT universality class and revisits a few conceptual frameworks for describing their free energy landscape, before introducing our new dynamical scheme for studying instantons and illustrating its effectiveness for a simple ferromagnetic Ising model; Sec.~\ref{sec:dynpotentialcalc} describes the dynamical potential calculation for prototypical spin glasses. Using the conceptual framework presented in Sec.~\ref{sec:RFOTsummary}, dynamical potential  and  simulation results are then analyzed in Secs.~\ref{sec:analyticalresults} and ~\ref{sec:simulationresults}, respectively. Section~\ref{sec:discu_specul} concludes with a broader discussion of the findings and speculations about how to surmount the remaining hurdles.

\section{Methods}
\label{sec:RFOTsummary}

In this work, we focus on the Langevin dynamics of the spherical $p$-spin model -- introduced by Kirkpatrick and Wolynes \cite{kirkpatrick1987} as a paradigm for glass-forming liquids -- which captures the key dynamics and metastability features of the RFOT universality class. By solving non-causal mean-field dynamical equations and by running extensive numerical simulations, we obtain insight into the dominant escape paths in finite-$N$ systems and interpret the results in terms of a refined free energy landscape description. 

In this section, we first review the model landscape considered in this work, which belongs to the RFOT universality class. We also briefly revisit the equilibrium phenomenology of this model as well as previous studies of the local free energy potential and of the underlying energy landscape. We then propose a physical description of the rare escape trajectories for leaving an equilibrium state in this landscape.

\subsection{The spherical $p$-spin glass}

One of the simplest mean-field models in the RFOT universality class is the pure spherical $p$-spin glass with
\begin{equation}\label{eq:Ham}
H_J(\mathbf{\sigma}) = \sum_{i_1< i_2< ... <i_p}J_{i_1 i_2 ... i_p}\sigma_{i_1} \sigma_{i_2} \cdots \sigma_{i_p} \ ,
\end{equation}
where $\sigma$ is a vector of $N$ real variables $\sigma_i$ that satisfies the spherical constraint 
\begin{equation}\label{eq::spherical_contraint}
    \sum_i \sigma_i^2 = N \, ,
\end{equation}
and the couplings $J_{i_1 i_2 ... i_p}$ are i.i.d.~Gaussian random variables with zero mean and variance $1/(2 N^{p-1} p!)$. 

For this model, we consider both the equilibrium Boltzmann--Gibbs distribution, $P_\mathrm{eq}(\sigma)=e^{-\beta H_J(\sigma)}/Z_J$, and the overdamped Langevin dynamics at temperature $T$,
\begin{equation}
\label{eq::Langevin}
    \dot{\sigma}_i=-\nabla_i H_J(\sigma) - \mu \sigma_i + \xi_i \ ,
\end{equation}
for $i=1,\dots,N$ and white noise $\langle \xi_i(t) \xi_j(s) \rangle = 2T\delta_{ij}\delta(t-s)$. (The term $-\mu\sigma_i$ enforces the spherical constraint given by Eq.~\eqref{eq::spherical_contraint}.)

As we describe below, at sufficiently low $T$, such a model is characterized by a large number of metastable states.
For a dynamics initiated in a typical equilibrium state taken from $P_\mathrm{eq}(\sigma)$, we are specifically interested in the probability and typical profile of rare trajectories that relax out of this state.

\subsection{Phase transitions within RFOT}

In the thermodynamic limit, the equilibrium Boltzmann--Gibbs distribution $P_\mathrm{eq}(\sigma)$ of mean-field models within the RFOT universality class presents three equilibrium regimes separated by two transitions: one dynamical at $T_\mathrm{d}$ and the other static at $T_\mathrm{s}$, with $T_\mathrm{d}>T_\mathrm{s}$~\cite{kirkpatrick1987,kirkpatrick1987p}. 
These phase transitions can be conveniently captured via the Franz--Parisi (FP) potential $V(q)$ \cite{franz1995}, which
is nothing but a thermodynamic potential associated with
 the overlap between configurations $\sigma$ and $\tau$, $q=\frac{1}{N}\sum_j \sigma_j\tau_j$. (The overlap measures
 the similarity of configurations, $q=1$ corresponding to
 identical configurations and $q=0$ to decorrelated -- orthogonal -- ones.)
 Using as reference $\tau$ drawn from the unconstrained equilibrium measure, $P_\mathrm{eq}(\tau)=e^{-\beta H_J(\tau)}/Z_J$, the FP potential is then
\begin{equation}
\label{eq:VqFP}
V(q) \equiv \overline{\sum_{\tau} P_\mathrm{eq}(\tau) \ln \sum_{\sigma} e^{-\beta H_J(\sigma)} \delta( N q- \sigma \cdot \tau)} \ ,
\end{equation}
where the average over the (quenched) disorder $J$ is denoted by $\overline{[\ldots]}$. Note that the overlap is the \emph{order parameter} of $V(q)$, and hence it is not optimized during the replica-based calculation of the quenched average; a different optimization is solved for each $q$. 

For the high-temperature {\it ergodic phase} ($T > T_\mathrm{d}$), 
the minimum of $V(q)$ lies at $q = 0$ and $V(q)$ is everywhere convex. As a result, the cost of keeping two configurations at $q>0$ is always positive; two independent configurations are typically uncorrelated.

The high-temperature ergodic phase undergoes an ergodicity breaking transition at $T_\mathrm{d}$, below which phase space is disconnected in states (also dubbed the \textit{clustered phase}). The FP potential then develops a secondary metastable minimum at a finite $q_\mathrm{eq}>0$. While it is always thermodynamically favorable to have uncorrelated equilibrium configurations, $q=0$, it is also possible to have a locally stable minimum with two configurations having finite overlap $q_\mathrm{eq}>0$, which is thus interpreted as the equilibrium overlap (Edwards--Anderson parameter) within a metastable state. Put differently, the Boltzmann--Gibbs distribution is then broken into many states, composed by groups of configurations with overlap $q_\mathrm{eq}>0$, and fully decorrelated from each other, i.e., with mutual overlap $q=0$.
The number of states $\mathcal{N}$ can be computed~\cite{franz1995,monasson1995structural} and it is so large as to give rise to a non-zero complexity, $\Sigma(T)=\frac{1}{N}\ln\mathcal{N}_N(T)$.
In this regime, because the height of the free energy barrier separating the two minima of the FP potential is extensive in system size, states are long-lived in the limit $N\rightarrow\infty$. Within the basin of attraction of a state, the value of the FP potential relative to its value at the local minimum therefore quantifies the probability of the corresponding overlap $q$. For finite-$N$ systems with $T_\mathrm{s}<T<T_\mathrm{d}$, however, the finiteness of those barriers -- together with the stable state having $q=0$ -- makes escaping a metastable cluster possible. Such activated escapes, whatever their form, need to surmount a free energy barrier to proceed, in analogy to instantons in simple systems.

\begin{ruledtabular}
\begin{table}[]
    \centering
    
    \begin{tabular}{c|c|c|c}
        Quantity & Value & Quantity & Value\\
        \colrule
        $E_\mathrm{d}$ & -0.8165 &
        $q_\mathrm{d}$ & 1/2 \\
        $E_\mathrm{s}$ & -0.8532 &
        $q_\mathrm{s}$ & 0.6540\\
        \hline
        $E_\mathrm{eq}$ & -0.8475 & $q_\mathrm{eq}$ & 0.6340 \\
        $E_\mathrm{mg}$ & -0.8354 &$q_\mathrm{mg}$ &  0.5134\\        
        $E_\mathrm{irr}$ &-0.8339 &$q_\mathrm{irr}$& 0.3234 \\
        $E_\mathrm{hub}$ &-0.8349 &$q_\mathrm{hub}$& 0.3792 \\
        $E_\mathrm{th}$ & -0.8312 & $q_\mathrm{th}$& 1/2
\end{tabular}

    \caption{Landscape parameters for the spherical $p$-spin glass model with $p=3$; for the bottom part,  $T=1/1.695$.}
    \label{tab:3spin1_695}
\end{table}
\end{ruledtabular}

 Finally, for $T<T_\mathrm{s}$ the minimum of $V(q)$ at $q_\mathrm{eq}>0$ becomes the absolute minimum, and hence nothing drives the system away from a given (equilibrium) state.
At $T_\mathrm{s}$ and below, $\Sigma(T_\mathrm{s})=0$, which gives rise to a thermodynamic phase transition toward an \textit{ideal glass} state at $T=T_\mathrm{s}$.  

For pure spherical $p$-spin glass models in particular, 
\beq
T_\mathrm{d}=\sqrt{\frac{p(p-2)^{p-2}}{2(p-1)^{p-1}}} \ ,
\eeq
with $q_\mathrm{d}=q_\mathrm{eq}(T_\mathrm{d})=(p-2)/(p-1)$ being the typical overlap of pairs of configurations belonging to the same state at $T_{\mathrm{d}}$.
The static transition temperature
$T_\mathrm{s}$ has a slightly more complex expression (see, e.g.,~\cite[Sec.~7.2.2]{leuzzi2007thermodynamics}), but because we here only consider the case $p=3$, it suffices to note that $T_\mathrm{d} = 0.612372$ and $T_\mathrm{s} = 0.586054$. Although all the calculations and analysis are here performed in the clustered phase at $T=1/1.695\approx0.5900$ -- closer to $T_\mathrm{s}$ than to $T_\mathrm{d}$ -- the overall presentation should be largely independent of that specific choice, as long as $T_\mathrm{s}<T<T_\mathrm{d}$. Various landscape parameters for this setup are summarized in Table~\ref{tab:3spin1_695}. 

Note that in the following, \textit{state} systematically denotes one of the sets of typical equilibrium configurations with mutual overlap $q_\mathrm{eq}$, and \textit{basin} denotes a wider set of (possibly non-equilibrium) configurations that share a common state as their dynamical attractor. In general, we assume readers to be familiar with the above picture. For the others, detailed reviews can be found in Refs.~\cite{castellani2005spin,cugliandolo2002dynamics,folena2020mixed}.

\subsection{Dynamical Potential Scheme}
\label{sec:dynpotscheme}
The strategy we propose for extracting instantons for systems in the RFOT universality class entails starting from a randomly selected equilibrium reference configuration $\tau$ at time $t_\mathrm{i}=0$, and then constraining its dynamical evolution such that at a time $t_\mathrm{f}$ later the system has a given overlap $q$ with $\tau$. 
(This dynamical construction is reminiscent yet distinct from the static FP potential given by Eq.~\eqref{eq:VqFP}, as discussed below.) 
Because this scheme is agnostic about the number of barrier jumps and the type of barriers crossed to decorrelate (either fully for $q=0$ or partially for $0<q<q_\mathrm{eq}$) from $\tau$, its results should robustly provide the optimal dynamical pathway taken by the system over a given $t_\mathrm{f}$. In the limits $t_\mathrm{f}\rightarrow\infty$ and $N\rightarrow\infty$ (in that order) for $q=0$, instantonic pathways should therefore be recovered. However, this limit is inaccessible in disordered systems. We are therefore constrained to consider the opposite limit $N\rightarrow\infty$ at fixed $t_\mathrm{f}$, and then extract information from the large deviation function $V_{t_\mathrm{f}}(q)$. 

Before proceeding with the analysis of this function for RFOT landscapes, we  pedagogically illustrate its construction in a ferromagnetic system for which instantons are well-understood. 

\subsubsection{Ferromagnetic Case}

Let us consider 
a mean-field ferromagnetic Ising model for $N$ spin variables $\sigma_i\pm1$~\cite{griffiths1966relaxation} with
\beq
H(\sigma)  =-\frac{1}{N}\sum_{ij} \sigma_i \sigma_j - h \sum_i \sigma_i 
=-\frac{M^2}{2N}- h M \ ,
\label{ham}
\eeq
using the magnetization $M=\sum_{i=1}^{N} \sigma_i$ as the order parameter.
Here, because the ground states are ordered with all spins either up or down, one can replace the 
overlap (scalar product between two typical configurations) with the magnetization
(scalar product between an equilibrium configuration and the ground state) with qualitatively similar results.

The free energy constrained to a fixed $M$, which is then the analog of the FP potential, is then
\begin{equation}
\begin{split}
    F(M)&=-\frac{1}{\beta}\ln\left[\sum_{\sigma|\sum_i\sigma_i=M} e^{-\beta H(\sigma)}\right]\\
    &=-\frac{1}{\beta}\ln\left[{N \choose (N+M)/2} e^{\frac{\beta}{2 N} M^2 + \beta h M} \right],
\end{split}
\end{equation}
which in the thermodynamic limit $N\to\infty$ can be written as a function of the average magnetization $m=M/N$,
\begin{align}\label{eq:freeFM}
\lim_{N\to\infty}&\frac{1}{N}F(mN)=f(m) = -\frac{m^2}{2} - m h \\
&+ \frac{1}{\beta}\left[ \frac{(1 + m)}{2} \ln(1 + m) + \frac{(1 - m)}{2} \ln(1 - m) \right] \ .\nonumber
\end{align}

Because the system Hamiltonian depends only on $M$, it follows that at any time $t$, the probability of obtaining a configuration $\sigma$ depends on its magnetization alone, i.e., $p_t(\sigma) = p_t [M(\sigma)]$. The evolution equation for the single spin-flip dynamics can then be expressed as~\cite{griffiths1966relaxation,mora2012transition}
\begin{align}
\label{MFmaster}
\LL p_t&=\partial_t p_t(M)= W_+(M-2) p_t(M-2)\\
&+ W_-(M+2) p_t(M+2) \nonumber \\ 
&- [W_-(M) + W_+(M)] p_t(M),\nonumber
\end{align}
with transition rates 
\begin{align}
W_+(M) &= \frac{N-M}2 e^{\beta (\frac{M+1}{N}+h)}, \\
W_-(M) &=\frac{N+M}2 e^{-\beta(\frac{M-1}{N}+h)},
\end{align}
consistent with the detailed balance condition $W_+(M)e^{-\beta F(M)}=W_-(M+2)e^{-\beta F(M+2)}$.
As for the static case given by Eq.~\eqref{eq:freeFM}, in the thermodynamic limit it is possible to obtain  the time-dependent large deviation function for a given $t_\mathrm{f}$ 
\begin{equation}
V_{t_\mathrm{f}}(m)=-\lim_{N\rightarrow\infty} \frac{1}{N}\ln p_{t_\mathrm{f}}(M),
\end{equation}
which corresponds to $p_{t_\mathrm{f}}(M) = e^{-N V_{t_\mathrm{f}}(M/N) + o(N)}$.    
To obtain an expression for $V_{t_\mathrm{f}}(m)$, we write Eq.~\eqref{MFmaster} by considering the continuous version of the detailed balance condition
\begin{align}
\frac{w_-(m)}{w_+(m)} &= \exp[2\b f'(m)]\,,
\end{align}
with rates 
\begin{equation}
w_\pm(m) \equiv \lim_{N\to \infty} \frac{1}{N} W_{\pm}(N m) =  \frac{1\mp m}2 e^{\pm \beta ( m +  h)} \ .    
\end{equation}
We therefore obtain, to leading asymptotic order in the limit $N\to\infty$~\cite{griffiths1966relaxation},
\begin{align}\label{eq:dynpot}
 [ e^{-\partial_t V_t(m)} -1 ] &= w_+(m) [ e^{2\partial_m V_t(m)} -1 ] \\
&+ w_-(m) [ e^{-2\partial_m V_t(m)} -1 ] \nonumber \ .
\end{align}
Considering stationary solutions, which correspond to the long-time dynamics, $\lim_{t_\mathrm{f}\rightarrow\infty}V_{t_\mathrm{f}}(m) = V(m)$, gives
\beq
0 = w_+(m) [ e^{2 V'(m)} -1 ] + w_-(m) [ e^{-2 V'(m)} -1 ] \ .
\eeq
A constant, $V(m)= B$, is a possible solution of this equation.
If $V(m)$ is not a constant, then we can equivalently write
\beq
\frac{e^{2 V'(m)} -1}{e^{-2 V'(m)} -1} = - \frac{w_-(m)}{w_+(m)} = - e^{2\b f'(m)} \ ,
\eeq
for which $V(m) = \b f(m) + A$ is also
a stationary solution. 

Combining these two functions and imposing that the solution be continuous and derivable at every point results in a stationary solution of the piecewise form (in the two-minimum regime)
\begin{equation}
V(m) = \begin{cases}
\b f(m)+A & m > m_{\rm max} \ ,\\
\b f(m_{\rm max}) + A & m_{\rm left-min} <m< m_{\rm max} \ ,\\
\b f(m)+B & m < m_{\rm left-min} \ ,
\end{cases}
\end{equation}
where $m_{\rm max}$ is the position of the intermediate maximum of $f(m)$ (the barrier position) and $m_{\rm left-min}$ is the position of the minimum of $f(m)$ to the left of that barrier. The constant $A$ is determined by imposing $\min V(m) = 0$, which follows from the normalization of $p_{t_\mathrm{f}}(M)$, while $B$ is chosen to match heights at $m_{\rm left-min}$, thus providing a continuous function. The stationary condition is then satisfied on both sides of the barrier and corresponds to the long-time dynamics starting from an initial condition concentrated around the right minimum.

\begin{figure}
    \centering
    \includegraphics[width=\columnwidth]{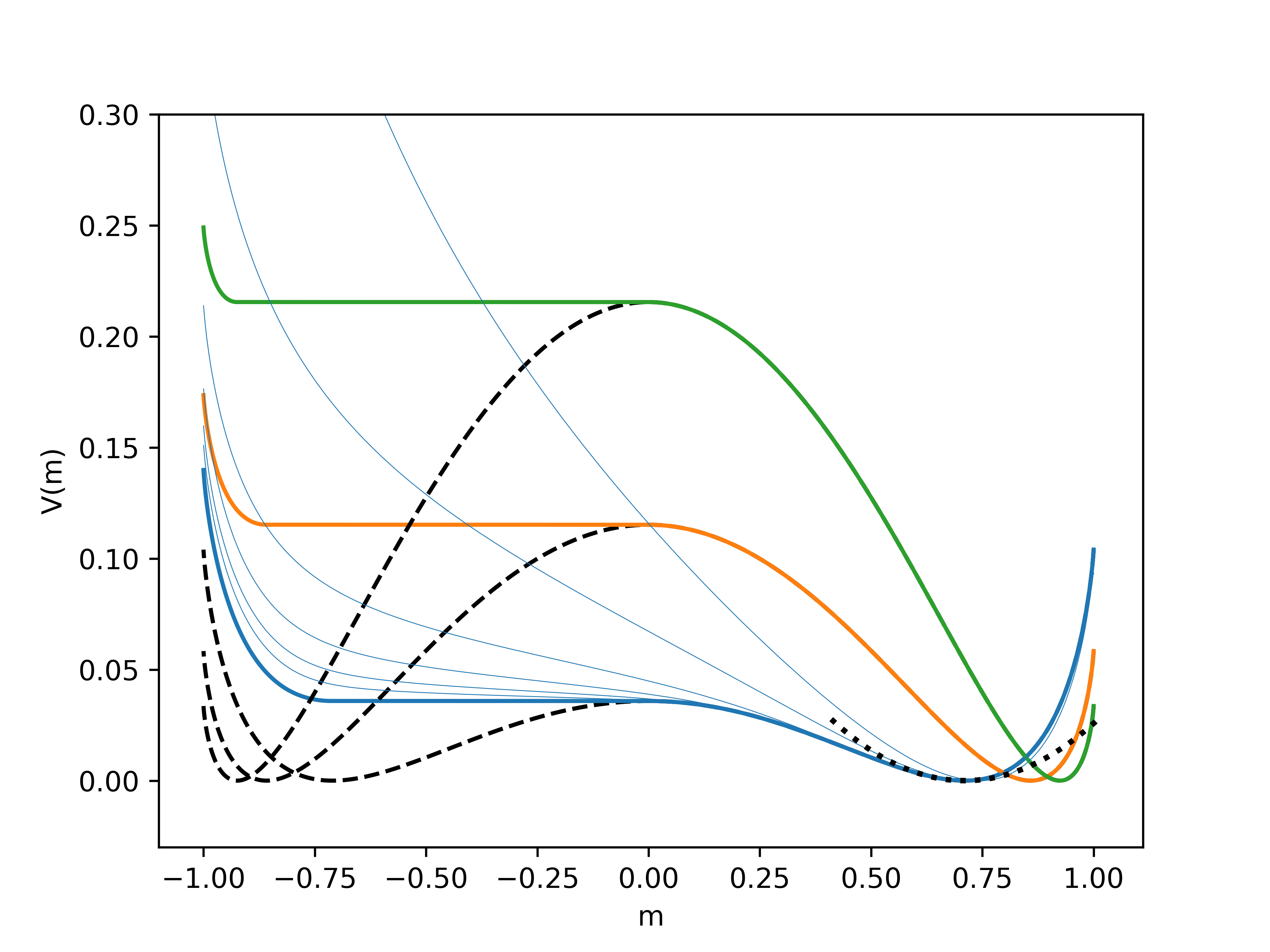}
    \caption{Long-time limit of the dynamical potential, $V(m)$ (colored lines), along with the Landau free energy $\beta f(m)$ (dashed lines), for the fully-connected Ising model with $h=0$ at various sub-critical $T=4/5$ (blue), 2/3 (orange), and 4/7 (green). (The critical temperature is $T_c=1$.) The finite-time $V_{t_\mathrm{f}}(m)$ for $t_\mathrm{f}=1, 2, 4, 6, 10, 15,$ and $20$ (pale blue lines, from top to bottom) for $T=4/5$ is obtained by numerically integrating Eq.~\eqref{eq:dynpot}. The harmonic approximation to the right minimum is given for $T=4/5$ (dotted line).}
    \label{fig:VmIsing}
    \end{figure}
    
    \begin{figure}
    \centering
    \includegraphics[width=0.95\columnwidth]{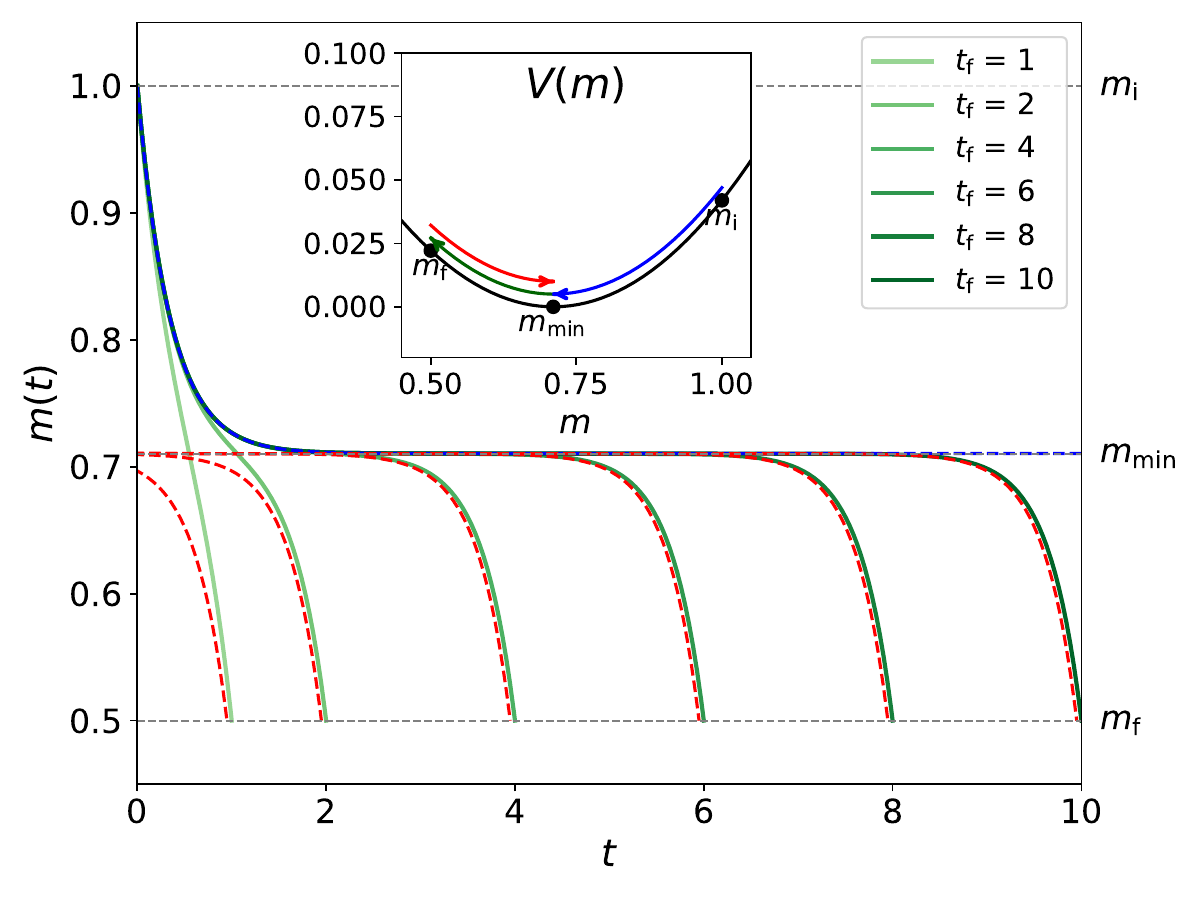}
    \caption{Optimal trajectories of the magnetization (in a harmonic well) around $m_\mathrm{min}\approx0.71$ for $T=4/5$. The dynamics is initiated at $m_\mathrm{i}=1$ and set to reach $m_\mathrm{f}=1/2$ at times $t_\mathrm{f}=1,2,..,10$ (green lines). For long enough $t_\mathrm{f}$, the optimal relaxation trajectories from $m_i$ to $m_\mathrm{min}$  (dashed blue line) and the mirrored ones for $t_\mathrm{f}-t$ at from $m_\mathrm{f}$ to $m_\mathrm{min}$ (red dashed lines) recapitulate approach to and depart from $m_\mathrm{min}$, respectively. 
    (Inset) Representation of the forward relaxation (blue arrow), the jump to the final magnetization $m_\mathrm{f}$ (green arrow), and its mirroring relaxation (red arrow) within the harmonic potential.}
    \label{fig:ferro}
\end{figure}



Without loss of generality, we consider the case $h=0$
and temperatures below the critical point, $T<T_c=1$. As expected, a time that grows exponentially with $N$ is then needed to escape a basin (Fig.~\ref{fig:VmIsing}).
Assuming that the system is initiated with $m=m_\mathrm{eq}$ at $t=0$,
two regimes relative to $m_\mathrm{max}=0$ emerge for $V(m)$: (i)  $m_\mathrm{max}<m<1$, and (ii) $-1<m<m_\mathrm{max}$. In (i), the dynamics remains close to the equilibrium value, corresponding to the spontaneous magnetization -- the local minimum of $f(m)$ -- for most of the trajectory, until a quick jump to $m$ takes place near the end of the allocated time, irrespective of the chosen (large enough) $t_\mathrm{f}$. This quick jump is completely captured by taking the mirror image of the relaxation to the nearest equilibrium starting from the end point, as is well known in the theory of instantons~\cite{lopatin1999instantons,Freidlin2012,grafke2019numerical}. (See also Sec.~\ref{sec:revdyn}.)
The relaxation curves therefore fully collapse for $t-t_\mathrm{f}$. In (ii), $p_{t_{\rm f}}$ is dominated by the barrier crossing process. After the barrier at $m_\mathrm{max}$ is crossed, relaxation to any $m>m_\mathrm{left-min}$ takes a time of order one. As a consequence, $V_{t_{\rm f}}$ is constant for $m_\mathrm{left-min}<m<m_{\rm max}$. Only for $m<m_\mathrm{left-min}$, i.e., upon approaching $m=-1$, is an additional exponential factor required.
Note that in absence of a magnetic field, the model considered is perfectly symmetric with $m_\mathrm{max}=0$. The dynamical results therefore perfectly match what the static $f(m)$ would have suggested. 

\subsubsection{Reversible regime}
\label{sec:revdyn}
Because the dynamical properties of regime (i) are particularly important for the analysis below, we also illustrate them for a ferromagnetic model. For the sake of analytical simplicity, we consider the case of Gaussian distributed spherical spins -- instead of Ising spins -- and perform a rotation such that the magnetization density obeys the Langevin equation,
\beq
\frac{dm}{dt}=-(\mu+\beta) m + \frac{1}{\sqrt{N}}\xi.
\eeq
Using a path-integral representation of the Langevin dynamics as discussed, e.g., in Ref.~\cite[Appendix]{rizzo2021}, we then obtain that the transition rate is given by
\begin{equation}
 T_{J,t_\mathrm{f}} (m_f | m_i) \propto \int [dm] e^{N \mathcal{L}} \ ,
\end{equation}
with
\begin{equation}
 \mathcal{L}=- \frac{A}{4} m_f^2+ \frac{A}{4} m_i^2-\frac{1}{4} \int_0^{t_\mathrm{f}} dt \left[ \left(\frac{dm}{dt}\right)^2 + A^2 m^2(t) \right]  \, ,
\end{equation}
where $A=\mu+\beta$.
Due to the factor $N$ in the exponential the integral over trajectories is dominated by the maximum. Extremization yields
\begin{equation}
    \frac{d^2 m}{d t^2}- A^2 m=0
\end{equation}
with boundary conditions $m(0)=m_\mathrm{i}$ and $m(t_\mathrm{f})=m_\mathrm{f}$, which has for solution
\begin{equation}
    m(t)= m_\mathrm{i} \cosh A \, t + (m_\mathrm{f} \,\mathrm{csch}\, A\, t_\mathrm{f}-m_\mathrm{i}\, \coth  A \,t_\mathrm{f}) \, \sinh A \, t \ .
    \end{equation}
Figure~\ref{fig:ferro} shows that for large $t_\mathrm{f}$ this solution behaves as described above.
The Lagrangian on the solution can then be computed,
\begin{multline}
    T_{J,t_\mathrm{f}} (m_f | m_i) \propto \exp  \left\{\, N A \, \left[ -\frac{m_\mathrm{f}^2}{2}+ \right.\right.
    \\
 \left. \left.  \frac{1-\coth A\,t_\mathrm{f}}{4}(m_\mathrm{f}^2+m_\mathrm{i}^2)-\frac{1}{2} m_\mathrm{f}\,m_\mathrm{i} \mathrm{csch}\, A\, t_\mathrm{f} \right]\right\}.
\end{multline}
In the limit $t_\mathrm{f}\rightarrow\infty$, we then have $T_{J,t_\mathrm{f}} (m_f | m_i) \to  e^{- N A  \frac{m_\mathrm{f}^2}{2}}$ irrespective of $m_\mathrm{i}$.
Note also that if $m_\mathrm{i}=0$ and $m(t_\mathrm{f})=m$ we have
\begin{equation}
    V_{t_f}(m)=B(t_\mathrm{f})\, m^2,
\end{equation}
where  $B(t_\mathrm{f}) \equiv A(1+\coth t_{\mathrm{f}})/4$ diverges as $1 /t_\mathrm{f}$ in the limit $t_\mathrm{f} \to 0$ and converges to $A/2$ in the limit $t_\mathrm{f} \to \infty$.

\section{Dynamical Potential Calculation}
\label{sec:dynpotentialcalc}
This section presents the analytical machinery needed for obtaining a dynamical counterpart to the FP potential in the sense illustrated in Sec.~\ref{sec:dynpotscheme} for a simple ferromagnetic model. Put succinctly, for the limit $N\rightarrow\infty$ we obtain a series of integro-differential equations for a set of order parameter functions. We here sketch how the analytical computation is done, referring to Appendix~\ref{sec::anal_calculations} for its technical details.  
We then describe how these equations were numerically evaluated. 

\subsection{Analytical approach}
\label{sec:gendynpot}

Starting from an initial equilibrium configuration $\tau$  -- drawn from the equilibrium distribution $P_\mathrm{eq}(\tau)$ -- we wish to determine the logarithm of the probability that, over a time interval $t_\mathrm{f}$, the Langevin dynamics given by Eq.~\eqref{eq::Langevin} leads to any configuration $\sigma$ having an overlap $q$ with $\tau$. This \emph{dynamical} analog of the static potential in Eq.~\eqref{eq:VqFP} is expressed as
\beq
\begin{split}
\label{eq:Vtfq}
V_{t_\mathrm{f}}(q_\mathrm{f}) &\equiv \overline{\sum_{\tau} P_\mathrm{eq}(\tau) \ln \mathcal{Z}_{J, t_{\mathrm{f}}}(\tau ; q_{\mathrm{f}})}
 \,, \\
 \mathcal{Z}_{J, t_{\mathrm{f}}}(\tau ; q_{\mathrm{f}}) &\equiv \sum_\sigma T_{J,t_\mathrm{f}} (\sigma | \tau) \delta( N q_\mathrm{f} - \sigma \cdot \tau)
\end{split}
\eeq
where 
$T_{J,t_\mathrm{f}}(\sigma | \tau)$ encodes the probability of following a given trajectory starting from $\tau$ and reaching $\sigma$ within a time $t_{\mathrm{f}}$. For convenience, we express the rate using a helper function 
\beq
\hat{T}_{J,t_\mathrm{f}}(\sigma,\tau) \equiv e^{{\beta \over 2} [H_J(\sigma)- H_J(\tau)]} T_{J,t_\mathrm{f}}(\sigma | \tau),
\eeq
which due to the detailed balance condition is  symmetric with respect to the exchange $\tau\leftrightarrow\sigma$. 

In order to calculate this dynamical potential one has to average over disorder, $J$. We do so by employing the replica method twice:  first to get rid of the logarithm in  Eq.~\eqref{eq:Vtfq}, and second to obtain the inverse partition function in $P_\mathrm{eq}(\tau)$ as $Z_J^{-1} = \lim_{m \to 0} Z^{m-1}_J$. We can then write the dynamical potential as
\beq
V_{t_\mathrm{f}}(q_\mathrm{f})
= \lim_{m \rightarrow 0} \lim_{n \rightarrow 0}{d \over dn}  \overline{ Z_{J,\hT}(m,n)},
\eeq 
where we have defined
\beq
\label{eq:ZT}
\begin{split}
&Z_{J,\hT}(m,n) \equiv  \\ &\sum_{\substack{\sigma_1 \dots \sigma_m \\ \tau_1 \dots \tau_{n}}} e^{-\beta \sum_{a=2}^m H_J(\tau_a)-\beta \left(1-{n \over 2}\right)H_J(\tau_1) -{\beta \over 2} \sum_{b=1}^{n}H_J(\sigma_b) } \\ 
&\times \prod_{b=1}^{n}\hT_{J,t_{\mathrm{f}}}(\sigma_b,\tau_1) \delta (N q_\mathrm{f}-\tau_1 \cdot \sigma_b) \ ,
\end{split}
\eeq
Note that for $T>T_\mathrm{s}$ an annealed (equilibrium) average coincides with a quenched average, and hence a single replica, $m=1$, suffices
\beq
\label{eq:RSdynapprox}
\lim_{m \rightarrow 0} \lim_{n \rightarrow 0}{d \over dn}\overline{   Z_{J,\hT}(m,n)} \approx {1 \over \overline{Z_J}}\lim_{n \rightarrow 0}{d \over dn}  \overline{Z_{J,\hT}(1,n)} \ .
\eeq 
The quantity $\hat{T}_{J,t_\mathrm{f}}(\sigma,\tau)$ can then be written as a path integral; see~\cite[Appendix]{rizzo2021} for a detailed derivation. As we show in Appendix~\ref{sec::anal_calculations}, introducing bosonic variables that behave as the product of two Grassmann variables 
compactly expresses this path integral in terms of superfields, in a form that closely resembles the standard static calculation for this model. Averaging over the quenched disorder can then be performed, and in the thermodynamic limit $N\rightarrow\infty$ a saddle point approach can be used to obtain the dynamical potential. Under a RS ansatz assumption (see Appendix~\ref{app:equations}), its expression depends on eight two-times order parameter functions that are respectively the \emph{correlation functions} $C(t, t')$ and  $C^\mathrm{dt}(t, t')$, the (time-symmetrized) \emph{response functions} $\hat R_1(t, t')$, $\hat R_1(t, t')$, $\hat R_2(t, t')$, and $\hat R_2^\mathrm{dt}(t, t')$, and the \emph{causal susceptibilities} $\hat \chi(t, t')$ and $\hat \chi^\mathrm{dt}(t, t')$ that satisfy self-consistent integro-differential equations. Note, however, that the RS condition breaks down in the fibered regime of the FP potential, as discussed in more detail in Sec.~\ref{sec:analyticalresults}.

At the saddle point, those quantities have a simple physical interpretation.   The function $C(t, t')$ represents the mean correlation between system configurations encountered along a single Langevin trajectory at two different times $t$ and $t'$; $C^\mathrm{dt}(t,t')$ quantifies the correlation between configurations sampled from distinct dynamical trajectories, hence the ``$\mathrm{dt}$'' superscript. More specifically,
\begin{subequations}
    \begin{align}
        C(t,t') &= \overline{ \left[ \langle s_i(t) s_i(t') \rangle \right] } \\
        C^\mathrm{dt}(t,t') &= \overline{ \left[ \langle s_i(t) \rangle \langle s_i(t') \rangle \right] },
    \end{align}
\end{subequations}
where $s_i(t)$ is used to denote the value of spin $i$ at time $t$, so as to avoid confusion with the initial and final configurations denoted by $\sigma$ and $\tau$, respectively.
The square brackets $\left[ \ldots \right]$ denote averaging over initial condition $\tau$, and the angular brackets $\langle \ldots \rangle$ denote averaging over the Langevin dynamics starting from $\tau$ and constrained to evolving, within a time window $t_\mathrm{f}$, to a configuration with overlap $q_\mathrm{f}$ from $\tau$. Similarly,
\begin{subequations}
\label{eq::physical_meaning_order_parameters}
    \begin{align}
        \hat R_1(t,t') &= \hat R_2(t',t) \\
        \hat R_1^\mathrm{dt}(t,t') &= \hat R_2^\mathrm{dt}(t',t) \\
        \hat R_2(t,t') &= \frac{1}{\beta}\left. \overline{ \left[ \left\langle \frac{\partial s_i(t)}{\partial h_i(t')} \right \rangle \right] }\right|_{h_i = 0} 
        \\
        \hat R_2^\mathrm{dt}(t,t') &= \frac{1}{\beta}\left. \overline{ \left[ \langle s_i(t) \rangle \frac{1}{\mathcal{Z}_{J, t_{\mathrm{f}}}} \frac{\partial \mathcal{Z}_{J, t_{\mathrm{f}}}}{\partial h_i(t')}  \right] }\right|_{h_i = 0} 
        \\
        \hat \chi(t,t') &= \frac{1}{\beta^2}\left. \overline{\left[  \frac{1}{\mathcal{Z}_{J, t_{\mathrm{f}}}} \frac{\partial^2 \mathcal{Z}_{J, t_{\mathrm{f}}}}{\partial h_i(t) \partial h_i(t')} \right] } \right|_{h_i = 0} 
        \\
        \hat \chi^\mathrm{dt}(t,t') &= \frac{1}{\beta^2}\left. \overline{ \left[ \frac{1}{\mathcal{Z}^2_{J, t_{\mathrm{f}}}} \frac{\partial \mathcal{Z}_{J, t_{\mathrm{f}}}}{\partial h_i(t)}  \frac{\partial \mathcal{Z}_{J, t_{\mathrm{f}}}}{\partial h_i(t')}  \right] }\right|_{h_i = 0} 
    \end{align}
\end{subequations}
where $h_i(t)$ is an instantaneous magnetic field switched on at time $t$ on site $i$. See  Appendix~\ref{sec::anal_calculations} for additional details. 
Together with the two-time functions, we have to account for two additional Lagrange multipliers, $\hat \mu(t)$ and $\mu(t)$, that impose the spherical constraint from Eq.~\eqref{eq::spherical_contraint}, $C(t, t) = 1$, and the value of the response function at equal times, 
$\hat R_2(t, t) = \frac{1}{2}$, respectively. Appendix~\ref{app::dynamical_equations_final_condition} shows how the dynamical constraint over the final condition results in additional constraints for the correlation and the response functions when either $t=t_\mathrm{f}$ or $t'=t_\mathrm{f}$. Appendix~\ref{app:equations} also reports the complete list of dynamical equations, together with the boundary conditions that the order parameters must satisfy. 
The free case ($\beta=0)$ can be solved analytically (see Appendix~\ref{sec::freecase}), thus validating these equations and offering insights into certain features of the solution. For finite $\beta$, however, specialized numerical methods are needed.

\subsection{Numerical solution of the dynamical equations}
Numerically evaluating the dynamical equations for finite $\beta$ is no simple task. Some of the technicalities involved are presented in this subsection.
\subsubsection{Discretization and initialization}

\noindent Defining the matrices
\begin{subequations}
\begin{align}
    \mathcal{C}(t, t') &= 
    \begin{pmatrix}
        C(t, t') & \hat R_2(t, t') \\
        \hat R_1(t', t) & \hat \chi(t, t')
    \end{pmatrix} \\
     \mathcal{C}^\mathrm{dt}(t, t') &= 
    \begin{pmatrix}
        C^\mathrm{dt}(t, t') & \hat R_2^\mathrm{dt}(t, t') \\
        \hat R_1^\mathrm{dt}(t', t) & \hat \chi^\mathrm{dt}(t, t')
    \end{pmatrix}
\end{align}
\end{subequations}
gives a compact expression for the dynamical equations 
\begin{equation}
\label{eq::compact_dyn_eq}
    \mathcal{F}(\mathcal{C}, \mathcal{C}^\mathrm{dt}, \mu, \hat \mu) = 0,
\end{equation}
where $\mathcal{F}$ is an integro-differential operator. Equation~\eqref{eq::compact_dyn_eq} 
can then be evaluated by discretizing the time interval $t \in [0, t_{\mathrm{f}}]$ in steps $\Delta t = t_{\mathrm{f}} / N_t$. The discretization process results in each of the eight unknown two-time dependent functions being represented by a matrix of size $(2N_t + 1) \times (2N_t + 1)$. The quantities $\mathcal{C}$ and $\mathcal{C}^\mathrm{dt}$ then become matrices of size $2(2N_t + 1) \times 2(2N_t + 1)$. The corresponding total number of variables is therefore $\mathcal{O}(32 N_t^2)$. Because $\mathcal{C}$ and $\mathcal{C}^\mathrm{dt}$ are symmetric, however, the total number of unknowns can be reduced by a factor two, i.e. $\mathcal{O}(16 N_t^2)$.

In our implementation, the operator $\mathcal{F}$ is discretized by approximating its integrals and derivatives. The integrals were computed using the trapezoidal rule, corresponding to errors of order $\Delta t^2$, while the first- and second-order derivatives were approximated using finite difference methods, yielding discretization errors of the same order.

As a starting point for the correlation and response functions with the equilibrium initialization, we conveniently use
\begin{subequations}
    \begin{align}
        C(t, t') &= C_{\mathrm{eq}}(|t-t'|) \\
        \hat R_2(t, t') &= 
        \sign(t-t') \frac{1}{2} \frac{d}{dt} C(t, t') \\
        \hat \chi(t, t') &= \frac{1}{4} \frac{d^2}{dt dt'} C(t, t') \\
        C^\mathrm{dt}(t, t') &= C_{\mathrm{eq}}(t+t') \\
        \hat R_2^\mathrm{dt}(t, t') &= 
        \sign(t-t') \frac{1}{2} \frac{d}{dt} C^\mathrm{dt}(t, t') \\
        \hat \chi^\mathrm{dt}(t, t') &= \frac{1}{4} \frac{d^2}{dt dt'} C^\mathrm{dt}(t, t') \\
        \hat \mu(t) &= 0 \\
        \mu(t) &= \mu_{\mathrm{eq}} = 1 + \frac{\beta^2}{2} f'(1)
    \end{align}
\end{subequations}
with $f(x) \equiv x^p$ and $C_{\mathrm{eq}}(t)$ the solution of the equilibrium dynamical equation
\begin{equation}
    \dot{C}_{\mathrm{eq}}(t) = - C_{\mathrm{eq}}(t) - \frac{\beta^2}{2} \int_0^t ds \, f'(C_{\mathrm{eq}}(t-s)) \, \dot{C}_{\mathrm{eq}}(s) \,.
\end{equation}

\subsubsection{Newton's method with Generalized Minimal Residual Jacobian inversion}

Equation~\eqref{eq::compact_dyn_eq} has been solved using Newton's method with a numerical scheme similar to that used in Ref.~\cite{rizzo2021}. Denoting $x_n$ the vector containing the current approximation to all the unknown variables at iteration $n$, 
 this corresponds to updating
\begin{equation}
\label{eq::Newton_Update}
    x_{n+1} = x_{n} - \mathcal{J}_{\mathcal{F}}^{-1}(x_{n}) \mathcal{F}(x_{n})
\end{equation}
where $\mathcal{J}_{\mathcal{F}}$ corresponds to the Jacobian of the operator $\mathcal{F}$. 

The primary difficulty of using Newton's update~\eqref{eq::Newton_Update} here lies in evaluating the inverse of the Jacobian, which is a matrix of size  $\mathcal{O}(16N_t^2 \times 16 N_t^2)$. Directly inverting such a matrix is computationally prohibitive for practical values of $N_t$.  To surmount this hurdle, we utilized the generalized minimal residual (GMRES) method, which provides an efficient way of approximately solving systems of linear equations $\mathcal{J}_{\mathcal{F}} (x_n) \Delta x_n = \mathcal{F}(x_n)$, where $\Delta x_n \equiv x_{n+1} - x_{n}$, without explicitly computing the inverse of $\mathcal{J}_{\mathcal{F}}$.

GMRES works by constructing an approximate solution within a sequence of Krylov subspaces of increasing dimension. Starting from an initial guess $\Delta x_n^{(0)}$, it iteratively refines the solution by building a subspace spanned by vectors of the form $\{r_0, \mathcal{J}_{\mathcal{F}} \, r_0, \mathcal{J}_{\mathcal{F}}^2 \, r_0 \,, \dots \}$,  where $r_0 \equiv \mathcal{F} - \mathcal{J}_{\mathcal{F}} \Delta x_n^{(0)}$ is the initial residual. Using an Arnoldi iteration for the Gram--Schmidt orthogonalization provides an orthonormal basis for this subspace. At each step, GMRES solves a least-squares regression within the subspace, thus minimizing the norm of the residual and iteratively improving the solution. This approach avoids explicitly computing or storing the full Jacobian by relying instead on computing the powers of the Jacobian on the starting residual. 

A couple other optimizations have also been implemented. First, to further refine the discretization, i.e., to increase $N_t$, 
we have adopted a matrix compression scheme. Because $\mathcal{C}$ and $\mathcal{C}^{\mathrm{dt}}$ are symmetric matrices, as in Ref.~\cite{rizzo2021}  we have only retained the upper diagonal during the GMRES routine. Second, because correlations decay faster near the diagonal, a decimation scheme for the Jacobian was also implemented. 
Because large $N_t$ are particularly important at large $t_\mathrm{f}$, for $t_\mathrm{f} = 40$ and $60$ we have used all the points for $t-t' < \tau_\mathrm{comp}$, but retained only half of them for $t-t' \ge \tau_\mathrm{comp}$, with $\tau_\mathrm{comp}=4.5$. 
Despite this approximation, 
the algorithm converges to fixed points that solve the equations with high accuracy in those particular cases as well. 


\section{Analytical Results}
\label{sec:analyticalresults}
\begin{figure*}[t]
\centering
\includegraphics[width=0.48\textwidth]{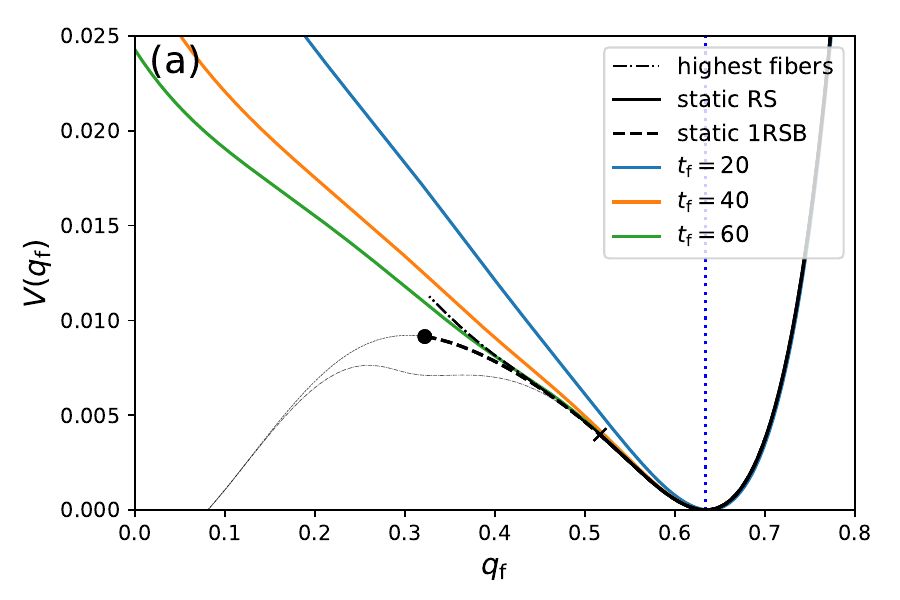}
\includegraphics[width=0.48\textwidth]{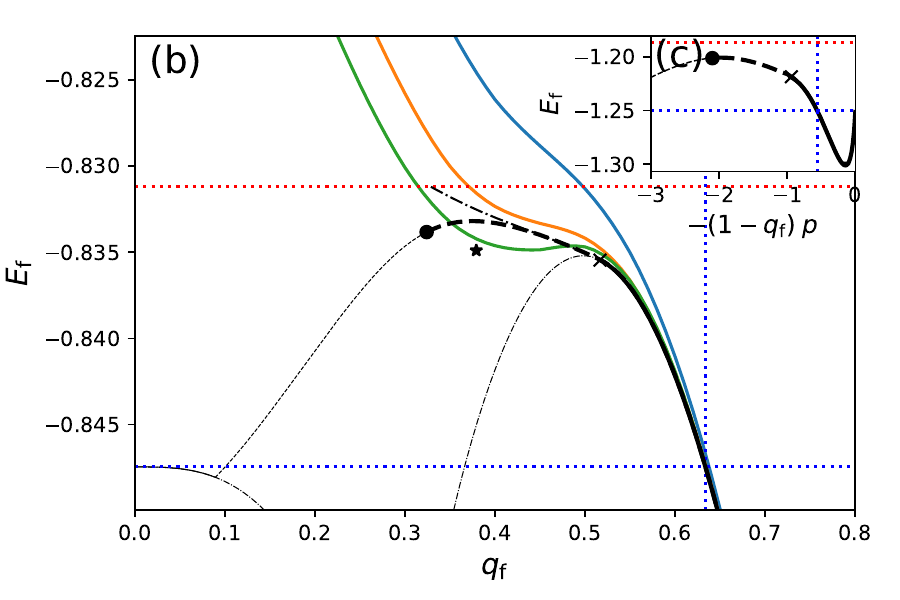}
\caption{\textbf{(a)} Dynamical potential $V(q_\mathrm{f})\equiv V_\mathrm{t_\mathrm{f}}(q)$ and \textbf{(b)} final energy $E_\mathrm{f}\equiv E(\mathrm{t_\mathrm{f}})$ at $T=1/1.695$ for $t_\mathrm{f}=10, 20, 40,$ and $60$ (solid lines). As reference, the (shifted) static FP potential  and the corresponding static energy (thick black lines) are included, along with the RS (thin dash-dotted line) and 1RSB (dashed thin line) extensions as well as the highest free-energy fibers (thick dash-dotted line). Between this last line and the 1RSB line lays the fibered phase of the basin, in which an exponential number (in $N$) of fibers is expected (see Appendix~\ref{sec::CompFib}). 
The equilibrium $E_\mathrm{eq}$ and $q_\mathrm{eq}$ 
(blue dotted lines),  the threshold energy $E_{\mathrm{th}}$ (dashed red line)  as well as the transition from the convex (RS) to the fibered (1RSB) regime at $q_\mathrm{mg}$ 
(cross) and the irreversibility onset at $q_\mathrm{irr}$ (dot) are also indicated. The second minimum of the three replica potential ($M_2$, star)~\cite{cavagna1997}, corresponds to the second replica being at $q_\mathrm{irr}$, thus defining the hub state. 
As expected, as $t_\mathrm{f}$ increases the dynamical potential converges to the FP potential in the convex (RS) regime. Already for $t_\mathrm{f}=60$, an unphysical discrepancy is noted for the energy in the fibered (1RSB) regime.
\textbf{(c):} Static FP energy profile  at $T = 1/2.5<T_\mathrm{d}=1/\sqrt{2e}$, within the clustered phase, for the $p$-spin model in the limit $p \to \infty$ (black line). Even in this case, which corresponds to the (spherical) REM, the 1RSB solution (dashed line) remains below the threshold energy (red dotted line).}
\label{fig:vq}
\end{figure*}




\label{sec:results}


As previously mentioned in Sec.~\ref{sec:schematiclandscape}, analytical results for the dynamical approach described in Sec.~\ref{sec:dynpotentialcalc} were obtained for the spherical $p$-spin glass with $p=3$ in the clustered phase at $T=1/1.695$. 
In order to obtain the profile of the correlation and energy of rare trajectories described by the dynamical potential in Eq.~\eqref{eq:Vtfq}, the dynamical equations in Appendix \ref{app:equations} were integrated for different final $t_\mathrm{f}$ 
and different $q_\mathrm{f}$. 
Figure~\ref{fig:vq} presents both the dynamical potential $V_{t_\mathrm{f}}(q_\mathrm{f})$ and the final energy of the trajectories $E_{\mathrm{f}} = E(t_\mathrm{f})$ against their final overlap $q_\mathrm{f}$. Like for the ferromagnetic case upon decreasing magnetization  in Fig.~\ref{fig:ferro}, the dynamical potential monotonically increases upon approaching $q_\mathrm{f} = 0$. As $t_\mathrm{f}$ increases, $V_{t_\mathrm{f}}(q)$ steadily converges to the static FP potential at high $q$, while for $q < q_\mathrm{irr}$ the two potentials obviously diverge. 

In the convex (RS) regime, $q_\mathrm{f} > q_{\mathrm{mg}}$, we observe a clear convergence of the dynamical results toward the static predictions for both the dynamical potential and the energy, with excellent agreement already at $t_\mathrm{f} = 60$. By contrast, in the fibered (RSB) regime, $q_\mathrm{irr} < q_\mathrm{f} < q_\mathrm{mg}$, there is no such convergence.  
Recall, however, that the integrated dynamical equations have a RS structure. Therefore, only in the convex regime are the results of these equations expected to converge to the static FP potential in the limit $t_\mathrm{f}\to\infty$. For instance, in the fibered regime the energy  artifactually curves towards the static RS approximation. 
Because of the reversibility in the dynamics in the fibered regime, we also know that a \emph{correct} dynamical evaluation of the energy of rare trajectories accounting for RSB should instead lead to the result described by the static FP potential in the limit $t_\mathrm{f}\to\infty$. Consequently, we conclude that the dynamical equations derived within the RS scheme in Sec.~\ref{sec:gendynpot} are insufficient to describe the whole dynamics. Separately, we note that the energy of the dominant fibers in the limit $t_\mathrm{f}\to\infty$ --given by the static FP potential with the RSB ansatz-- remains nearly constant and \emph{well below the threshold energy}. We therefore also conclude from a purely static calculation that the energy of escape trajectories --since it should match the one dominant fibers-- is expected to remain well below the threshold energy in the limit $t_\mathrm{f}\rightarrow\infty$. Interestingly, this result holds for any $p\geq3$, including in the limit $p\rightarrow\infty$ (see Fig.~\ref{fig:vq}c). Even for the minimal (spherical) REM is state escape expected to follow a non-trivial pathway that does not involve the threshold states. 



In the rest of this section, we first discuss the behavior of the correlation and of the energy of rare trajectories when the overlap is in the range $q_\mathrm{f}>q_\mathrm{irr}$, distinguishing the convex, $q_\mathrm{f}>q_\mathrm{mg}$, from the fibered, $q_\mathrm{irr}<q_\mathrm{f}<q_\mathrm{mg}$, regimes. The instantonic regime with $q_\mathrm{f}<q_\mathrm{irr}$, in which dynamics is irreversible is then considered. As described in Sec.~\ref{sec:schematiclandscape}, both $q_\mathrm{mg}$ and $q_\mathrm{irr}$ are obtained by static calculations; they correspond to the rupture of replica symmetry from RS to 1RSB \cite{franz1995} and to the appearance of the second minimum in the three replica potential \cite{cavagna1997}, respectively. 


\begin{figure}[t]
\centering
\includegraphics[width=9cm]{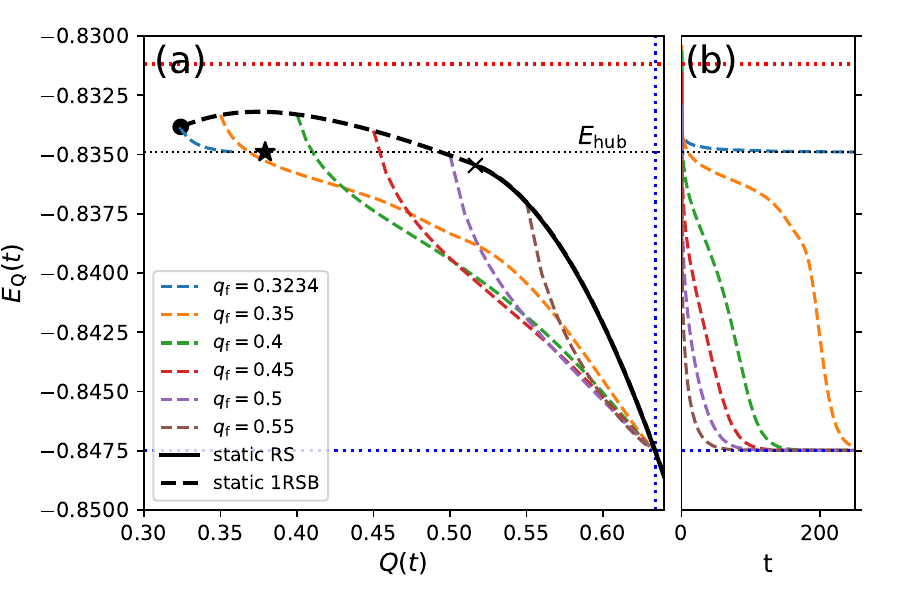}
\caption{\textbf{(a):} Time parametric plot of $E_\mathrm{Q}(t)$ and $Q(t)$ with the reference equilibrium configuration $\tau$, for initial configurations sampled from the 
static FP potential (thick black line) \cite{Barrat1998} at different overlap with $\tau$, $q_\mathrm{f}=q_\mathrm{irr}, 0.35, 0.4, 0.45, 0.5,$ and $0.55$ (dashed colored lines). For all $q_\mathrm{f}>q_\mathrm{irr}$ the trajectory remains within the basin of attraction of $\tau$, with $q_\mathrm{eq}$ and $E_\mathrm{eq}$ (dotted blue lines). \textbf{(b):} Dynamical evolution of  $E(t)$ under the same conditions. Note that the relaxation time diverges as $q_\mathrm{f}$ approaches $q_\mathrm{irr}$, and that the dynamics then remains stuck in a metastable state (star) of energy $E_\mathrm{Q}(\infty)=E_\mathrm{hub} > E_\mathrm{eq}$.
}
\label{fig:rev_dyn}
\end{figure}

\begin{figure*}[t]
\centering
\includegraphics[width=0.99\textwidth]{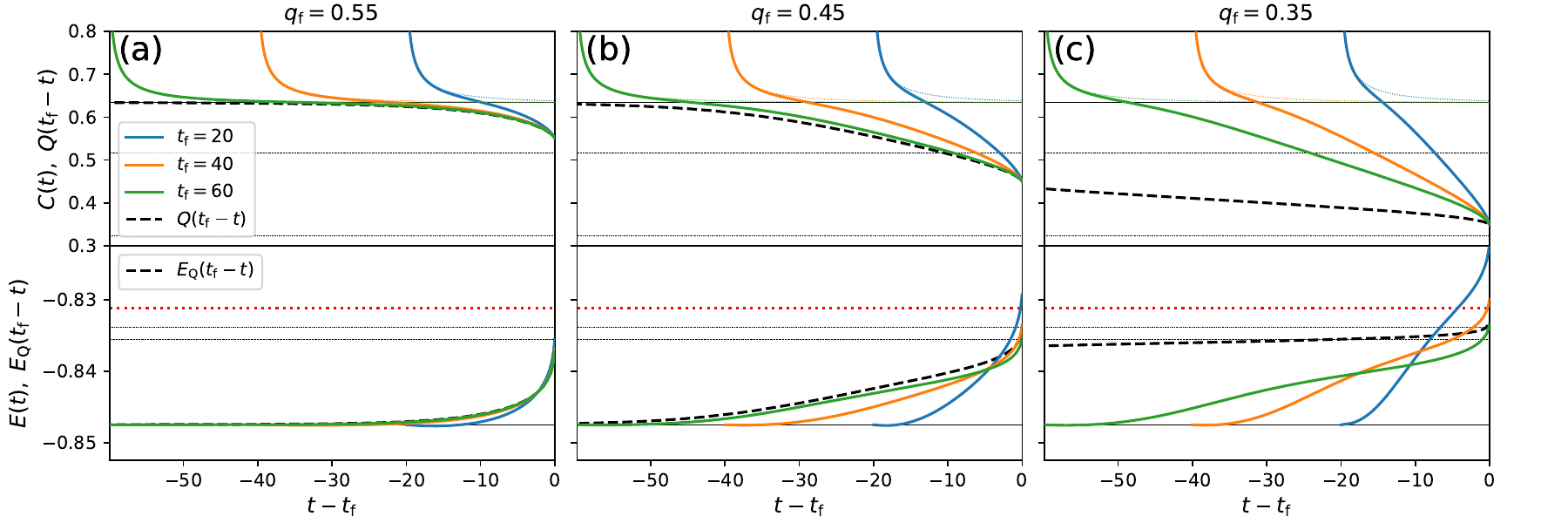}
\caption{Comparison between (bottom) the energy $E(t)$  and (top) the correlations  for forward rare trajectories $C(t)$ (colored lines) and backward relaxation trajectories $Q(t_\mathrm{f}-t)$ (dashed black line) for three different overlaps: \textbf{(a)} $q_\mathrm{f}=0.55$, \textbf{(b)} $0.45$, and \textbf{(c)} $0.35$. As reference, the equilibrium $C(t)$ (pale colored lines) as well as $q_\mathrm{eq},E_\mathrm{eq}$ (solid black lines),  $q_\mathrm{mg},E_\mathrm{mg}$ (dashed black lines), $q_\mathrm{irr},E_\mathrm{irr}$ (dotted black lines)  and the threshold energy $E_\mathrm{th}$ (red dotted line) are provided. In \textbf{(a)}, backward and forward dynamics nearly perfectly match for $t_\mathrm{f}=40$ already. In \textbf{(b)}, a good agreement is observed for $t_\mathrm{f}=60$, even though the forward solution uses an approximate (RS instead of 1RSB) dynamics. In \textbf{(c)}, because the relaxation time of the backward dynamics, $t_\mathrm{Q}(q_\mathrm{f})\approx 200$ (see Fig.~\ref{fig:rev_dyn}), is much larger than $t_\mathrm{f}=60$, the matching regime is not reached.}
\label{fig:REV}
\end{figure*}

\subsection{Reversible Regime}
\label{sec:RevReg}

For $q_\mathrm{f}>q_\mathrm{irr}$ the dynamics is confined to the basin of the reference configuration $\tau$. Therefore, in this regime, the dynamical potential, Eq.~\eqref{eq:Vtfq}, is dominated by a single trajectory in the limit $t_\mathrm{f}\to\infty$, as in the ferromagnetic case. Two important remarks follow from this identification: (i) the trajectory is \textit{self-averaging} (in the thermodynamic limit), and (ii) the most probable trajectory reaching $q_\mathrm{f}$ (\textit{forward trajectory}) is time reversed relative to the relaxational trajectory starting from $q_\mathrm{f}$ sampled with the static FP measure (\textit{backward trajectory}). This backward trajectory for $q>q_\mathrm{irr}$ can be independently constructed using the approach of Barrat and Franz \cite{Barrat1998}. 

Figure~\ref{fig:rev_dyn} presents a parametric plot of the energy and  overlap for the time-reversed trajectories sampled following the Barrat--Franz approach at different $q_\mathrm{f}$. Each one equilibrates within the state defined by the reference configuration $\tau$. As a result, the overlap $Q(t)$ with $\tau$ evolves from $Q(0)=q_\mathrm{f}$ to $Q(\infty)=q_\mathrm{eq}$, and so does the energy, with $E_\mathrm{Q}(0)=E_\mathrm{f}$ and $E_\mathrm{Q}(\infty)=E_\mathrm{eq}$. The relaxation time $t_\mathrm{Q}(q_\mathrm{f})$ diverges as $q_\mathrm{f}$ approaches $q_\mathrm{irr}$. Interestingly, a system initialized at $q_\mathrm{irr}$ becomes asymptotically trapped in a metastable state at an energy higher than the equilibrium one. Somewhat counterintuitively, this metastable \textit{hub} state lies closer to the reference configuration $\tau$, with $q_\mathrm{hub} > q_\mathrm{irr}$. For more details,  see Appendix \ref{sec::CompFib}.

For the dynamical potential, we expect that at large $t_\mathrm{f}$ the correlation $C(t)$ with the initial condition first decays to $q_\mathrm{eq}$ and then goes from $q_\mathrm{eq}$ to the target value $q_\mathrm{f}$ through a process time-reversed relative to the backward relaxation, as discussed in Sec.~\ref{sec:schematiclandscape}. As a check, Fig.~\ref{fig:REV} compares forward (colored lines) and backward (black dashed lines) profiles at different $q_\mathrm{f}$. As expected, a near perfect agreement between $C(t) \rightarrow  Q(t_\mathrm{f}-t)$ is observed for $q_\mathrm{f}=0.55$, in the convex regime. As $q_\mathrm{f}\to q_\mathrm{irr}$, however, agreement worsens for two reasons. First, for the forward and backward trajectories to converge, one would need at least $t_\mathrm{f} > t_\mathrm{Q}(q_\mathrm{f})+t_\mathrm{eq}$, the last being the equilibrium relaxation time, which lies beyond computational reach as $t_\mathrm{Q}(q_\mathrm{f})$ increases. Second, for $q<q_\mathrm{mg}$ the backward relaxation is computed using 1RSB dynamics, while the forward one uses RS dynamics, and hence do not match even asymptotically.

\begin{figure}[t]
\centering
\includegraphics[width=\columnwidth]{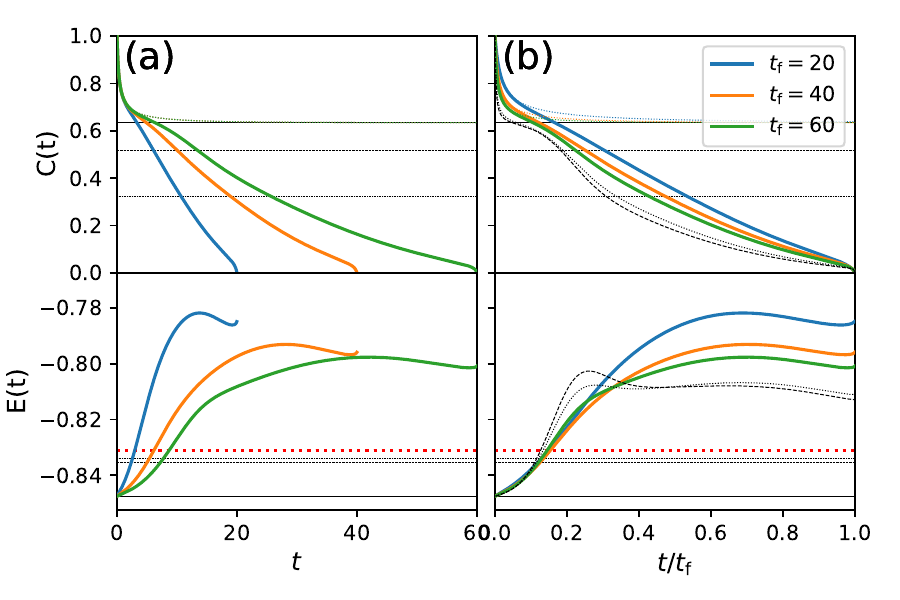}
\caption{\textbf{(a):} Time evolution of (top) the correlation with the initial configuration $C(t)$ and (bottom) the energy $E(t)$  for $q_\mathrm{f}=0$ and $t_\mathrm{f}=20$, $40$, and $60$. Equilibrium $C(t)$ results (colored dotted lines) are provided as reference. 
\textbf{(b):} Same results as in \textbf{(a)} rescaled as $t/t_\mathrm{f}$. This time rescaling appears to collapse results as $t_\mathrm{f}$ grows. Extrapolations to the limit $t_\mathrm{f}\to\infty$ using linear (dotted black lines) and quadratic (dash-dotted lines) forms are also attempted. In both cases, the energy rises from equilibrium (solid black line) well above the threshold energy $E_\mathrm{th}$ (red dotted line), and hence of $E_\mathrm{irr}$ (dashed lines), thus further validating their nonphysical nature, as in Ref.~\cite{rizzo2021}. 
 }
\label{fig:Etq0}
\end{figure}

\subsection{Instantonic Irreversible Regime}
\label{iir}
Although the above analysis reveals the quantitative inadequacy of $V_{t_\mathrm{f}}(q)$ -- as computed here -- for $q<q_\mathrm{mg}$, we nevertheless expect its predictions to be qualitatively close to the true physical behavior for $q>q_\mathrm{irr}$, as can be validated by comparing with the FP potential. However, for $q<q_\mathrm{irr}$ -- as for the Ising case for $m<m_\mathrm{max}$ -- $V_{t_\mathrm{f}}$ then markedly differs from the FP potential. Analyzing the dynamical results is then somewhat trickier. 


We nevertheless note a key qualitative difference from what happens for $q>q_\mathrm{irr}$. It seems that the correlation $C(t)$ and energy $E(t)$ then evolve on a scale that grows with $t_\mathrm{f}$. The difference in scaling is particularly clear for $q=0$, as can be seen in
Fig.~\ref{fig:Etq0}. The results seemingly follow a scaling of the form $E(t)=E_u(t/t_\mathrm{f})$ at large $t_\mathrm{f}$, as was also observed in Ref.~\cite{rizzo2021}. 

In principle, this scaling could be explained by a quasi-equilibrium dynamics in which the system is in equilibrium on finite timescales inside states that are slowly evolving on large $O(t_\mathrm{f})$ timescales. (Note that this is  precisely what happens in the aging dynamics \cite{cugliandolo1993analytical} of this model.) However, quasi-equilibrium dynamics would require both the presence of states at the given energy and that these states be marginal.  This last condition is at odds with the fact that, much as in Ref.~\cite{rizzo2021}, extrapolations of $E_f$ for $t_\mathrm{f} \to \infty$ tend to energies much higher than the threshold, where states are typically marginal. The extrapolated energy is in fact so high that we do not expect to find states at all at that point. 

The above discussion nevertheless offers a possible resolution to this conundrum. In the instantonic regime, RSB dynamics is required; RS energies lead to inconsistent results. The same likely happens in Ref.~\cite{rizzo2021}. Unfortunately, extending the dynamical computation to RSB would be quite involved, and it is therefore left for future work.





\begin{figure}[t]
\centering
\includegraphics[width=\columnwidth]{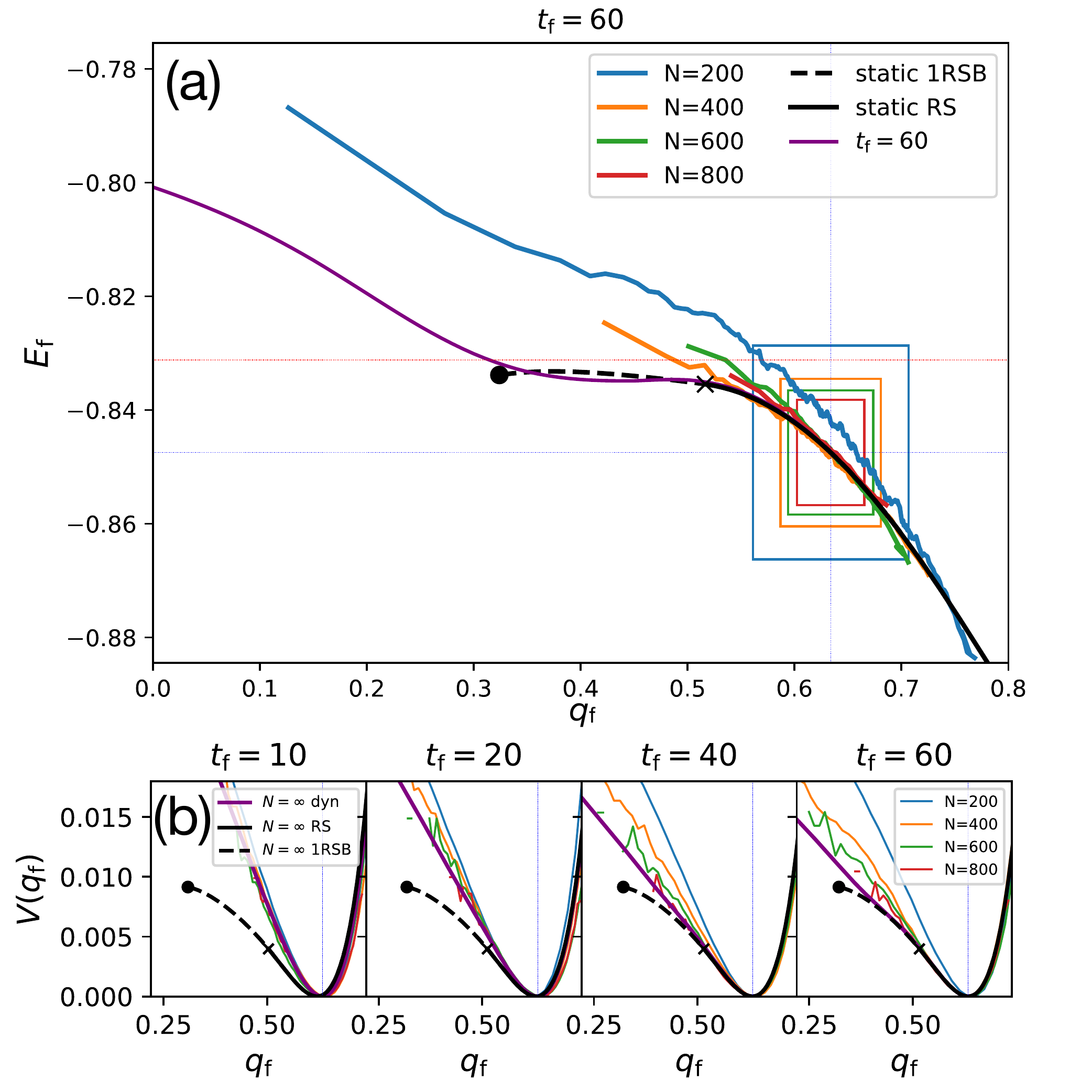}
\caption{\textbf{(a):} Parametric plot of the energy $E(t_\mathrm{f})$ and overlap $C(t_\mathrm{f})$ for trajectories with $t_\mathrm{f}=60$. Simulation results for $N=200,400,600$, and $800$ at $T=0.5900$ (full colored lines) are compared with the theory results for dynamics (purple) and statics (full black line for RS solution; dashed line for the 1RSB solution) at $T=1/1.695\approx0.5900$. As reference, the marginal 
(cross), irreversibility 
(filled circle), and equilibrium 
(dotted blue lines) points as well as the threshold energy $E_\mathrm{th}$ 
(dotted red line) are provided. The range of equilibrium values pre-selected (one standard deviation around $q_\mathrm{eq}$ and $E_\mathrm{eq}$) for initial configurations (colored rectangles) is also given. As $N$ increases the simulations results cleanly converge to the theoretical predictions. \textbf{(b):} Dynamical potential $V_{t_\mathrm{f}}(q)$ for $t_\mathrm{f}=10,20,40$, and $60$ from theory (purple line) and simulations with different $N$ (colored lines). As reference, the static FP potential (solid and dashed black line) up to $q_\mathrm{irr}$ is provided. Up to $t_\mathrm{f}=40$, the simulation results for the largest system are essentially indistinguishable from the theoretical results.}
\label{fig:Fig1Sim}
\end{figure}

\section{Simulations Results}
\label{sec:simulationresults}
Given the inherent limitation of RS dynamics for $q<q_\mathrm{mg}$ identified in Sec.~\ref{sec:results}, numerical simulations offer an important validation and complement. The equilibrium dynamics of the spherical $3$-spin glass model is therefore also simulated for equilibrated systems of size $N=200, 400, 600$, and $800$ at $T_\mathrm{s}<T=0.5900\approx1/1.695<T_\mathrm{d}$, starting from configurations obtained by quiet planting \footnote{Quiet planting consists of first choosing a reference configuration uniformly at random and then sampling the quenched disorder so that the chosen configuration is sampled with its proper equilibrium weight for the system.} 
as in Ref.~\cite{Folena_2021}. 
As N increases, we reduce the number of planted configurations as 8000, 4000, 2000, and 1000. Because planted systems are subject to $N^{-1/2}$ sample-to-sample fluctuations, initial states are pre-selected to speed convergence to the thermodynamic limit, $N\to\infty$. For each planted configuration, we evaluate the distribution of overlap $q$ and energy $E$ at long times, $\approx t_\mathrm{eq}$, over 100 trajectories (see \cite{Folena_2022}), and only configurations with $q$ and $E$ within one standard deviation of $q_\mathrm{eq}$ and $E_\mathrm{eq}$ (for $N\to\infty$) are kept. For each of these, 100 trajectories are evolved up to $t_\mathrm{e}=100$. The resulting dataset is used for sampling conditional averages at different final times $t_\mathrm{f}$ and overlaps $q_\mathrm{f}$.

Figure~\ref{fig:Fig1Sim}(a) shows the resulting $E(t_\mathrm{f})$ 
for $t_\mathrm{f}=60$. 
As expected, curves converge to the thermodynamic result as $N$ increases.\footnote{Such convergence is thanks to pre-selection; without pre-selection, numerical results are more steeply inclined than the equilibrium curve.} However, the range of accessible $q$ then also markedly shrinks, thus limiting the extent of the validation.
Figure~\ref{fig:Fig1Sim}(b) more convincingly shows the agreement between simulation and analytical results of the dynamical potential for different $t_\mathrm{f}$. From this comparison, we also see that systems with $N=200$ and $400$ are sufficiently large to limit finite-$N$ corrections yet present a sufficiently broad range of $q$ so as to attain the fibered and instantonic regimes. We get back to this issue below.


\begin{figure*}[t]
\centering
\includegraphics[width=0.8\textwidth]{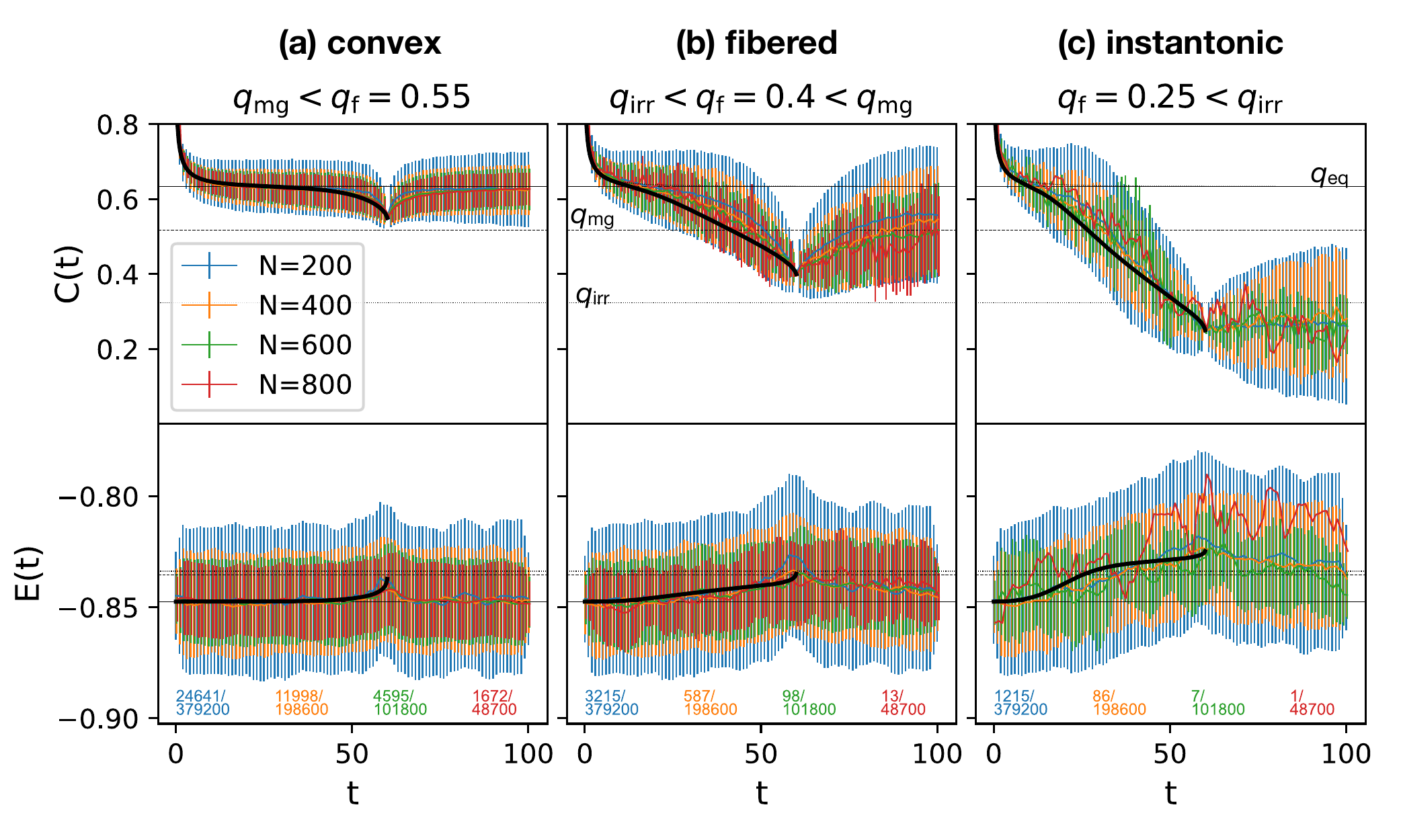}
\caption{(top) Correlation with the reference configuration $C(t)$  and (bottom) energy $E(t)$ of finite $N$ simulations (colored lines) and analytical dynamics (thick black lines) with $t_\mathrm{f}=60$ for overlaps \textbf{(a)} $q_\mathrm{f}=0.55$, \textbf{(b)} $q_\mathrm{f}=0.4$, and \textbf{(c)} $q_\mathrm{f}=0.25$, which fall in the \textit{convex},  \textit{fibered}, and \textit{instantonic} regimes, respectively. These conditional averages are computed through direct sampling of trajectories. For each $N$, the colored bars denote  the standard deviation over trajectories. (The fractions at the bottom of each panel give the ratio of the number of trajectories to the number of pre-selected ones.) \textbf{(a):} In the convex (RS) regime, the free energy landscape is indistinguishable from that of a ferromagnetic model, and so is its dynamics. The correlation quickly relaxes to $q_\mathrm{eq}$, and only upon approaching $t_\mathrm{f}$ does it jump to  $q_\mathrm{f}$. It subsequently relaxes back to its equilibrium value. \textbf{(b):} In the fibered (1RSB) regime, 
replica symmetry breaking implies that 
only a few fibers dominate the probability measure, and that as a result fluctuations are markedly increased. Results for $C(t)$ are particularly illustrative in this respect. Because the dynamics remains reversible, however, after reaching $q_\mathrm{f}$ (and $E_\mathrm{f}$) the system relaxes back to equilibrium. \textbf{(c):} 
In the instantonic regime, the system does not relax back to an equilibrium state, and remains at a larger finite distance from the reference equilibrium configurations, with a markedly higher energy.}
\label{fig:Fig2Sim}
\end{figure*}

Figure~\ref{fig:Fig2Sim} shows the average shape of $E(t)$ and $C(t)$ for $t_\mathrm{f}=60$ and different $N$ in the (a) \textit{convex}, (b) \textit{fibered}, and (c) \textit{instantonic} regimes. Note that these regimes are defined (are only meaningful) in the ordered limit $t_\mathrm{f}\to\infty$ before $N\to\infty$. The dynamical potential, by contrast, is valid for any $t_\mathrm{f}$ in the $N\to\infty$ limit. Note also that the correlation of rare trajectories is \textit{not self-averaging}. In the $N\to\infty$ limit, sample-to-sample fluctuations remain relevant, especially in the 1RSB regime.

In the convex regime, the dynamics is expected to be reversible and self-averaging. The average correlation indeed rapidly relaxes to the equilibrium overlap $q_\mathrm{eq}$ and jumps to $q_\mathrm{f}$ only near $t_\mathrm{f}$. Then, for $t>t_\mathrm{f}$, the system relaxes back to the initial state (mirroring trajectory), as does the energy. 
In the fibered regime, the dynamics is also expected to be reversible, but sample-to-sample fluctuations can produce \textit{non self-averaging} trajectories. While the average correlation steadily relaxes back to the initial state, its fluctuations do not appear to scale as $1/\sqrt{N}$, as a self-averaging quantity would. 
In the instantonic regime, by contrast, trajectories instead of relaxing back to the reference state with $q_\mathrm{eq}$ or to a distinct equilibrium state with $q=0$ remain stuck in new states with an overlap near $q_\mathrm{f}$ and an energy near $E_\mathrm{f}$. 

\begin{figure}[t]
\centering
\includegraphics[width=\columnwidth]{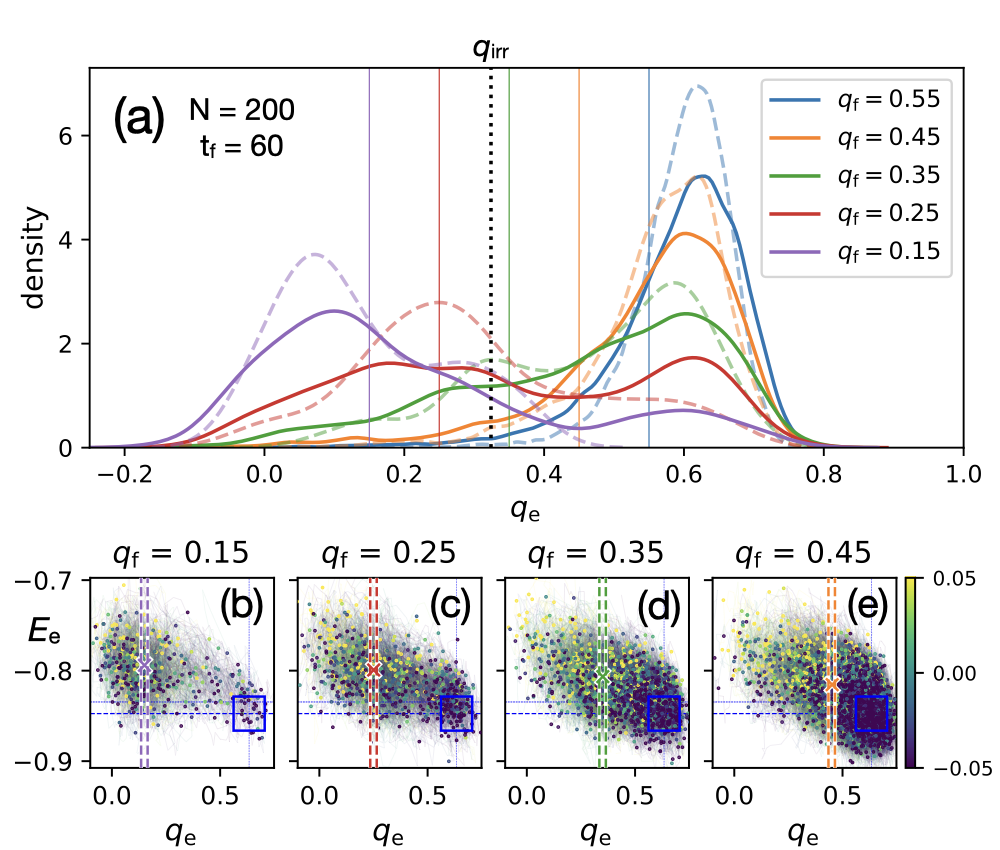}
\caption{Dynamical (ir)reversibility analysis of the simulation trajectories in Fig.~\ref{fig:Fig2Sim} for $t_\mathrm{f}=60$, considering their end values (at $t_\mathrm{e}=100$), $q_\mathrm{e}$ and $E_\mathrm{e}$. \textbf{(a):} Distributions of $q_\mathrm{e}$ for trajectories with different  $q_\mathrm{f}$ 
for  $N=200$ (full lines) and $N=400$ (dashed lines). For $q_\mathrm{f}>q_\mathrm{irr}$, the dynamics is typically reversible and relaxes back to equilibrium (right-hand peak); for $q_\mathrm{f}<q_\mathrm{irr}$, the dynamics remains stuck away from the reference equilibrium configuration. For $q_\mathrm{f}=0.35$, which is nearest to $q_\mathrm{irr}$, 
the distinction is most ambiguous. Given the slow equilibrium relaxation time in the intermediate regime, a significant fraction of the trajectories persist in that vicinity. Note that for $q_\mathrm{f}=0.25$ the peak is also near $q_\mathrm{f}$, but for $q_\mathrm{f}=0.15$ the peak shifts to smaller overlap. 
\textbf{(b)-(e):} scatter plot of $E_\mathrm{e}$ and $q_\mathrm{e}$ for the trajectories with $N=200$ in (a) with \textbf{(b)} $q_\mathrm{f}=$  0.15, \textbf{(c)} 0.25, \textbf{(d)} 0.35, and \textbf{(e)} 0.45. As reference, the pre-selection range (square blue box), the $q_\mathrm{f}$ selection range $t_\mathrm{f}=60$ (dashed vertical lines) as well as the average $E_\mathrm{f}$ (cross) are provided. The color of each point (scale reported in colored bar) indicates the jump in energy $E_\mathrm{e}-E_\mathrm{f}$, dark blue means relaxing to lower energy. In \textbf{(d)} and \textbf{(e)} the clouds concentrates around the equilibrium pre-selection box, while in \textbf{(c)} and \textbf{(b)} the cloud lies around $(q_\mathrm{f},E_\mathrm{f})$, which implies the presence of \textit{hub} states at that overlap and energy.} 
\label{fig:Fig3Sim}
\end{figure}

In order to better understand the fate of trajectories and the role of reversibility, Fig.~\ref{fig:Fig3Sim}(a) -- for the same trajectories as in Fig.~\ref{fig:Fig2Sim} -- presents the distribution of the \emph{ending} overlap $q_\mathrm{e}$ at time $t_\mathrm{e}=100$ for trajectories with different $q_\mathrm{f}$ at $t_\mathrm{f}=60$. For $q>q_\mathrm{irr}$ the distribution is peaked around the equilibrium overlap, while for $q<q_\mathrm{irr}$ it is peaked at  $q\lesssim q_\mathrm{f}$. As expected, for $N=400$ compared to $N=200$, the distribution concentrates and the irreversibility is more pronounced. Insets show the scatter plot of the end energies $E_\mathrm{e}$ and overlaps $q_\mathrm{e}$ for different $q_\mathrm{f}$. For  $q<q_\mathrm{irr}$, not only does $q_\mathrm{e}$ peak around $q_\mathrm{f}$, but $E_\mathrm{e}$ also remains higher in the landscape. In other words, the system seemingly gets temporarily trapped in some high energy states before relaxing again to other equilibrium states. We interpret these states as \emph{hubs} of the equilibrium relaxations. However, it remains unclear which of these hubs is most relevant in the limit $N\to\infty,t_\mathrm{f}\to\infty$. Interestingly, for the lowest $q_\mathrm{f}$ considered, systems relax, which could be understood as the hub having been then exited. This interpretation, however, remains fairly tentative.

\section{Discussion and Speculations}
\label{sec:discu_specul}

In this work, we have analyzed the rare trajectories that enable a system to escape equilibrium states for a $p$-spin spherical glass, a prototypical model for the RFOT universality class. We validated in a fully dynamical setting the existence of two key transitions that define the basin of attraction of an equilibrium state, previously suggested via a static analysis. The basin is convex for overlaps $q > q_\mathrm{mg}$ and then fibered as the overlap decreases until $q_\mathrm{irr}$, at which point the system exits the basins of attraction and the dynamics becomes irreversible. Within basins, the static and dynamical descriptions are perfectly consistent, and the backward relaxation dynamics accurately describes the behavior of rare trajectories in the limit $t_\mathrm{f} \to \infty$. However, beyond $q_\mathrm{irr}$, the fate of the dynamics remains unclear. Tackling this question directly would require a 1RSB (or fullRSB) dynamical calculation over much longer timescales than is computationally feasible at this time.\footnote{Even the current numerical integrations required substantial resources, using up to 516 GB of virtual memory.}

\begin{figure}[t]
\centering
\includegraphics[width=0.7\columnwidth]{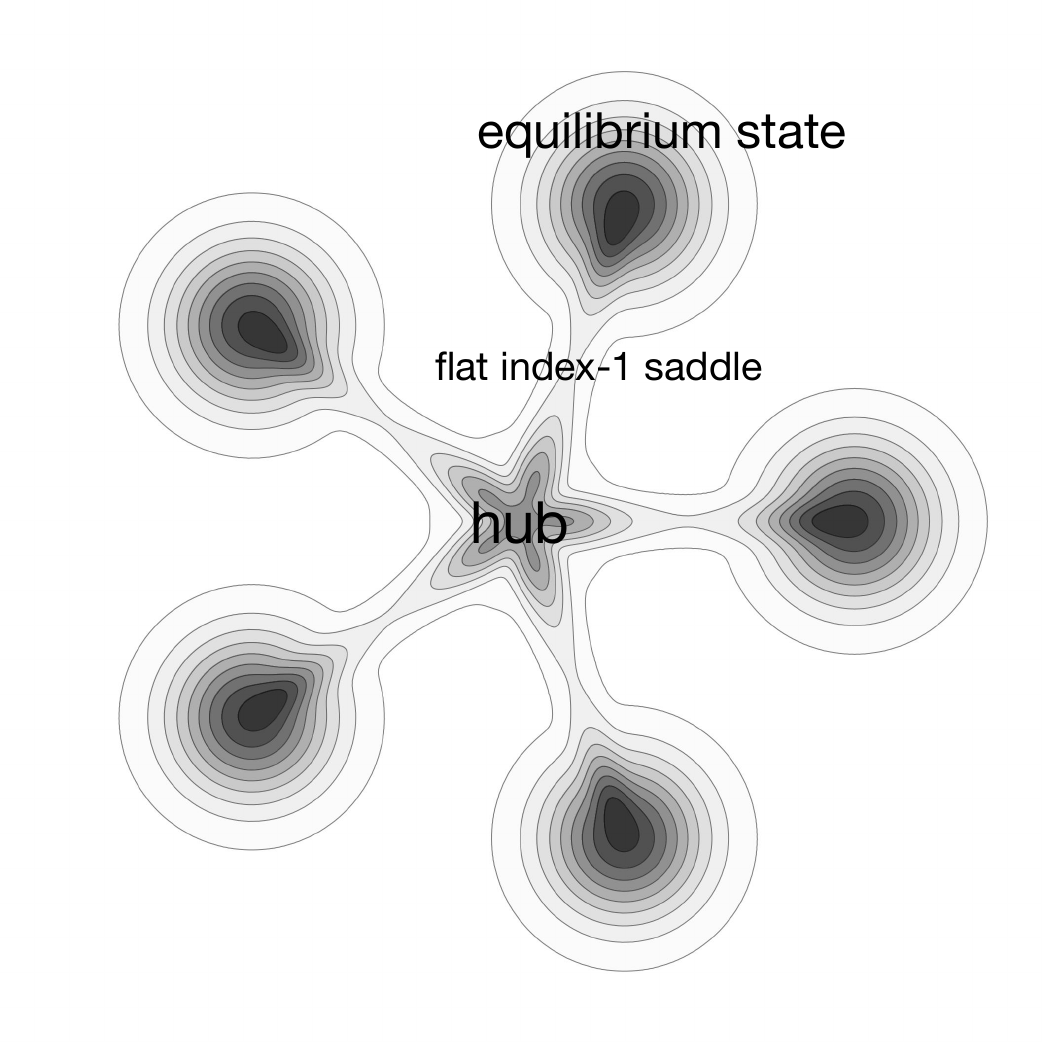}
\caption{Schematic connection between equilibrium states to a central hub state through flat index-1 saddles in the TAP free energy landscape. This proposal naturally follows from the observations in this work, but remains to be fully validated. Note that this sketch should be drawn on a curved surface in order to respect the condition $q_\mathrm{hub} > q_\mathrm{irr}$.} 
\label{fig:hubs}
\end{figure}

Our analysis nevertheless highlights the fundamental role played by the irreversible overlap $q_\mathrm{irr}$ in the relaxation dynamics. At this distance from a reference equilibrium configuration $\tau$, we identify the nearest fiber saddle that lies below threshold and thus facilitates escape from the basin. We observed that this is valid for general $p$, even in the limit $p\to \infty$ which corresponds to the (spherical) REM. As schematized in Fig.~\ref{fig:hubs}, this saddle appears to connect to a high-energy metastable \emph{hub} state, which corresponds to the second minimum of the three-replica potential $M_2$.

At $T \to 0$, escape rates can be determined by analyzing saddles of the energy landscape \cite{Berglund2013}. At finite $T$, the Thouless--Anderson--Palmer (TAP) free energy landscape plays a comparable role, as all stationary points, not just minima, define the long-time dynamics. (See, e.g., \cite[Eq.~57]{Biroli1999}.) The analysis of saddles in the TAP free energy landscape would therefore allow a Kramers-like calculation of escape probabilities.

The metastable state reached during relaxation from $q_\mathrm{irr}$ corresponds to a local minimum of the TAP free energy landscape, which is connected via an index-1 saddle to another TAP state representing the initial equilibrium. This identification motivates a direct investigation of the TAP landscape through Kac--Rice constrained complexity calculations, thus revisiting the energy landscape analyses presented in Refs.~\cite{Folena_2020,ros2019b,jaron2024}. The objective of this effort would be to determine whether these metastable states indeed act as dominant \textit{hubs} in the relaxation process by linking exponentially many equilibrium states (see Fig.~\ref{fig:hubs})\footnote{A detailed analysis of the landscape geometry around (zero-temperature) ground states in related models of jamming and constrained optimization,  such as the negative perceptron~\cite{Franz_2016} and the one-hidden-layer neural networks~\cite{Star2023,zambon2025sampling}, finds a star-shaped structure strikingly similar to that conjectured in Fig.~\ref{fig:hubs} for the finite-$T$ states of pure $p$-spin models. To what extent this structure is universal in different optimization problems remains an open question.}. While these saddles are expected to be index-1, with a single unstable direction correlated with the reference equilibrium state (as seen in previous energy landscape studies), the hub states--although closer to the equilibrium states ($q_\mathrm{hub} > q_\mathrm{irr}$)--are expected to be directionally uncorrelated with them. This lack of alignment further supports a curved fiber geometry. Hubs are expected to mainly facilitate connections between numerous equilibrium states. If this hypothesis is correct, then the long-time dynamics is expected to resemble a jump process between random pairs of equilibrium states mediated by these hubs, resulting in a trap model-like dynamics \cite{bouchaud1992}, in which the traps exhibit nontrivial sample-to-sample fluctuations \cite{Franz_2020}.

This picture is further complicated by the fact that, based on the backward dynamics results, hub states are reached on a diverging timescale, i.e., $t_\mathrm{Q}(q_\mathrm{irr}) = \infty$. In practice, this divergence arises due to the thermodynamic limit $N \to \infty$, whereas for finite $N$, there is a $N$-dependent cutoff to this timescale: $t^N_\mathrm{Q}(q_\mathrm{irr}) = O(N^a)$. This divergence is probably related to the near perfect flatness of the index-1 TAP saddles that allow escape. Such flatness introduces a $N^{a}$-dependent prefactor into the Kramers escape rate with a power law $a=1/4$ \cite{Berglund2008}. From this analysis, we expect the typical instantons in RFOT models to be \textit{anomalous} in that their inherent timescale diverges with $N$. Their behavior would then be strongly dependent on system size, thus necessitating a thorough scaling analysis. In other words, the optimal relaxation trajectory of an RFOT system involves exponentially rare system-wide rearrangements, which require a time of $O(N^a)$ to occur. However, we cannot rule out the possibility that non-anomalous instantons might exist in other RFOT models. This could be the case, for instance, if $q_\mathrm{irr}>q_\mathrm{mg}$. In such a scenario, the basin might still be fibered, but the dominant fibers would no longer correspond to the lowest energy ones (which are also the flattest ones). 

What happens below the static temperature, $T<T_\mathrm{s}$, remains open. We venture to speculate that because the TAP free energy landscape is then tilted towards the dominant state, even if there are fibers that allow that state to reach a metastable minimum -- a proto-hub -- no forward lower free energy states could then be found. The system would then relax back to the dominant state. This repeated back-and-forth motion across multiple fibers may bear some connection to the two-level systems observed in low-energy glasses \cite{folena2022marginal}. 

This work opens a path toward a first-principles understanding of the relaxation dynamics of structural glasses. The mean-field RFOT physics can be interpreted in terms of either particle number or spatial dimension. In the former, exemplified by the $p$-spin model, the finite-range analogue leads to nucleation, which in disordered systems corresponds to the mosaic picture~\cite{cavagna2009}. Our hub/fiber framework provides a first mathematical basis for this phenomenology. In the latter, exemplified by the random Lorentz gas (RLG), relaxation appears as the escape of a particle from its cage of neighbors \cite{Biroli2021}, which in multiparticle systems can lead to dynamical facilitation~\cite{Charbonneau2014}. (This effect displays the same mean-field mode-coupling transition as RFOT systems \cite{PerrupatoRizzo2025}.) Extending our first-principle analysis to the RLG should therefore supply a dynamical structure based on the hub/fiber organization of single-particle escapes, which play a determining role in glass relaxation~\cite{ScallietGuiselinBerthier2022,OzawaBiroli2023}. 
In addition, our work introduces the dynamical potential as an observable for simulations of structural glass formers. Just as the static Franz--Parisi potential has been adapted to understand static correlations~\cite{Guiselin2022}, the dynamical potential could be used to identify how dynamical heterogeneity affects structural relaxation. 


To conclude, our analysis not only unifies and clarifies various earlier works on activated processed in RFOT model, but also opens a window -- fully based on dynamics -- on the fundamental mechanisms of relaxation in structural glasses. Several promising future directions then emerge. These include a detailed investigation of the TAP landscape constrained by equilibrium and hub states, as well as an exploration of various mixed $p$-spin models to determine whether different types of instantons, both anomalous and non-anomalous, could be identified. The implementation of these findings to structural glass models would bring about the most challenging part of the glass problem within a first-principle grasp.

$\\$
\paragraph*{Acknowledgements.}
We acknowledge stimulating discussions with Riccardo Cipolloni, Silvio Franz, Giorgio Parisi, and Federico Ricci-Tersenghi. E.M.M. acknowledges the MUR-Prin 2022 funding Prot. 20229T9EAT, financed by the European Union (Next Generation EU). G.F. acknowledges partial support from a postdoctoral fellowship from the Duke Center on Computational Thinking. P.C. acknowledges partial support from the Simons Foundation (Grant No.~454937). P.C. also thanks the Chimera group for their hospitality during part of this work.

\paragraph*{Data availability statement.}
The data that support the findings of this article are openly available~\cite{RDR} 

\bibliography{biblio}

\clearpage

\appendix

\begin{widetext}

\section{Analytical Calculations}~\label{sec::anal_calculations}
As shown in the main text, the dynamical potential can be written as
\beq
V_{t_\mathrm{f}}(q_{\mathrm{f}}) \equiv \sum_{\tau} P_\mathrm{eq}(\tau) \ln \sum_\sigma T_{t_\mathrm{f}} (\sigma | \tau) \delta( N q_{\mathrm{f}} - \sigma \cdot \tau) = \lim_{m \rightarrow 0} \lim_{n \rightarrow 0}{d \over dn}  \overline{ Z_{J, \hT}(m,n)} ,
\eeq
where
\beq
\label{eq::starting_point}
Z_{J,\hT}(m,n) \equiv \sum_{\substack{\sigma_1 \dots \sigma_n \\ \tau_1 \dots \tau_{m}}} \exp\left[-\beta \sum_{a=2}^m H_J(\tau_a)-\beta \left(1-{n \over 2}\right)H_J(\tau_1) -{\beta \over 2} \sum_{b=1}^{n}H_J(\sigma_b) \right]\prod_{b=1}^{n}\hT_{J,t_{\mathrm{f}}}(\sigma_b,\tau_1) \delta (N q_\mathrm{f}-\tau_1 \cdot \sigma_b) \,.
\eeq
For $T>T_\mathrm{s}$, annealed averages coincide the equilibrium averages and we can consider the case with only one replica, $m=1$,
\beq
\lim_{m \rightarrow 0} \lim_{n \rightarrow 0}{d \over dn}\overline{   Z_{\hT}(m,n)} \approx {1 \over \overline{Z}}\lim_{n \rightarrow 0}{d \over dn}  \overline{Z_{\hT}(1,n)} \ .
\eeq

\subsection{Path integral expression of  $\hT_{J,t_{\mathrm{f}}}$}
As shown in Ref.~\cite[Appendix]{rizzo2021}, the symmetric $\hat{T}_{J,t_\mathrm{f}}(\sigma,\tau)$ can be written as a standard path integral
\begin{equation}
    \hT_{J,t_{\mathrm{f}}}(\sigma,\tau) = \int^{s(t_{\mathrm{f}}) = \sigma}_{s(0) = \tau} \mathscrsfs{D} s \mathscrsfs{D} \hat s \, e^{- \mathscrsfs{L}(s, \hat s)},
\end{equation}
where 
\begin{equation}
    \mathscrsfs{L}(s, \hat s) = \sum_i \int dt \left[ \frac{1}{4} \left( \frac{d s_i}{dt}\right)^2 - \hat s_i^2(t) + \beta \hat s_i(t) \frac{d H_J}{d s_i} - \frac{\beta}{2} \sum_i \frac{d^2 H_J}{d s_i^2} \right].
\end{equation}
The spherical constraint over the dynamical configuration $s(t)$ can be imposed by adding a quadratic term to the Hamiltonian 
\begin{equation}
    H_J \to H_J + \frac{1}{2N\epsilon} \left( \sum_i s_i^2(t) - N \right)^2
\end{equation}
and taking the limit $\epsilon \to 0$. As shown in Ref.~\cite{rizzo2021}, this construction is equivalent to having two additional Lagrange multiplier, $\mu$ and $\hat \mu$.
For a multi-linear interaction model, we then have 
\begin{subequations}
\begin{align}
    \label{eq::path_integral}
    \hT_{J,t_{\mathrm{f}}}(\sigma,\tau) &= \int^{s(t_{\mathrm{f}}) = \sigma}_{s(0) = \tau} \mathscrsfs{D} s \mathscrsfs{D} \hat s \mathscrsfs{D} \mu \mathscrsfs{D} \hat \mu\, e^{- \mathscrsfs{L}(s, \hat s, \mu, \hat \mu)} \\
    \label{eq::L}
    \mathscrsfs{L}(s, \hat s, \mu, \hat \mu) &= \sum_i \int dt \left[ \frac{1}{4} \left( \frac{d s_i}{dt}\right)^2 - \hat s_i^2(t) + \beta \hat s_i(t) \frac{d H_J}{d s_i} \right] \\
    &\nonumber + \frac{1}{2} \int dt \, \hat \mu(t) \left( \sum_i s_i^2(t) - N \right) + \int dt \, \mu(t) \left( \sum_i s_i(t) \hat s_i(t) - \frac{N}{2}\right) .
\end{align}
\end{subequations}

\subsection{Introducing replicas and superfields}

The previous expression can be written in considerably more compact form by writing it in terms of an additional bosonic variable $\eta$ that behaves as the product of two Grassmann variables~\cite{rizzo2021}, i.e., $\eta^2 = 0$, $\int \eta \, d\eta  = 1$, $\int d\eta = 0$. This formulation is convenient because it simplifies the average over disorder. Next, we define a new coordinate $1 \equiv (t, \eta)$ and the two fields
\begin{subequations}
    \begin{align}
        s_i(1) &\equiv s_i(t) + \hat s_i(t) \eta \\
        \mu(1) &\equiv \mu(t) + \hat \mu(t) \eta\,.
    \end{align}
\end{subequations}
With those definitions, we can write Eq.~\eqref{eq::L} as
\begin{equation}
    \mathscrsfs{L} = \frac{1}{2} \sum_i \int d1 d2 \, s_{i}(1) \mathcal{M}(1, 2) s_i(2) + \beta \int d1 \, H_J(s(1)) - \frac{N}{2} \int d1 \, (1+\eta) \mu(1),
\end{equation}
where we have introduced the notation $\int d1 \equiv \int dt d\eta$, and where we have defined the operator
\begin{subequations}
\begin{align}
    \mathcal{M}(1, 2) &\equiv \mathcal{M}_{\mathrm{kin}}(1,2) + \mathcal{M}_{\mathrm{sph}}(1,2) \\ 
    \mathcal{M}_{\mathrm{kin}}(1,2) &\equiv 
    - \frac{1}{2} \eta_1 \eta_2 \delta(t_1 - t_2) \frac{d^2}{dt_2^2} - 2 \delta(t_1 - t_2) \\
    \mathcal{M}_{\mathrm{sph}}(1,2) &\equiv \hat \mu(t_1) \delta(t_1 - t_2) \eta_1 \eta_2 + \mu(t_1) \delta(t_1 - t_2) (\eta_1 + \eta_2) .
\end{align}
\end{subequations}
We then proceed by replicating $n$ times the path integral expression in Eq.~\eqref{eq::path_integral}. The delta function over the overlap between the initial and final configuration is treated separately later on, as it requires a precise characterization of how the kinetic operator acts. 

We now introduce  the vectorized variable $\boldsymbol{s}(\boldsymbol{1}) \equiv (\tau_1 \,, \sigma_1 \,, \dots \,, \sigma_n\,, s_{1}(1) \,, \dots\,, s_{n}(1))$, which fuses the \emph{static} part (corresponding to the initial and final conditions to the dynamics) with the ($n$-dimensional) dynamical variables. Before averaging over  disorder, we have one such variable for each site $i$, which we denote $\boldsymbol{s}_i(\boldsymbol{1})$. We also define the corresponding integration rule as
\begin{equation}
\label{eq::integration_rule_superfields}
    \int d \boldsymbol{1} \, A(\boldsymbol{s}(\boldsymbol{1})) \equiv \left(1 - \frac{n}{2} \right) A(\tau_1) + 
    \frac{1}{2} \sum_{b=1}^n A(\sigma_b) + \sum_{b=1}^n \int d 1 \, A(s_b(1)).
\end{equation}
The replicated partition function $Z_{J,\hT}(1,n)$ can therefore be written as
\begin{equation}
    Z_{J,\hT}(1,n) \equiv \sum_{\substack{\sigma_1 \dots \sigma_n \\ \tau_1 
    }}  \prod_{b=1}^{n} \int^{s_b(t_{\mathrm{f}}) = \sigma_b}_{s_b(0) = \tau_1} \mathscrsfs{D} s_b \mathscrsfs{D} \hat s_b \mathscrsfs{D} \mu_b \mathscrsfs{D} \hat \mu_b \, e^{- \frac{1}{2} \sum_{i} \int d \boldsymbol{1} d\boldsymbol{2} \, \boldsymbol{s}(\boldsymbol{1}) \boldsymbol{\mathcal{M}}(\boldsymbol{1}, \boldsymbol{2}) \boldsymbol{s}(\boldsymbol{2}) - \beta \int d \boldsymbol{1} \, H_J(\boldsymbol{s}(\boldsymbol{1})) +  \frac{N}{2} \sum_{b=1}^n \int d1 \, (1+\eta) \mu_b(1) } ,
\end{equation}
where we have defined $\boldsymbol{\mathcal{M}}(\boldsymbol{1}, \boldsymbol{2})$ to be non zero only in the dynamical block; it is then equal to $\mathcal{M}(1, 2)$.

\subsection{Averaging over disorder}

The average over disorder can now be done by using the relation
\begin{equation}
    \overline{e^{-\beta \int d \boldsymbol{1} \, H_J(\boldsymbol{s}(\boldsymbol{1}))}} = e^{\frac{N \beta^2}{4} \int d\boldsymbol{1} d \boldsymbol{2} \, f(\boldsymbol{Q}(\boldsymbol{1}, \boldsymbol{2}))},
\end{equation}
where in the pure $p$-spin model that we analyze in this paper $f(Q) \equiv Q^p$. We have further defined the overlap in the superfield space
\begin{equation}
    \boldsymbol{Q}(\boldsymbol{1}, \boldsymbol{2}) \equiv \frac{1}{N} \sum_i \boldsymbol{s}_i(\boldsymbol{1}) \boldsymbol{s}_i(\boldsymbol{2}) \,.
\end{equation}
We can impose this overlap with a delta function. We then get
\begin{equation}
    \overline{Z_{J,\hT}(1,n)} = \int \prod_{a=1}^n  \mathscrsfs{D} \mu_a \mathscrsfs{D} \hat \mu_a \int \mathscrsfs{D} Q \mathscrsfs{D} \Lambda \, e^{\frac{N}{2} \mathcal{S}},
\end{equation}
where the action is
\begin{equation}
    \mathcal{S} = \int d \boldsymbol{1} d \boldsymbol{2} \, \boldsymbol{Q}(\boldsymbol{1}, \boldsymbol{2}) \boldsymbol{\Lambda}(\boldsymbol{1}, \boldsymbol{2}) + \frac{\beta^2}{2} \int d\boldsymbol{1} d\boldsymbol{2} \, f(\boldsymbol{Q}(\boldsymbol{1}, \boldsymbol{2})) - \ln \det \left[ \boldsymbol{\mathcal{M}}(\boldsymbol{1}, \boldsymbol{2}) + \boldsymbol{\Lambda}(\boldsymbol{1}, \boldsymbol{2}) \right] + \sum_{b=1}^n \int d1 \, (1+\eta) \mu_b(1).
\end{equation}

\subsection{Final condition}~\label{sec::final_condition}

We deal now with the final condition constraint in the starting expression of the dynamical potential, Eq.~\eqref{eq::starting_point}. 
The delta function over the overlap between the initial and final configuration can be written using an integral representation over a variable $\epsilon$, thus leading to a term of the form $\epsilon\, \sum_i s_i(0)s_i(t_\mathrm{f}) $ in the action. As usual, this term can be interpreted as favoring dynamical trajectories in which the final configuration is less correlated with the initial configuration. This term can further be written in terms of superfields with a quadratic form
\begin{subequations}
\begin{align}
    \epsilon \sum_{i} \sum_{b = 1}^n \tau_{1 i} \sigma_{ib} &= - \frac{1}{2} \sum_{i} \int d\boldsymbol{1} d\boldsymbol{2} \, \boldsymbol{s}_i(\boldsymbol{1}) \boldsymbol{\mathcal{M}}_{\epsilon}(\boldsymbol{1}, \boldsymbol{2}) \boldsymbol{s}_i(\boldsymbol{2}) \\
    \boldsymbol{\mathcal{M}}_{\epsilon}(\boldsymbol{1}, \boldsymbol{2}) &= - 2 \epsilon \left( \delta_{\boldsymbol{1}0}\delta_{\boldsymbol{2} t_\mathrm{f}}+\delta_{\boldsymbol{1} t_\mathrm{f}}\delta_{\boldsymbol{2}0} \right)   
\end{align}
\end{subequations}
having used the integration rule in Eq.~\eqref{eq::integration_rule_superfields}. (Note that every time the final condition is involved, the differential leads to an additional factor of 1/2, as written in Eq.~\eqref{eq::integration_rule_superfields}.) We have here denoted by $\delta_{\boldsymbol{1}0}$ and $\delta_{\boldsymbol{1} t_\mathrm{f}}$ the Kronecker delta functions that select the initial and the final conditions, respectively, on the supervariable $\boldsymbol{1}$, i.e.,
\begin{subequations}
\begin{align}
    \int d \boldsymbol{1} \, \boldsymbol{s}(\boldsymbol{1}) \, \delta_{\boldsymbol{1} 0} &= \left(1 - \frac{n}{2} \right) \tau_1 \,,\\
    \int d \boldsymbol{1} \, \boldsymbol{s}(\boldsymbol{1}) \, \delta_{\boldsymbol{1} t_\mathrm{f}} &= \frac{1}{2} \sum_{b=1}^n s_b(t_\mathrm{f}) \equiv \frac{1}{2} \sum_{b=1}^n \sigma_b \,.
\end{align}
\end{subequations}
Moving forward, it is also important to specify how the kinetic operator acts if one of the two variables is constrained to the final condition. The operator then involves first-order instead of second-order derivatives. The reason can be intuitively understood from the underlying integration by parts:
\begin{equation}
    \int_{0}^{t_\mathrm{f}} \left( {ds_i \over dt}\right)^2 dt = s_i(t)  \left. {ds_i \over dt} \right|_{t=t_\mathrm{f}}-s_i(t)  \left. {ds_i \over dt} \right|_{t=0}  -  \int_{0}^{t_\mathrm{f}} s_i(t) {d^2 s_i \over dt^2} dt  .  
\end{equation}
As a result, we have to account for a new term in the kinetic operator when considering the initial or the final condition. Because the initial condition is totally independent of the dynamics, we will not focus on the expression for this operator when either of the two supervariables is set to the initial time. However, at variance with Ref.~\cite{rizzo2021}, the final condition actually impacts the dynamics, since it is not extracted from equilibrium, but induced by the overlap constraint. The first term on the right-hand side of the previous equation, when replicated, can then be written as
\begin{subequations}
\begin{align}
    -\frac{1}{4} \sum_i \sum_b s_{ib}(t_\mathrm{f})  \left. {ds_{ib} \over dt} \right|_{t=t_\mathrm{f}} &= -\frac{1}{2} \sum_i \int d \boldsymbol{1} d\boldsymbol{2} \, \boldsymbol{s}_i(\boldsymbol{1}) 
    \boldsymbol{\mathcal{M}}_{\mathrm{kin}}^{t_\mathrm{f}}(\boldsymbol{1},\boldsymbol{2}) \boldsymbol{s}_i(\boldsymbol{2}) \\
    \boldsymbol{\mathcal{M}}_{\mathrm{kin}}^{t_{\mathrm{f}}}(\boldsymbol{1}, \boldsymbol{2}) &= \delta_{\boldsymbol{1} t_\mathrm{f}} \delta(t_2 - t_\mathrm{f})\eta_2 \frac{d}{dt_2} 
\end{align}
\end{subequations}
Because both terms we have analyzed are quadratic in the superfields, we can absorb them into the definition of $\boldsymbol{\mathcal{M}}(\boldsymbol{1}, \boldsymbol{2})$, which then reads
\begin{equation}
    \boldsymbol{\mathcal{M}}(\boldsymbol{1}, \boldsymbol{2}) \equiv \boldsymbol{\mathcal{M}}_{\mathrm{kin}} (\boldsymbol{1},\boldsymbol{2}) + \boldsymbol{\mathcal{M}}_{\mathrm{kin}}^{t_\mathrm{f}}(\boldsymbol{1}, \boldsymbol{2}) + \boldsymbol{\mathcal{M}}_{\mathrm{sph}}(\boldsymbol{1},\boldsymbol{2}) + \boldsymbol{\mathcal{M}}_{\epsilon}(\boldsymbol{1}, \boldsymbol{2})
\end{equation}

\beq
\epsilon(t_\mathrm{f} c)Q(cb)= - 2 \epsilon  Q(0 b)
\eeq

\subsection{Saddle point equations}
The saddle point for $\boldsymbol{Q}$ and $\boldsymbol{\Lambda}$ reads
\begin{subequations}
    \begin{align}
        \boldsymbol{\Lambda}(\boldsymbol{1}, \boldsymbol{2}) &= - \frac{\beta^2}{2} f'(\boldsymbol{Q}(\boldsymbol{1}, \boldsymbol{2})) \\
        \boldsymbol{Q}(\boldsymbol{1}, \boldsymbol{2}) &= \left[\boldsymbol{\mathcal{M}} + \boldsymbol{\Lambda} \right]^{-1}(\boldsymbol{1}, \boldsymbol{2})
    \end{align}
\end{subequations}
or 
\begin{subequations}
\label{eq::saddle_point_equations}
    \begin{align}
        \boldsymbol{\Lambda}(\boldsymbol{1}, \boldsymbol{2}) &= - \frac{\beta^2}{2} f'(\boldsymbol{Q}(\boldsymbol{1}, \boldsymbol{2})) \\
        \int d \boldsymbol{3} \left(\boldsymbol{\mathcal{M}}(\boldsymbol{1}, \boldsymbol{3}) + \boldsymbol{\Lambda}(\boldsymbol{1}, \boldsymbol{3}) \right) \boldsymbol{Q}(\boldsymbol{3}, \boldsymbol{2}) &= \boldsymbol{\delta}(\boldsymbol{1}, \boldsymbol{2}) \,.
    \end{align}
\end{subequations}
The dynamical potential can then be obtained from the function $\epsilon(q)$ by integrating
\beq
\epsilon = { d \, V_t(q) \over d\, q}
\eeq

\section{RS ansatz and the dynamical equations}
\label{app:equations}

We now impose an RS ansatz on the $n$ dynamical replicas
\begin{equation}
    \boldsymbol{Q}(\boldsymbol{1}, \boldsymbol{2}) = \delta_{ab} Q(1, 2) + (1-\delta(ab)) Q^\mathrm{dt}(1,2) \,.
\end{equation}
and over the spherical constraints variables
\begin{equation}
    \mu_a(1) = \mu(1) \,, \qquad a = 1, \dots, n \,.
\end{equation}
When we select the dynamical block in the indices $\boldsymbol{1}$ and $\boldsymbol{2}$ of the saddle point given by Eq.~\eqref{eq::saddle_point_equations}, we can repeat the same steps as in~\cite[Appendix B]{rizzo2021}, starting by expressing the dynamical overlaps $Q(1, 2)$ and $Q^\mathrm{dt}(1,2)$ in terms of two-time order parameters
\begin{subequations}
\label{eq::order_parameters}
    \begin{align}
        Q(1,2) &= C(t, t') + \eta_2 \hat R_2(t,t') + \eta_1 \hat R_1(t,t') + \eta_1 \eta_2 \hat \chi(t, t') \\
        Q^\mathrm{dt}(1,2) &= C^\mathrm{dt}(t, t') + \eta_2 \hat R_2^\mathrm{dt}(t,t') + \eta_1 \hat R_1^\mathrm{dt}(t,t') + \eta_1 \eta_2 \hat \chi^\mathrm{dt}(t, t')
    \end{align}
\end{subequations}
We will denote the corresponding quantities related to $\Lambda(1, 2)$ and $\Lambda^\mathrm{dt}(1,2)$ as
\begin{subequations}
    \begin{align}
        \Lambda(1,2) &= C_{\Lambda}(t, t') + \eta_2 \hat R_{2, \Lambda}(t,t') + \eta_1 \hat R_{1, \Lambda}(t,t') + \eta_1 \eta_2 \hat \chi_{\Lambda}(t, t') \\
        \Lambda^\mathrm{dt}(1,2) &= C_{\Lambda^\mathrm{dt}}(t, t') + \eta_2 \hat R_{2, \Lambda^\mathrm{dt}}^\mathrm{dt}(t,t') + \eta_1 \hat R_{1, \Lambda^\mathrm{dt}}^\mathrm{dt}(t,t') + \eta_1 \eta_2 \hat \chi_{\Lambda^\mathrm{dt}}(t, t')
    \end{align}
\end{subequations}
Note that the order parameters functions~\eqref{eq::order_parameters} can be written in terms of the variable $s$ and $\hat s$ as
\begin{subequations}
    \begin{align}
        C(t,t') &= \overline{ \left[ \langle s_i(t) s_i(t') \rangle \right] } \\
        C^\mathrm{dt}(t,t') &= \overline{ \left[ \langle s_i(t) \rangle \langle s_i(t') \rangle \right] } \\
        \hat R_1(t,t') &= \hat R_2(t',t) \\
        \hat R_1^\mathrm{dt}(t,t') &= \hat R_2^\mathrm{dt}(t',t) \\
        \hat R_2(t,t') &= 
        \overline{ \left[ \langle s_i(t) \hat s_i(t') \rangle \right] } 
        \\
        \hat R_2^\mathrm{dt}(t,t') &= 
        \overline{ \left[ \langle s_i(t) \rangle \langle \hat s_i(t') \rangle \right] }
        \\
        \hat \chi(t,t') &= 
        \overline{ \left[ \langle \hat s_i(t) \hat s_i(t') \rangle \right] }  
        \\
        \hat \chi^\mathrm{dt}(t,t') &= 
        \overline{ \left[ \langle \hat s_i(t) \rangle \langle \hat s_i(t') \rangle \right] }
    \end{align}
\end{subequations}
where we remind that the square brackets $\left[ \ldots \right]$ denote averaging over initial condition $\tau$, and the angular brackets $\langle \ldots \rangle$ denote averaging over the Langevin dynamics starting from $\tau$ and constrained to evolving, within a time window $t_\mathrm{f}$, to a configuration with overlap $q_\mathrm{f}$ from $\tau$. It is easy to show that those relations are related to derivatives of the constrained partition function $\mathcal{Z}_{J, t_\mathrm{f}}$ in terms of magnetic fields as shown in~\eqref{eq::physical_meaning_order_parameters}. 

The order parameters in~\eqref{eq::order_parameters} satisfy the same equations as in~\cite{rizzo2021}, with the difference that we have to use Eq.~\eqref{eq::integration_rule_superfields} instead of~\cite[Eq.~(A8)]{rizzo2021}. 
To be as close as possible to Ref. \cite{rizzo2021}, we choose the final time at time $\Tau \equiv t_\mathrm{f}/2$ and the initial time at time $-\Tau$. We then obtain

\beqa
0 & = & -2 \hR_1(t,t')+\mu(t)C(t,t') +
\nonumber
\\
& + &\int_{-\Tau}^{+\Tau} \left(C_\Lambda(t,t'')\hR_{1}(t'',t')+\hR_{2,\Lambda}(t,t'')C(t'',t')\right)dt'' +
\nonumber
\\
& + &(n-1)\int_{-\Tau}^{+\Tau} \left(C_{\Lambda^\mathrm{dt}}(t,t'')\hR_{1}^\mathrm{dt}(t'',t')+\hR_{2,{\Lambda^\mathrm{dt}}}(t,t'')C^\mathrm{dt}(t'',t')\right)dt'' +
\nonumber
\\
& + & \left( 1-{n \over 2}\right)C_\Lambda(t,-\Tau)C(-\Tau,t')+{1 \over 2} C_\Lambda(t,\Tau)C(\Tau,t') +{n-1 \over 2} C_{\Lambda^\mathrm{dt}}(t,\Tau)C^\mathrm{dt}(\Tau,t') \ .
\label{exp1}
\eeqa

\beqa
\delta(t-t') & = & -{1 \over 2}{d^2 \over dt^2} C(t,t')+ \mu(t)\hR_1(t,t')+\hat{\mu}(t)C(t,t') +
\nonumber
\\
& + &\int_{-\Tau}^{+\Tau} \left(\hat{R}_{1,\Lambda}(t,t'')\hR_{1}(t'',t')+\hX_\Lambda(t,t'')C(t'',t') \right)dt''+
\nonumber
\\
& + &(n-1) \int_{-\Tau}^{+\Tau} \left(\hat{R}_{1,\Lambda^\mathrm{dt}}(t,t'')\hR_{1}^\mathrm{dt}(t'',t')+\hX_{\Lambda^\mathrm{dt}}(t,t'')C^\mathrm{dt}(t'',t') \right)dt''+
\nonumber
\\
& + & \left( 1-{n \over 2}\right)\hR_{1,\Lambda}(t,-\Tau)C(-\Tau,t')+{1 \over 2} \hR_{1,\Lambda}(t,\Tau) C(\Tau,t') +{n-1 \over 2} \hR_{1,\Lambda^\mathrm{dt}}(t,\Tau)C^\mathrm{dt}(\Tau,t') \ .
\label{exp2}
\eeqa

\beqa
\delta(t-t') & = & -2 \hX(t,t')+\mu(t)\hR_2(t,t') +
\nonumber
\\
& + &\int_{-\Tau}^{+\Tau}  \left(\hat{R}_{2,\Lambda}(t,t'')\hR_{2}(t'',t')+C_\Lambda(t,t'')\hX(t'',t') \right)dt''+
\nonumber
\\
& + &(n-1) \int_{-\Tau}^{+\Tau}  \left(\hat{R}_{2,\Lambda^\mathrm{dt}}(t,t'')\hR_{2}^\mathrm{dt}(t'',t')+C_{\Lambda^\mathrm{dt}}(t,t'')\hX^\mathrm{dt}(t'',t') \right)dt''+
\nonumber
\\
& + & \left( 1-{n \over 2}\right)C_\Lambda(t,-\Tau)\hR_2(-\Tau,t')+{1 \over 2} C_\Lambda(t,\Tau)\hR_2(\Tau,t') +{n-1 \over 2} C_{\Lambda^\mathrm{dt}}(t,\Tau)\hR_2^\mathrm{dt}(\Tau,t') \ .
\label{exp3}
\eeqa

\beqa
0 & = & -{1 \over 2}{d^2 \over dt^2} \hR_2(t,t')+ \mu(t)\hX(t,t')+\hat{\mu}(t)\hR_2(t,t')+
\nonumber
\\
& + &  \int_{-\Tau}^{+\Tau} \left(\hat{R}_{1,\Lambda}(t,t'')\hX(t'',t')+\hX_\Lambda(t,t'')\hR_{2}(t'',t') \right)dt'' +
\nonumber
\\
& + &(n-1)  \int_{-\Tau}^{+\Tau} \left(\hat{R}_{1,\Lambda^\mathrm{dt}}(t,t'')\hX^\mathrm{dt}(t'',t')+\hX_{\Lambda^\mathrm{dt}}(t,t'')\hR_{2}^\mathrm{dt}(t'',t') \right)dt'' +
\nonumber
\\
& + & \left( 1-{n \over 2}\right)\hR_{1,\Lambda}(t,-\Tau)\hR_2(-\Tau,t')+{1 \over 2} \hR_{1,\Lambda}(t,\Tau)\hR_2(\Tau,t') +{n-1 \over 2} \hR_{1,\Lambda^\mathrm{dt}}(t,\Tau)\hR_2^\mathrm{dt}(\Tau,t') \ .
\label{exp4}
\eeqa

%
%
%

\beqa
0 & = & -2 \hR_1^\mathrm{dt}(t,t')+\mu(t)C^\mathrm{dt}(t,t') +
\nonumber
\\
& + &\int_{-\Tau}^{+\Tau} \left(C_{\Lambda^\mathrm{dt}}(t,t'')\hR_{1}(t'',t')+\hR_{2,{\Lambda^\mathrm{dt}}}(t,t'')C(t'',t')\right)dt'' +
\nonumber
\\
& + &\int_{-\Tau}^{+\Tau} \left(C_{\Lambda}(t,t'')\hR_{1}^\mathrm{dt}(t'',t')+\hR_{2,{\Lambda}}(t,t'')C^\mathrm{dt}(t'',t')\right)dt''  +
\nonumber
\\
& + &(n-2)\int_{-\Tau}^{+\Tau} \left(C_{\Lambda^\mathrm{dt}}(t,t'')\hR_{1}^\mathrm{dt}(t'',t')+\hR_{2,{\Lambda^\mathrm{dt}}}(t,t'')C^\mathrm{dt}(t'',t')\right)dt'' +
\nonumber
\\
& + & \left( 1-{n \over 2}\right)C_\Lambda(t,-\Tau)C(-\Tau,t')+{1 \over 2} C_{\Lambda^\mathrm{dt}}(t,\Tau)C(\Tau,t')+{1 \over 2} C_\Lambda(t,\Tau)C^\mathrm{dt}(\Tau,t') \nonumber \\
&+& {n-2 \over 2} C_{\Lambda^\mathrm{dt}}(t,\Tau)C^\mathrm{dt}(\Tau,t') \ .
\label{exp1dt}
\eeqa

\beqa
0  & = & -{1 \over 2}{d^2 \over dt^2} C^\mathrm{dt}(t,t')+ \mu(t)\hR_1^\mathrm{dt}(t,t')+\hat{\mu}(t)C^\mathrm{dt}(t,t') +
\nonumber
\\
& + & \int_{-\Tau}^{+\Tau} \left(\hat{R}_{1,\Lambda^\mathrm{dt}}(t,t'')\hR_{1}(t'',t')+\hX_{\Lambda^\mathrm{dt}}(t,t'')C(t'',t') \right)dt''+
\nonumber
\\& + &\int_{-\Tau}^{+\Tau} \left(\hat{R}_{1,\Lambda}(t,t'')\hR_{1}^\mathrm{dt}(t'',t')+\hX_{\Lambda}(t,t'')C^\mathrm{dt}(t'',t') \right)dt''+
\nonumber
\\
& + &(n-2) \int_{-\Tau}^{+\Tau} \left(\hat{R}_{1,\Lambda^\mathrm{dt}}(t,t'')\hR_{1}^\mathrm{dt}(t'',t')+\hX_{\Lambda^\mathrm{dt}}(t,t'')C^\mathrm{dt}(t'',t') \right)dt''+
\nonumber
\\
& + & \left( 1-{n \over 2}\right)\hR_{1,\Lambda}(t,-\Tau)C(-\Tau,t')+{1 \over 2} \hR_{1,\Lambda^\mathrm{dt}}(t,\Tau)C(\Tau,t')+{1 \over 2} \hR_{1,\Lambda}(t,\Tau)C^\mathrm{dt}(\Tau,t') \nonumber\\
&+&{n-2 \over 2} \hR_{1,\Lambda^\mathrm{dt}}(t,\Tau)C^\mathrm{dt}(\Tau,t')  \ .
\label{exp2dt}
\eeqa

\beqa
0 & = & -2 \hX^\mathrm{dt}(t,t')+\mu(t)\hR_2^\mathrm{dt}(t,t') +
\nonumber
\\
& + & \int_{-\Tau}^{+\Tau}  \left(\hat{R}_{2,\Lambda^\mathrm{dt}}(t,t'')\hR_{2}(t'',t')+C_{\Lambda^\mathrm{dt}}(t,t'')\hX(t'',t') \right)dt''+
\nonumber\\
& + & \int_{-\Tau}^{+\Tau}  \left(\hat{R}_{2,\Lambda}(t,t'')\hR_{2}^\mathrm{dt}(t'',t')+C_{\Lambda}(t,t'')\hX^\mathrm{dt}(t'',t') \right)dt''+
\nonumber
\\
& + &(n-2) \int_{-\Tau}^{+\Tau}  \left(\hat{R}_{2,\Lambda^\mathrm{dt}}(t,t'')\hR_{2}^\mathrm{dt}(t'',t')+C_{\Lambda^\mathrm{dt}}(t,t'')\hX^\mathrm{dt}(t'',t') \right)dt''+
\nonumber
\\
& + & \left( 1-{n \over 2}\right)C_\Lambda(t,-\Tau)\hR_2(-\Tau,t')+{1 \over 2} C_{\Lambda^\mathrm{dt}}(t,\Tau)\hR_2(\Tau,t') +{1 \over 2} C_\Lambda(t,\Tau)\hR_2^\mathrm{dt}(\Tau,t') \nonumber \\
& + & {n-2 \over 2} C_{\Lambda^\mathrm{dt}}(t,\Tau)\hR_2^\mathrm{dt}(\Tau,t') \ .
\label{exp3dt}
\eeqa

\beqa
0 & = & -{1 \over 2}{d^2 \over dt^2} \hR_2^\mathrm{dt}(t,t')+ \mu(t)\hX^\mathrm{dt}(t,t')+\hat{\mu}(t)\hR_2^\mathrm{dt}(t,t')+
\nonumber
\\
& + &  \int_{-\Tau}^{+\Tau} \left(\hat{R}_{1,\Lambda^\mathrm{dt}}(t,t'')\hX(t'',t')+\hX_{\Lambda^\mathrm{dt}}(t,t'')\hR_{2}(t'',t') \right)dt'' +
\nonumber
\\& + &  \int_{-\Tau}^{+\Tau} \left(\hat{R}_{1,\Lambda}(t,t'')\hX^\mathrm{dt}(t'',t')+\hX_{\Lambda}(t,t'')\hR_{2}^\mathrm{dt}(t'',t') \right)dt'' +
\nonumber
\\
& + &(n-2)  \int_{-\Tau}^{+\Tau} \left(\hat{R}_{1,\Lambda^\mathrm{dt}}(t,t'')\hX^\mathrm{dt}(t'',t')+\hX_{\Lambda^\mathrm{dt}}(t,t'')\hR_{2}^\mathrm{dt}(t'',t') \right)dt'' +
\nonumber
\\
& + & \left( 1-{n \over 2}\right)\hR_{1,\Lambda}(t,-\Tau)\hR_2(-\Tau,t')+{1 \over 2} \hR_{1,\Lambda^\mathrm{dt}}(t,\Tau)\hR_2(\Tau,t') +{1 \over 2} \hR_{1,\Lambda}(t,\Tau)\hR_2^\mathrm{dt}(\Tau,t') \nonumber \\
& + & {n-2 \over 2} \hR_{1,\Lambda^\mathrm{dt}}(t,\Tau)\hR_2^\mathrm{dt}(\Tau,t') \ .
\label{exp4dt}
\eeqa

\subsection{The dynamical equations involving the final condition}\label{app::dynamical_equations_final_condition}

The equations obtained in the previous subsection, at variance with those in Ref.~\cite{rizzo2021}, are insufficient to determine the solutions. The coupling $\epsilon$ with the final condition, in particular, does not then appear. We then also need equations obtained from the saddle point Eq.~\eqref{eq::saddle_point_equations} by imposing that the supervariable $\boldsymbol{1}$ is in the static block corresponding to the final condition $t = \Tau$. As a result, because of the kinetic term derived in Sec.~\ref{sec::final_condition}, these equations involve first-order instead of second-order derivatives.

In particular, we obtain the following equation valid for $t' < t = +\Tau$:
\beqa
0 & = & \left. {d \over dt} C(t,t')\right|_{t=\Tau}+\mu(\Tau)C(\Tau,t') +
\nonumber
\\
& + &\int_{-\Tau}^{+\Tau} \left(C_\Lambda(\Tau,t'')\hR_{1}(t'',t')+\hR_{2,\Lambda}(\Tau,t'')C(t'',t')\right)dt'' +
\nonumber
\\
& + &(n-1)\int_{-\Tau}^{+\Tau} \left(C_{\Lambda^\mathrm{dt}}(\Tau,t'')\hR_{1}^\mathrm{dt}(t'',t')+\hR_{2,{\Lambda^\mathrm{dt}}}(\Tau,t'')C^\mathrm{dt}(t'',t')\right)dt'' +
\nonumber
\\
& + & \left( 1-{n \over 2}\right)C_\Lambda(\Tau,-\Tau)C(-\Tau,t')+{1 \over 2} C_\Lambda(\Tau,\Tau)C(\Tau,t') +{n-1 \over 2} C_{\Lambda^\mathrm{dt}}(\Tau,\Tau)C^\mathrm{dt}(\Tau,t') +
\nonumber
\\
& - & 2 \, \epsilon \, C(-\Tau, t')
\label{bord1}
\eeqa

\noindent Subtracting from  Eq.~\eqref{exp1} evaluated at $t=\Tau$ from this result, we obtain
\beq
0  =  \left. {d \over dt} C(t,t')\right|_{t=\Tau}+2 \hR_1(\Tau,t')+\Delta \mu \, C(\Tau,t')- 2 \, \epsilon \, C(-\Tau, t'),
\label{bb1}
\eeq
where we have accounted for the parameter $\mu(t)$ being possibly discontinuous at $t=\Tau$ by the definition:
\beq
\Delta \mu \equiv \mu(\Tau)-\mu(\Tau^-)
\eeq
Similarly, we obtain an equation from which subtracting Eq.~\eqref{exp3} evaluated at $t=\Tau$ we obtain 
\beq
0  =  \left. {d \over dt} \hR_2(t,t')\right|_{t=\Tau}+2 \hX(\Tau,t')+\Delta \mu\, \hR_2(\Tau,t')-2 \epsilon \, \hR_2(-\Tau, t').
\label{bb2}
\eeq
The above equations are valid for $t'<\Tau$. The expression for $t'=\Tau$ can be obtained noting that  Eqs.~(\ref{exp2}, \ref{exp3}, \ref{exp4}) imply that the discontinuities of ${d \over dt} C(t,t')$ for $t=t'$ is equal to $-2$ while the discontinuity of ${d \over dt'} \hR_2(t',t)$ is
\beq
\left. {d \over dt'} \hR_2(t',t)\right|_{t'=t^+}-\left. {d \over dt'} \hR_2(t',t)\right|_{t'=t^-}=-\mu(t).
\eeq
For the different trajectory functions we obtain
\begin{subequations}
\begin{align}
0 &= \left. {d \over dt} C^\mathrm{dt}(t,t')\right|_{t=\Tau}+2 \hR_1^\mathrm{dt}(\Tau,t')+\Delta \mu \, C^\mathrm{dt}(\Tau,t')- 2 \, \epsilon \, C^\mathrm{dt}(-\Tau, t')
\label{bb1dt} \\
0  &=  \left. {d \over dt} \hR_2^\mathrm{dt}(t,t')\right|_{t=\Tau}+2 \hX^\mathrm{dt}(\Tau,t')+\Delta \mu\, \hR_2^\mathrm{dt}(\Tau,t')-2 \epsilon \, \hR_2^\mathrm{dt}(-\Tau, t')
\label{bb2dt}
\end{align}
\end{subequations}

\subsection{The dynamical equations on the corners}
A number of symmetries that we are going to discuss in the following imply that the above equations are only needed on the corners of the integration domain.
We start observing that Eq.~\eqref{exp3} implies 
\beq
 \hX(t,t')=-{1 \over 2}\delta(t-t')+\delta \hX(t,t'),
\eeq
where $\delta \hX(t,t')$ is a regular function.  Because all replicas has the same initial condition, we must also have
\begin{subequations}
    \begin{align}
        C(-\Tau,t)&=C^\mathrm{dt}(-\Tau,t) \,,         \\
        \hR_1(t,-\Tau)&=\hR_1^\mathrm{dt}(t,-\Tau) = \hR_2(-\Tau,t) =\hR_2^\mathrm{dt}(-\Tau,t) \,.
    \end{align}
\end{subequations}
The difference between Eq.~\eqref{exp1} and Eq.~\eqref{exp1dt} evaluated for $t=-\Tau^+$ and the difference between  
Eq.~\eqref{exp4} and \eqref{exp4dt} evaluated for $t=-\Tau^+$
 additionally lead to 
\begin{subequations}
    \begin{align}
        \hR_1(-\Tau^+,t) &= \hR_1^\mathrm{dt}(-\Tau^+,t) \,, \\
        \hR_2(t,-\Tau^+) &= \hR_2^\mathrm{dt}(t,-\Tau^+) \,.
\end{align}
\end{subequations}
Note also that Eqs.~\eqref{exp2} and \eqref{exp2dt} reduce to the same equation  for $t=-\Tau^+$, and the same applies for Eqs.~\eqref{exp4} and \eqref{exp4dt}.

The difference between Eq.~\eqref{exp3} and Eq.~\eqref{exp3dt} evaluated at $t=-\Tau^+$ yields:
\beqd
\delta\hX(-\Tau^+,t)=\hX^\mathrm{dt}(-\Tau^+,t)\ ,
\eeqd
in turn leading to
\beq
\delta\hX(- \Tau^+,- \Tau^+)=\hX^\mathrm{dt}(- \Tau^+,-\Tau^+)={\mu_\mathrm{eq} \over 4}={1 \over 4}+{\beta^2 \over 8}f'(1) \ .
\eeq
Setting $t=t'=-\Tau^+$ in Eq.~\eqref{exp1} together with the above relationship we obtain  $\mu(-\Tau)=\mu_\mathrm{eq}$.


Due to the above symmetries, the eight equations (\ref{exp1}-\ref{exp4dt}) can be solved provided the corners values are given.
For $(t,t')=(-\Tau,-\Tau)$ we have:
\begin{subequations}
    \begin{align}
        C(-\Tau,-\Tau) &= C^\mathrm{dt}(-\Tau,-\Tau) = 1 \,,\\
        \hR_1(-\Tau,-\Tau) &= \hR_2(-\Tau,-\Tau) =\hR_1^\mathrm{dt}(-\Tau,-\Tau)=\hR_2^\mathrm{dt}(-\Tau,-\Tau)=\frac{1}{2}\,, \\ 
        \delta\hX(-\Tau,-\Tau) &= \hX^\mathrm{dt}(-\Tau,-\Tau) = {1 \over 4}+{\beta^2 \over 8} f'(1)
    \end{align}
\end{subequations}
For $(t,t')=(\Tau,\Tau)$  the unknowns are $C$, $\hR_1$, $\hR_2$, $\hX$ $C^\mathrm{dt}$, $\hR_1^\mathrm{dt}$,$\hR_2^\mathrm{dt}$ $\hX^\mathrm{dt}$ $\mu$ and $\Delta \mu$.  For these ten unknowns we have Eqs.~(\ref{exp1},\ref{exp3},\ref{exp1dt},\ref{exp3dt}) evaluated at $t=t'=\Tau$ and Eqs.~(\ref{bb1},\ref{bb2},\ref{bb1dt},\ref{bb2dt}) evaluated at $t'=\Tau$ plus the two equations 
\begin{subequations}
    \begin{align}
        C(\Tau,\Tau) &= 1 \,, \\ 
        \hR_1(\Tau,\Tau)+\hR_2(\Tau,\Tau)&=1
    \end{align}
\end{subequations}
For $(t,t')=(-\Tau,\Tau)$,  there are eight unknowns: $C$, $\hR_1$, $\hR_2$, $\hX$ $C^\mathrm{dt}$, $\hR_1^\mathrm{dt}$,$\hR_2^\mathrm{dt}$ $\hX^\mathrm{dt}$ that can be determined by Eqs.~(\ref{exp1},\ref{exp3},\ref{exp1dt},\ref{exp3dt}) evaluated at $t=-\Tau, t'=\Tau$ and Eqs.~(\ref{bb1},\ref{bb2},\ref{bb1dt},\ref{bb2dt}) evaluated at $t'=-\Tau$. In practice, Eqs.~(\ref{exp1dt},\ref{exp3dt},\ref{bb1dt},\ref{bb2dt}) are redundant using the properties derived earlier:
\begin{subequations}
    \begin{align}
        C(-\Tau,\Tau)&=C^\mathrm{dt}(-\Tau,\Tau)\,, \\
        \hR_1(-\Tau,\Tau)&=\hR_1^\mathrm{dt}(-\Tau,\Tau) \,, \\  
        \hR_2(-\Tau,\Tau) &= \hR_2^\mathrm{dt}(-\Tau,\Tau) \,, \\ 
        \delta \hX(-\Tau,\Tau)&=\hX^\mathrm{dt}(-\Tau,\Tau)
    \end{align}
\end{subequations}
and thus we only need the four Eqs.~(\ref{exp1},\ref{exp3},\ref{bb1},\ref{bb2}).

\subsection{The Energy}

The instantaneous energy along the trajectory reads:
\beq
E(t)\equiv \sum_{p=1}^{\infty} \sum_{i_1 < \dots < i_p} \overline{ J_{i_1\dots i_p}[\langle s_{i_1}(t)\dots s_{i_p}(t) \rangle ]}\ .
\eeq
The expression for this quantity in terms of the order parameter can be obtained as in~\cite[Appendix F]{rizzo2021}.  The only difference is due to the difference between \cite[Eq.~(A8)]{rizzo2021} and Eq.~\eqref{eq::integration_rule_superfields}.
In particular, exploiting the fact that the coupling constants $J$ are Gaussian random variables, by integrating by parts:
\beqa
e(t)\equiv {E(t) \over N} & =&-{\beta  \over 2}\left[  \left( 1-{n \over 2}\right)C_{f[C]}(t,-\Tau)+{1 \over 2}C_{f[C]}(t,\Tau)+ {n-1 \over 2}C_{f[C^\mathrm{dt}]}(t,\Tau)  + \right.
\nonumber
\\
&  & + \left.  \int_{-\Tau}^{+\Tau}  \hR_{2,f[C]}(t,t') dt'+(n-1)\int_{-\Tau}^{+\Tau} \hR_{2,f[C^\mathrm{dt}]}(t,t') dt'\right]
\label{ene}
\eeqa
In the special case of the pure $p$-spin {\it i.e.} $f(x)=x^p$, this result leads to a simple relationship between the energy and $\mu(t)$:
\beq
e(t)={1 \over p\, \beta}(1-\mu(t))\ .
\eeq
The above can be shown using Eq.~\eqref{exp1} at equal times and using that
\beq
f'(x)+f''(x)\, x=p \,f'(x)\, ,\  \mathrm{for}\ \ f(x)=x^p\ ,
\eeq
to make a connection with Eq.~\eqref{ene}.
Note that the energy is continuous in $t=\pm \Tau$, as it should.
\end{widetext}

\subsection{The Free case}
\label{sec::freecase}
\begin{figure*}[t]
\centering
\includegraphics[width=0.48\textwidth]{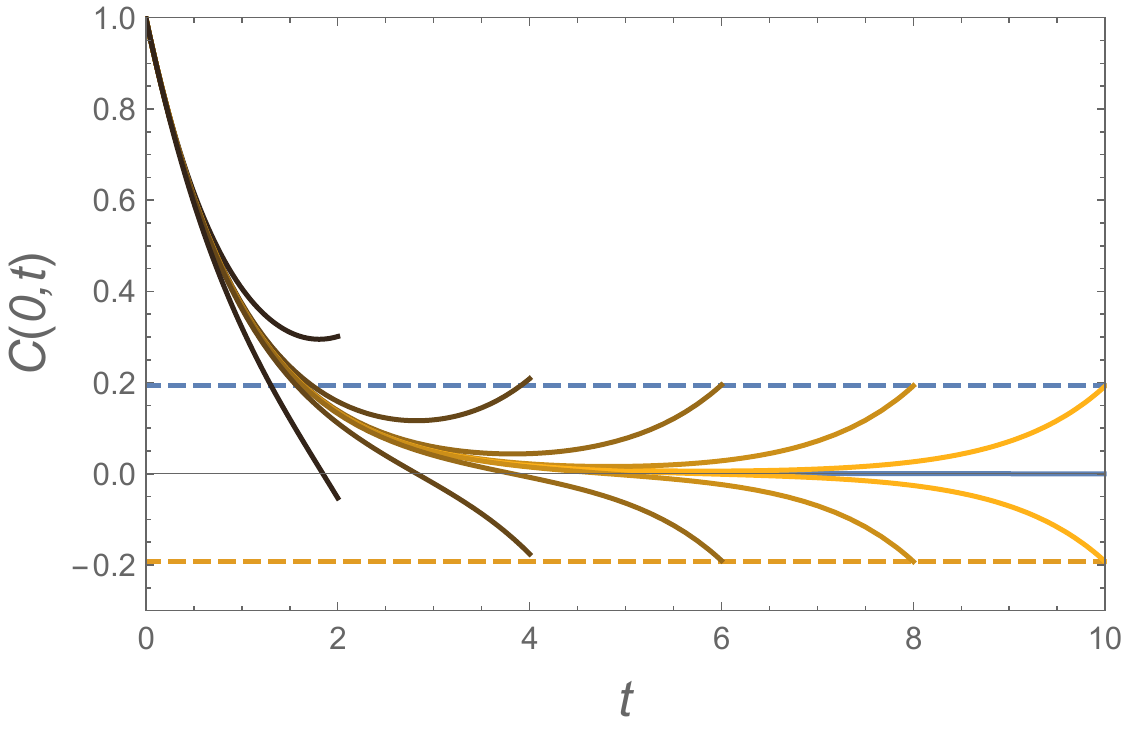}
\includegraphics[width=0.48\textwidth]{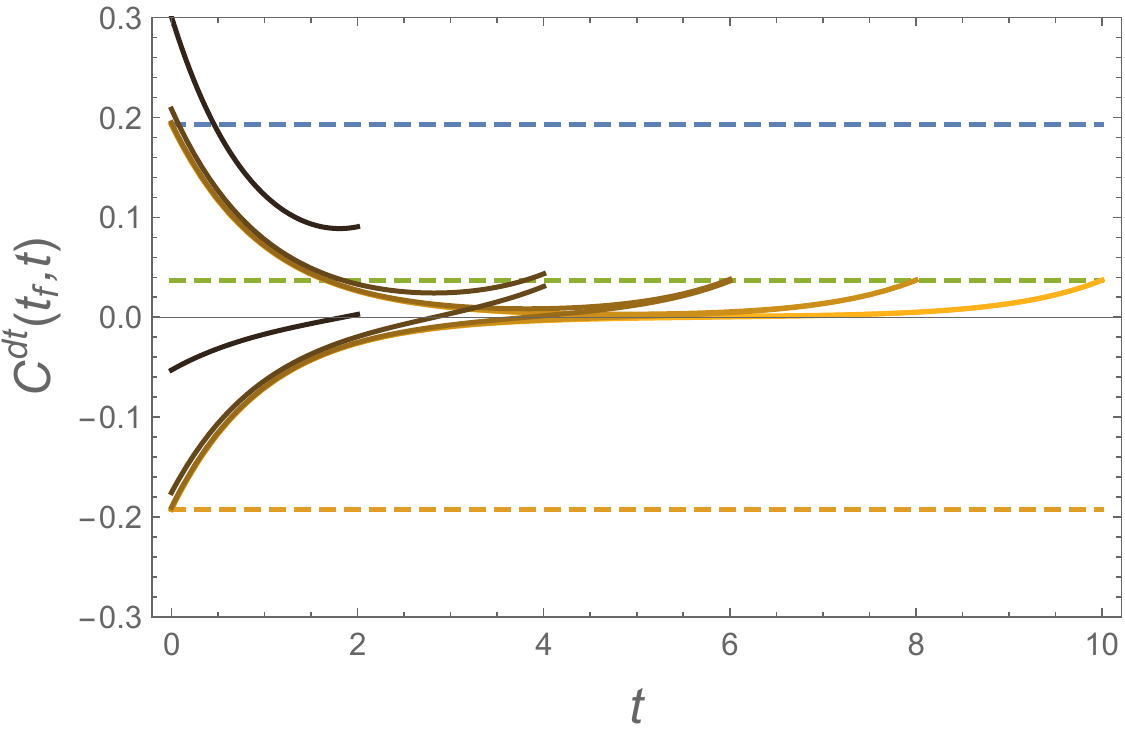}
\caption{Correlation functions \textbf{(left)} $C(t,0)$ and \textbf{(right)} $C^\mathrm{dt}(t,t_\mathrm{f})$ in the free case $T\rightarrow\infty$ for (left to right) $t_\mathrm{f}=2,4,6,8,$ and $10$, for (top curves) $\epsilon=0.2$ and (bottom curves) $\epsilon=-0.2$. At large $t_\mathrm{f}$, $C(0,t_\mathrm{f})$ and $C^\mathrm{dt}(0,t_\mathrm{f})$ converge to $q(\epsilon)$ from the static FP potential (dashed lines). See text for more details. Similarly $C^\mathrm{dt}(t_\mathrm{f},t_\mathrm{f})$ converges to the overlap between two configurations at overlap $q(\epsilon)$ with the reference configuration in the FP potential.
}
\label{fig:C-Cdt-free}
\end{figure*}

The free case $\beta = 0$ results in considerable simplifications, and can therefore help validate the overall scheme (see Fig.~\ref{fig:C-Cdt-free}). In particular, 
\beq
\mu(t)=1  \ \ \ \forall t < \Tau \, .
\eeq 
Then, Eq.~\eqref{exp1}, the symmetries $\hR_2(t,t')=\hR_1(t',t)$, $C(t,t')=C(t',t)$,  and Eq.~\eqref{exp3} lead to
\beqa
\hR_1(t,t') & = & {1 \over 2} C (t,t')
\\
\hR_2(t,t') & = & {1 \over 2} C (t,t')
\\
\hX(t,t') & = & -{1 \over 2} \delta(t-t')+{1 \over 4}C(t,t') \ .
\eeqa
The above relationships lead to:
\beqd
-{1 \over 2}{d^2 \over dt^2} C(t,t')+ \left({1 \over 2}+\hat{\mu}(t)\right) C(t,t')  =  \delta(t-t').
\eeqd
This equation has the same solution as the free case studied in Ref.~\cite{rizzo2021} with $\hat{\mu}(t)=\hat{\mu}$ determined by the value of $C(-\Tau, +\Tau)$.
\beq
C(t,t')=C(|t-t'|) \,, \ \  \hat{\mu}(t)=\hat{\mu} \, ,
\eeq
where
\beqa
a &   \equiv  &  1 + 2 \, \hat{\mu}
\\
C(x) & = &  \cosh \sqrt{a} x - {1 \over \sqrt{a}} \sinh  \sqrt{a} x \ .
\label{freesol}
\eeqa
Then, Eq.~\eqref{bb1} evaluated at $t'=\pm \Tau$ leads to:
\beq
0  =  \left. {d \over dt} C(t,-\Tau)\right|_{t=\Tau}+C(\Tau,-\Tau)+\Delta \mu \, C(\Tau,-\Tau)-2 \epsilon 
\eeq
\beq
0 =  \Delta \mu - 2 \epsilon \, C(-\Tau, +\Tau)
\eeq
Plugging the solution Eq.~\eqref{freesol} into the above equations we obtain two equations that determine $a$ and $\Delta \mu$ as a function of $\epsilon$ and $\Tau$:
\begin{widetext}
\beqa
0 & = & \frac{(a-\Delta \mu -1) \sinh \left(\sqrt{a}\,  t_\mathrm{f}\right)}{\sqrt{a}}+\Delta \mu \,  \cosh \left(\sqrt{a}\,
   t_\mathrm{f}\right)- 2 \epsilon
\\
0 & = & \sqrt{a} \, \Delta \mu \,  \sinh \left(\sqrt{a}\, t_\mathrm{f}\right)+(a-\Delta \mu -1) \, \cosh \left(\sqrt{a}\,
   t_\mathrm{f}\right)- 2 \epsilon,
\eeqa
\end{widetext}
where $t_\mathrm{f}=2 \Tau$.
The above equation can be solved numerically.  At large $t_\mathrm{f}$, the following asymptotic behaviors are obtained:
\beqa
(a -1)e^{t_\mathrm{f}} & \rightarrow &   {8 \epsilon \over  (1+ \sqrt{1+4 \epsilon^2})}  =  4 \, \epsilon +O(\epsilon^3)
\\
\Delta \mu & \rightarrow  &  = \sqrt{1+4 \epsilon^2}-1 =2  \epsilon^2+O(\epsilon^4)   
\\
C(-\Tau,\Tau) & \rightarrow  &   {\epsilon \over {1 \over 2} (1+ \sqrt{1+4 \epsilon^2})}= \epsilon+ O(\epsilon^3)  .
\eeqa
One can check that the asymptotic result for $C(-\Tau,\Tau)$ 
is precisely what the FP potential gives.

In the limit $\epsilon \rightarrow \infty$, the final configuration coincides with the initial configuration,  consistently we have that $a(\epsilon,t_\mathrm{f})$ approaches a finite limit
\beq
a_\infty(t_\mathrm{f}) \equiv \lim_{\epsilon \rightarrow \infty} a(\epsilon,t_\mathrm{f})
\eeq
that can be written in terms of $t_\mathrm{f}$ by imposing that $C(t_\mathrm{f})=1$:
\beq
1 =   \cosh \sqrt{a_\infty} t_\mathrm{f} - {1 \over \sqrt{a_\infty}} \sinh  \sqrt{a_\infty} t_\mathrm{f} \ .
\eeq
One can see that for $\epsilon \rightarrow \infty $, $\Delta \mu$ diverges linearly with $\epsilon$ while $a$ approaches its limit with a power-low correction:
\beq
\Delta \mu \approx 2 \, \epsilon-1+o(1) \, , \ \ \ \epsilon  \rightarrow \infty
\eeq

\beq
a \approx a_\infty+ \frac{c_1}{\epsilon}\, ,\ \  \epsilon  \rightarrow \infty
\eeq
where

\beq
c_1=-\frac{(a_\infty-1) a_\infty^{3/2}}{2
   \sqrt{a_\infty}+(a_\infty-1) \log
   \left(\frac{2}{\sqrt{a_\infty}-1}+1\right)}.
\eeq

\comment{

Note that the equation on the whole border becomes
\beq
2\,\delta_{t'\Tau}  = \left. {d \over dt} C(t,t')\right|_{t=\Tau}+C(\Tau,t')+\Delta \mu \,C(\Tau,t')+2 \epsilon \, C(-\Tau, t')
\eeq
These can be written as two equations for $\cosh a^{1/2} (\Tau-t')$ and  $\sinh a^{1/2} (\Tau-t')$,  transforming  $C(-\Tau, t')$ with the formulas for addition and subtraction of trigonometric functions.
This implies that the solution can be found imposing the value of $C(-\Tau,\Tau)$ and then one needs the two equations to determine $\Delta \mu$ and $\epsilon$.

In the general case,   at fixed $\epsilon$ you have an equation that fixes $\Delta \mu$ and one that fixes $C(-\Tau,+\Tau)$. Than the equation for $\hR_2(-\Tau,+\Tau)$ on the border evaluated at $t'=\pm \Tau$ gives two additional equations that fix the values of  $\hR_2(-\Tau,+\Tau)$ and $\hX(-\Tau,+\Tau)$.
}

For the observable corresponding to different trajectories we have:
\beqa
\hR_1^\mathrm{dt}(t,t') & = & {1 \over 2} \,C^\mathrm{dt} (t,t')
\\
\hR_2^\mathrm{dt}(t,t') & = & {1 \over 2}\, C^\mathrm{dt} (t,t')
\\
\hX^\mathrm{dt}(t,t') & = &  {1 \over 4}\,C^\mathrm{dt}(t,t') 
\eeqa
and $C^\mathrm{dt}(t,t')$ obeys

\beq
-{1 \over 2}{d^2 \over dt^2} C^\mathrm{dt}(t,t')+ \left({1 \over 2}+\hat{\mu}\right) C^\mathrm{dt}(t,t')  =  0,
\eeq
which is to be solved with boundary conditions $C^\mathrm{dt}(-\Tau,t)=C^\mathrm{dt}(t,-\Tau)=C(-\Tau,t)$ and the conditions on the borders:
\begin{widetext}
\beq
0 =  \left. {d \over dt} C^\mathrm{dt}(t,t')\right|_{t=\Tau}+C^\mathrm{dt}(\Tau,t')+\Delta \mu \, C^\mathrm{dt}(\Tau,t')- 2 \, \epsilon \, C^\mathrm{dt}(-\Tau, t')
\eeq
The solution then reads:
\beq
C^\mathrm{dt}(t,t')=\frac{e^{-\sqrt{a} t} \left(\sqrt{a}
   \cosh \left(\sqrt{a}
   t'\right)-\sinh
   \left(\sqrt{a}
   t'\right)\right)
   \left(\left(\sqrt{a}+\Delta \mu
   +1\right) e^{2 \sqrt{a} t_\mathrm{f}}-2
   \epsilon  e^{\sqrt{a} t_\mathrm{f}}+2
   \epsilon  e^{\sqrt{a} \left(t_\mathrm{f}+2
   t\right)}+\left(\sqrt{a}-\Delta
   \mu -1\right) e^{2 \sqrt{a}
   t}\right)}{\sqrt{a}
   \left(\left(\sqrt{a}+\Delta \mu
   +1\right) e^{2 \sqrt{a}
   t_\mathrm{f}}+\sqrt{a}-\Delta \mu
   -1\right)}
\eeq
and we obtain in the limit $t_\mathrm{f} \rightarrow \infty$
\beqa
C^\mathrm{dt}(\Tau,\Tau) & \rightarrow  &   {\epsilon^2 \over {1 \over 4} (1+ \sqrt{1+4 \epsilon^2})^2}= \epsilon^2+ O(\epsilon^4)  
\eeqa
again in agreement with the static FP result. 

\end{widetext}

\section{Complexity and Irreversibility of Fibers}~\label{sec::CompFib}

This appendix presents static calculations concerning the complexity of fibers and their associated irreversibility thresholds. As discussed in the main text, at each final overlap $q_\mathrm{f}$, the static measure is dominated by the lowest free-energy fibers. We have also argued that in the limits $N\rightarrow\infty$ and  $t_\mathrm{f}\rightarrow\infty$, the irreversible dynamics (for $q_\mathrm{f} < q_\mathrm{irr}$) is controlled solely by a small number of extremely deep fibers that develop a saddle at $q_\mathrm{irr}$ and subsequently relax into a metastable state, \emph{the hub}.

However, for finite $N$ and $t_\mathrm{f}$, the heterogeneity among fibers is fundamentally important. In what follows, we characterize these higher free-energy fibers, which are exponentially numerous in $N$ and exhibit a broad distribution of irreversibility thresholds (saddles) and associated metastable states (hubs). In order to evaluate the number of such fibers, we apply the Monasson formalism~\cite{monasson1995structural} to the FP potential. Instead of optimizing over the rupture variable $x$ (which corresponds to the Parisi block size), we treat it as a parameter to probe the complexity. Note that the FP potential imposes a radial constraint on these free-energy fibers, effectively turning them into minima of the potential. 

There exists an exponential (in $N$) number of such minima $\mathcal{N}$, corresponding to a positive complexity, $\Sigma = \frac{1}{N} \log \mathcal{N}$.
We here compute the complexity $\Sigma$ of these minima (fibers) as a function of their typical energy $E_\mathrm{f}$, basing our analysis on the 1RSB formulation of the FP potential
\begin{widetext}
\begin{equation}\label{eq:V1rsb}
    V^{(2)}_\mathrm{1RSB} = -\frac{1}{2\beta} \left[
2 \beta^2 f(q_\mathrm{f}) 
+ \beta^2 (x - 1) f(q_1) 
- \beta^2 x f(q_0) 
+ \log(1 - q_1) 
+ \frac{1}{x} \log\left(1 + \frac{x (q_1 - q_0)}{1 - q_1} \right) 
+ \frac{q_0 - q_\mathrm{f}^2}{1 - q_1 + x (q_1 - q_0)}
\right]
\end{equation}
and corresponding energy 
\begin{equation}
E^{(2)}_\mathrm{1RSB} = -\frac{1}{2} \left[ 
2 \beta f(1) 
+ 2 \beta f(q_\mathrm{f}) 
+ 2 \beta (x - 1) f(q_1) 
- 2 \beta x f(q_0) 
\right] \ .
\end{equation}
\end{widetext}
This potential must be extremized with respect to $q_0$ and $q_1$ at a given $x$, obtaining $V_\mathrm{1RSB}(x; q_\mathrm{f})=\text{extr}_{q_0,q_1}  V^{(2)}_\mathrm{1RSB}$. The complexity and relative (free-)energy of the state then follow as (see, e.g., \cite{folena2020mixed})
\begin{equation*}
\begin{cases}
\Sigma_{q_\mathrm{f}}(x) = x^2 \frac{d}{dx} V_\mathrm{1RSB}(x; q_\mathrm{f})\\
f_{q_\mathrm{f}}(x) = \frac{d}{dx} \left( x  V_\mathrm{1RSB}(x; q_\mathrm{f}) \right)\quad \text{or}\quad E_{q_\mathrm{f}}(x) = E_\mathrm{1RSB}(x; q_\mathrm{f})
\end{cases}
\end{equation*}

\begin{figure*}[t]
\centering
\includegraphics[width=0.49\textwidth]{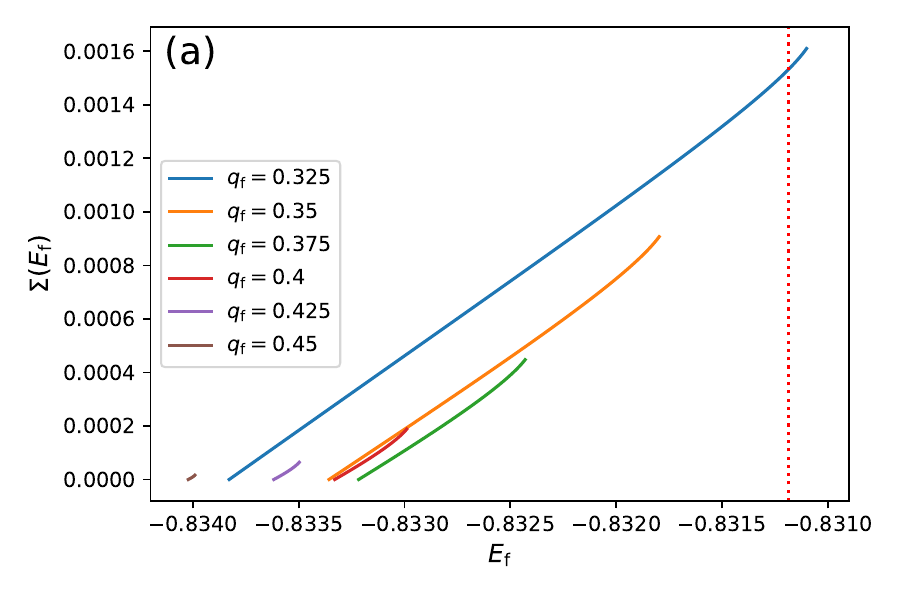}
\includegraphics[width=0.49\textwidth]{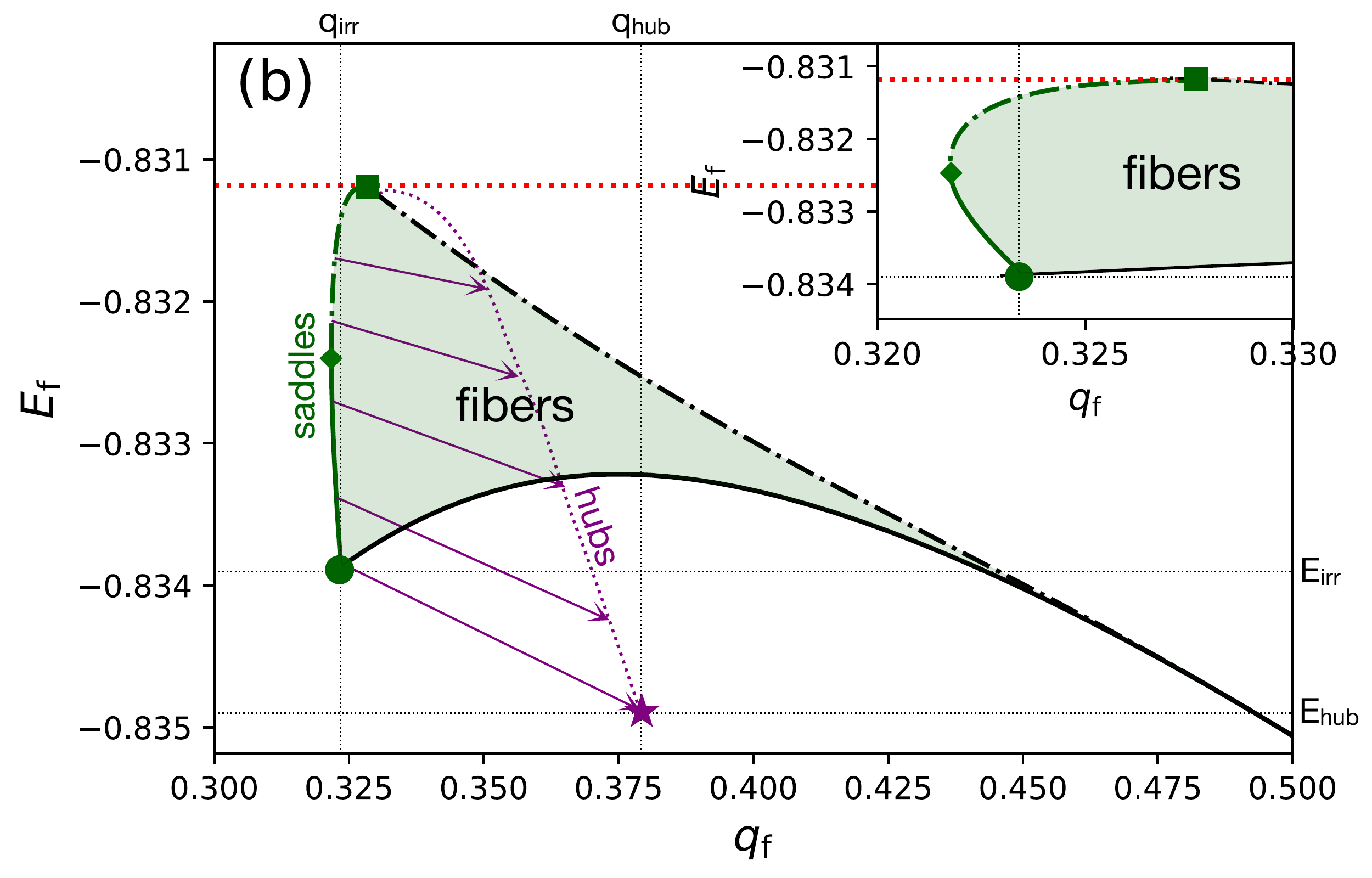}
\caption{\textbf{(a)} Fiber complexity as a function of the energy of each fiber, parametrically in $x$, for different final overlaps in the range $q_\mathrm{irr} < q_\mathrm{f} < q_\mathrm{mg}$, for the $3$-spin spherical model -- with $f(q) \equiv q^3/2$ -- at $T = 1/1.695$. Note that for $q_\mathrm{f} = 0.325$, the complexity curve intersects the threshold energy $E_{\mathrm{th}}$ (red line, see text). \textbf{(b)} The regime of reversible, finite-complexity fibers (light green) emerges between roughly $q_\mathrm{irr}$ and $q_\mathrm{mg}$, between the lowest (solid black line) and the highest (dash-dotted line) free-energy fibers. The latter is given by the marginality condition in Eq.~\eqref{eq:marg}. The onset of saddle points (green line) is where the $M_2$ minimum first appears in the three-replica potential. Within the light-green regime, fibers remain reversible, and hence the BF dynamics relaxes back to the reference equilibrium state. By contrast, initiating BF dynamics from the saddles relaxes toward asymptotic metastable states or hubs (dotted purple line). The deepest of these metastable states -- the dominant hub -- (purple star) is reached from the lowest free-energy fibers (large green dot) and is expected to dominate the dynamics in the limits $N \to \infty$ and $t_\mathrm{f} \to \infty$.
}
\label{fig:ComplIrr}
\end{figure*}

Figure~\ref{fig:ComplIrr}(a) shows the complexity as a function of the energy of each fiber, parametrically in $x$. 
Interestingly, the right endpoint of each curve is determined by the marginality condition
\begin{equation}\label{eq:marg}
    -1 + \beta^2 (1 - q_1)^2 f''(q_1) = 0,
\end{equation}
but we have not explicitly verified the full set of 1RSB stability conditions. It is important to emphasize that this calculation does not follow individual fibers across different $q_\mathrm{f}$. Whether the isocomplexity fibers at different $q_\mathrm{f}$ are the same or whether an “overlap chaos" of bifurcating and merging fibers emerges therefore remains an open question. 

For $q_\mathrm{f} = 0.325$, the complexity curve intersects the threshold energy $E_{\mathrm{th}}$, at which point the unconstrained landscape becomes dominated by free-energy saddles. At first sight, this observation might seem contradictory, but recall that the FP potential imposes a local spherical constraint and therefore artificially stabilize these saddles. As we will show below, the minima that lie above this threshold in fact correspond to fibers that have already crossed their irreversibility threshold and therefore no longer support reversible dynamics.

To identify the irreversibility threshold of the fibers, we employ the three-replica potential introduced in Ref.~\cite{cavagna1997}, whose second minimum $M_2$ corresponds to the asymptotic state reached by the Barrat--Franz (BF) dynamics~\cite{Barrat1998}. As a simple extension, instead of considering the second replica $\sigma$ as lying within a typical state (i.e., the deepest free-energy fiber) at distance $q_\mathrm{f}$\footnote{Notice that this overlap is called $q_{12}$ in Refs.~\cite{cavagna1997,Barrat1998} and $q_\mathrm{f}$ is called $\overline{q}$} from the reference configuration, we allow it to reside in an atypical, higher free-energy fiber.




For the first two replicas, we again  extremize Eq.~\eqref{eq:V1rsb} for a given $x$. The third replica potential follows from~\cite[Eqs.~(3.10–3.12)]{cavagna1997}. However, since we are only interested in locating the emergence of the $M_2$ minimum, we consider the RS version of Eq.~(3.10), with the additional simplification $w_{23} = q_{23}$. Under this assumption, the potential becomes
\begin{widetext}
\begin{equation}
V^{(3)}_{\mathrm{RS}} = 
-\frac{1}{2\beta} \left[ 2 \beta^2 f(q_{13}) + 
2 \beta^2 x f(q_{23}) - 
2 \beta^2 x f(z_{23}) + 
\beta^2 (-1) f(s_1) + 
\log(1 - s_1) + 
\frac{s_1 - y}{1 - s_1}\right]
\end{equation}
 with corresponding energy
 \begin{equation}
 E^{(3)}_{\mathrm{RS}} = -\frac{1}{2} \left[
2 \beta f(1) + 
2 \beta f(q_{13}) + 
2 \beta x f(q_{23}) - 
2 \beta x f(z_{23}) - 
2 \beta f(s_1)
\right],
 \end{equation}
 and Eq.~(3.11) becomes
 \begin{equation}
 y =  q_{13}^2 
+ \frac{2 q_\mathrm{f} q_{13} x (q_{23} - z_{23})}
       {-1 + q_1 + q_0 x - q_1 x} 
+ \frac{x (q_{23} - z_{23}) 
         \left( q_{23} - q_{23} q_1 + q_{23} (q_\mathrm{f}^2 - 2 q_0 + q_1) x 
         + z_{23} + q_1 (-1 + x) z_{23} - q_\mathrm{f}^2 x z_{23} \right)}
       {\left( 1 + q_1 (-1 + x) - q_0 x \right)^2}.
 \end{equation}
\end{widetext}
For any given $q_\mathrm{f}$ and $x$, we first determine the parameters $q_0$ and $q_1$ by extremizing the two-replica potential $V^{(2)}_\mathrm{1RSB}$. These values are then used to partially evaluate the variable $y$, which is subsequently inserted into the three-replica potential $V^{(3)}_{\mathrm{RS}}$. This potential is then extremized with respect to the remaining variables $q_{13}, q_{23}, z_{23}, s_1$.

Figure~\ref{fig:ComplIrr}(b) presents the regime of reversible, finite-complexity fibers. Following the above procedure, we find that a non-trivial solution $M_2$ -- a second minimum -- with $q_{13} \neq q_\mathrm{f}$ emerges for $q_\mathrm{f}$ to the left of the green line. Interestingly, two distinct mechanisms underlie the emergence of $M_2$. As better seen in the inset, the dash-dotted line corresponds to the marginality condition
\begin{equation}
    -1 + \beta^2 (1 - s_1)^2 f''(s_1) = 0,
\end{equation}
while the solid green line denotes a first-order transition.
The dynamical distinction between these two types of transitions within the Barrat--Franz framework remains an open question and is left for future investigation. 

For now, we interpret the line marking the appearance of $M_2$ as the set of points (overlap vs energy) at which fibers reach a saddle point. This point corresponds to the onset of dynamical irreversibility. We further conjecture that the lower portion of this line (solid curve) corresponds to index-1 saddles. After crossing this saddle, each fiber flows -- following the Barrat--Franz construction -- toward its corresponding asymptotic metastable \emph{hubs} state, identified by the overlap $q_{13}$ and the energy of the $M_2$ minimum. As noted in the main text, these metastable states have a higher overlap with the reference configuration $\tau$ than their associated saddles, implying that the unstable direction of the index-1 saddles must be orthogonal to the radial direction toward $\tau$. It would be interesting to explore whether this somewhat surprising \emph{hub geometry}, with fibers bending back toward $\tau$, is a general feature of RFOT models or merely a peculiarity of the present one.

We also observe that the first fibers to reach their saddle points are not the deepest (lowest energy) ones, but rather the highest ones (green square in the inset). This featuree complicates the dynamical interpretation. If we constrain the dynamical potential with $q_\mathrm{f}>q_\mathrm{irr}$ but lower than the overlap of these early fibers, we expect the system to be irreversible, but such irreversibility would be atypical (exponentially suppressed in $N$). It remains unclear how this \emph{fiber geometry} affects the inverted limit $t_\mathrm{f}\to\infty$ followed by $N\to\infty$, and whether  threshold states could dominate as a result.

A related calculation was carried out in Ref.~\cite{ros2019b}, though at the level of the energy landscape, where the reference configuration corresponds to an energy minimum. Comparing Fig.~\ref{fig:ComplIrr}(b) with their Fig.~1, we observe a correspondence between our $(q_\mathrm{irr}, E_\mathrm{irr})$ and their $(q^*, \epsilon^*)$. However, we also note a key difference. Aside from very high energy fibers, in our case the first index-1 saddles appear at the lowest energies as the overlap decreases, whereas in their case saddles appear at a mid-range energy $\epsilon_M$. Understanding the origin of this inversion remains an open question.

Finally, in the model we considered -- based on static calculations -- we expect the relaxation dynamics to be dominated by the deepest fibers. This suggests that, up to the irreversibility threshold, the system explores a relatively simple basin structure, with only a few escape fibers, not too different from ferromagnetic basins. It would therefore be very interesting to find evidence (perhaps in some mixed $p$-spin models) of a phase in which the fibers dominating the measure are not the deepest ones, even in the thermodynamic limit. In such a phase, the dominant fibers would have finite complexity, implying an exponential number of escape directions and thus a much richer and even more intricate dynamical behavior.

\end{document}